\documentclass[a4paper,11pt,DIV=13,abstract=true,numbers=noenddot,titlepage=false,twocolumn=false,draft=false]{scrartcl}
\pdfoutput=1

\makeatletter
\DeclareOldFontCommand{\rm}{\normalfont\rmfamily}{\mathrm}
\DeclareOldFontCommand{\sf}{\normalfont\sffamily}{\mathsf}
\DeclareOldFontCommand{\tt}{\normalfont\ttfamily}{\mathtt}
\DeclareOldFontCommand{\bf}{\normalfont\bfseries}{\mathbf}
\DeclareOldFontCommand{\it}{\normalfont\itshape}{\mathit}
\DeclareOldFontCommand{\sl}{\normalfont\slshape}{\@nomath\sl}

\usepackage[usenames,dvipsnames]{color}
  \definecolor{hgreen}{rgb}{0,0.3,0}
  \definecolor{hred}{rgb}{0.3,0,0}
  \definecolor{hblue}{rgb}{0,0,0.3}
  \definecolor{LightGray}{gray}{0.95}
  \definecolor{gray}{gray}{0.6}
\usepackage[
	    colorlinks=true,
	    linkcolor=hblue,
	    citecolor=hgreen,
	    filecolor=hblue,
	    urlcolor=hred
	    ]{hyperref}
\usepackage{mathrsfs}
\usepackage[intlimits]{amsmath}
\usepackage{amssymb}
\usepackage{slashed}
\usepackage[titletoc,title]{appendix}
\usepackage[affil-it]{authblk}
\usepackage{libertine}
\usepackage[numbers,sort&compress]{natbib}
\usepackage{graphicx}
\usepackage{microtype}
\usepackage{float}
\usepackage[table]{xcolor}

\setcapindent{1em}
\setkomafont{captionlabel}{\bfseries}
\setkomafont{caption}{\itshape}
\usepackage{fancyhdr}

\fancypagestyle{preprint}{
\fancyhead[LE,RO]{CP3-Origins-2019-38 DNRF90 \\ TTP19-036 \\ DO-TH 21/22}

}

\usepackage{comment}
\numberwithin{equation}{section}

\newcommand{\Verr}[3]{#1^{+#2}_{-#3}}

\newcommand{\DOS}{\cellcolor{blue!7} }
\newcommand{\OS}{\cellcolor{red!7} }
\newcommand{\SOS}{\cellcolor{green!7} }
\newcommand{\BACK}{\cellcolor{gray!7} }

\newcommand{\Lag}{\mathscr{L}}
\newcommand{\Vc}{\mathbb{V}}
\newcommand{\V}{{\cal V}}
\newcommand{\Vp}{ \bar{\cal V} }

\newcommand{\matel}[3]{\langle #1|#2|#3\rangle}
\newcommand{\al}{\alpha}
\newcommand{\be}{\beta}
\newcommand{\ga}{\gamma}
\newcommand{\de}{\delta}
\newcommand{\la}{\lambda}
\newcommand{\eps}{\epsilon}

\newcommand{\GeV}{\mbox{GeV}}
\newcommand{\MeV}{\mbox{MeV}}

\newcommand{\state}[1]{|#1\rangle}
\newcommand{\vev}[1] {\langle #1 \rangle}

\newcommand{\FTVb}{\overline{T}_\perp}
\newcommand{\FTAb}{\overline{T}_\parallel}

\newcommand{\LL}{{\mathbb{ L}}}

\newcommand{\RR}{{R}} 
\newcommand{\mBq}{ m_{B_q}  }

\newcommand{\fBq}{ f_{B_q}  }
\newcommand{\Corr}{\Pi}

\newcommand{\ksub}{k_0^2}
\newcommand{\ksubtwo}{k_1^2}
\newcommand{\kOPE}{2}
\newcommand{\kmat}{\omega}

\newcommand{\pB}{p} 

\newcommand{\PFF}{P}

\newcommand{\FP}{V_P}
\newcommand{\FL}{V_\LL}
\newcommand{\FV}{V_\perp}
\newcommand{\FA}{V_\parallel}
\newcommand{\FVA}{V_{\perp,\parallel}}

\newcommand{\FTL}{T_\LL}
\newcommand{\FTV}{T_\perp}
\newcommand{\FTA}{T_\parallel}

\newcommand{\mi}{\!-\!}
\newcommand{\bP}{b_{\rm P}}
\newcommand{\bV}{b_{\rm V}}
\newcommand{\bT}{b_{\rm T}}

\newcommand{\qsqlow}{(4.9 \GeV)^2}

\newcommand{\TAB}{Table}
\newcommand{\FIG}{Figure}

\begin{document}
\renewcommand\Authands{, }


\title{{\boldmath
Flavoured Axions in the  Tail  of $B_{q}\to\mu^+\mu^-$} \\ and $B \to \ga^*$ Form Factors}


\date{\today}
\author[a]{Johannes Albrecht%
        \thanks{\texttt{johannes.albrecht@cern.ch}}}
\author[a,b]{Emmanuel Stamou%
\thanks{\texttt{emmanuel.stamou@tu-dortmund.de}}}
\author[c,d]{\\ Robert Ziegler%
        \thanks{\texttt{robert.ziegler@cern.ch}}}
\author[e]{Roman Zwicky%
        \thanks{\texttt{roman.zwicky@ed.ac.uk}}}

\affil[a]{{\large Fakult\"at f\"ur Physik, TU Dortmund, D-44221 Dortmund, Germany}}
\affil[b]{{\large Institut de Th\'eorie des Ph\'enomenes Physiques, EPFL, Lausanne, Switzerland}}
\affil[c]{{\large Theoretical Physics Department, CERN, 1211 Geneva 23, Switzerland}}
\affil[d]{{\large Institut f\"ur Theoretische Teilchenphysik, KIT, 76131 Karlsruhe, Germany}}
\affil[e]{{\large Higgs Centre for Theoretical Physics, School of Physics and Astronomy, \newline
University of Edinburgh, Edinburgh EH9 3JZ, Scotland
}}

\maketitle
\thispagestyle{preprint}

\begin{abstract}
We discuss how LHC di-muon data collected to study $B_q \to
\mu \mu$ can be used to constrain light particles with flavour-violating
couplings to $b$-quarks. Focussing on the case of a
flavoured QCD axion, $a$,
we compute the decay rates for $B_q \to \mu \mu a$ and the
SM background process $B_q \to \mu \mu \gamma$ near the kinematic endpoint.
These rates depend on non-perturbative $B_q \to \gamma^{(*)}$ form factors
with on- or off-shell photons.
The off-shell form factors ---relevant for generic searches for beyond-the-SM particles---
are discussed in full generality and computed with QCD sum rules for the first time.
This includes an extension to the low-lying resonance region using a multiple subtracted dispersion relation.
With these results, we analyse available LHCb data  to
obtain the sensitivity on $B_q \to \mu \mu a$ at present and future
runs. We find that the full LHCb dataset alone will allow to probe
axion-coupling scales of the order of $10^6$ GeV for both $b\to d$ and
$b \to s$ transitions. As a spin-off application of the off-shell form factors we further
analyse the case of light, Beyond the Standard Model, vectors. 

\end{abstract}
\newpage

\pdfbookmark[1]{Table of Contents}{tableofcontents}
\setcounter{page}{1}

\tableofcontents

\section{Introduction and Motivation \label{sec:intro}}

Open questions in particle physics and cosmology may well be addressed
by very light particles that interact only feebly
with the Standard Model (SM). The prime example is the QCD
axion~\cite{WW1,WW2}, which is not only predicted within the Peccei--Quinn (PQ)~\cite{PQ1,PQ2}
solution to the strong CP Problem, but which can also explain the Dark Matter abundance
if it is sufficiently lighter than the meV scale~\cite{AxionDM1, AxionDM2, AxionDM3}.
In the past years much activity has been devoted towards experimental searches for
the QCD axion, and multiple proposals for new experiments are underway to
complement ongoing efforts to discover the axion, see Ref.~\cite{axionsearches} for a review.

While most axion searches rely on axion couplings to photons, the axion also couples
to SM fermions if they are charged under the PQ symmetry. Generically, these charges
constitute new sources of flavour violation, which induce flavour-violating axion
couplings to fermions, which can thus be probed by precision flavour experiments.
For instance, this situation arises naturally when the PQ symmetry is identified with
a flavour symmetry that shapes the hierarchical structure of the SM
Yukawas~\cite{Wilczek, Axiflavon, Japs, NardiFred}, therefore, connecting the strong CP
problem with the SM flavour puzzle. Even in the absence of such a connection, axion
models with flavour non-universal PQ charges can be easily constructed and motivated
by, e.g., stellar cooling anomalies that require suppressed axion couplings to
nucleons~\cite{astrophobic, astrophobic2, Saikawa:2019lng}.

In the absence of explicit models, the couplings of the axion to different flavours
are \emph{a priori} unrelated, and are parametrised by a model-independent
effective Lagrangian for Goldstone bosons.
The flavour-violating couplings in the various quark and lepton
sectors can then be constrained by experimental data, see Ref.~\cite{CPVZZ} for a recent
assessment of the relevant bounds in the quark sector using mainly hadron decays with
missing energy. In this article we explore a novel direction to
probe flavour-violating axion couplings involving $b$-quarks using the present and future
LHC data collected to study $B_q \to \mu \mu$.

We therefore focus on flavour-violating
$b \to q$ transitions, which are described by the Lagrangian
\begin{align}
	\Lag & = \frac{\partial_\mu a}{2 f_a} \overline{b} \gamma^\mu \left( C^V_{b q} + C^A_{b q} \gamma_5 \right) q +\text{h.c.}
	\equiv  \partial_\mu a \, \overline{b} \gamma^\mu \left( \frac{1}{F^V_{b q}} + \frac{\gamma_5}{F^A_{b q}}  \right)  q +\text{h.c.}\,,
\label{eq:Leff}
\end{align}
where $F_{bq}^{V/A}$ are  parity odd/even couplings, $q = d,s$ and $a$ denotes the
derivatively coupled QCD axion, whose mass is
inversely proportional to the axion decay constant, $f_a$, which suppresses all axion couplings.
The decay constant has to be much larger than the electroweak scale to
sufficiently decouple the axion from the SM in order to satisfy experimental
constraints~\cite{Peccei:1988ci, Turner:1989vc}.
This implies that the axion is light, with a mass much below an eV,  and stable even on
cosmological scales.

Therefore, two-body $B$-meson decays with missing energy,
which closely resemble the very rare SM decays with
final-state neutrino pairs that have been looked for at B-factories,
stringently constrain the couplings in Eq.~\eqref{eq:Leff}.
The resulting constraints on the vector couplings $F_{bq}^{V}$ (from $B \to K/\pi a$ decays)
and the axial-vector couplings $F_{bq}^{A}$ (from $B \to K^*/\rho a$ decays and $B_q$ mixing)
have been given in Refs.~\cite{CPVZZ} (see also Refs.~\cite{Murayama, King}) and are summarised
in \TAB~\ref{tab:presentbounds}.
\begin{table}[h]
\centering
\begin{tabular}{cll}
	& $\boldsymbol{F^V_{b q}}$ [GeV] & $\boldsymbol{F^A_{b q}}$ [GeV]  \\[0.2em]
\hline\hline\\[-0.8em]
$\boldsymbol{bd}$  & $1.2 \cdot 10^{8}$ ($B \to \pi a$)  & $4.8 \cdot 10^{6}$ ($B-\bar{B}$ mixing)\\
$\boldsymbol{bs}$  & $3.1 \cdot 10^{8}$ ($B \to K a$)    & $1.3 \cdot 10^{8}$ ($B \to K^* a$)\\
\hline
\end{tabular}
\caption{\label{tab:presentbounds}
Lower bounds on $F^{V,A}_{b q}$  at 90\% CL from $B$-decays and $B_q$-mixing, taken from Ref.~\cite{CPVZZ}.}
\end{table}
Note that constraints from neutral meson mixing are typically much weaker than the ones from
decays to vector mesons, except in the case of $b\to d$ transitions. This is mainly
due to the lack of experimental data on $B \to \rho \nu \overline{\nu}$ suitable for the
two-body recast.

In the present work, we investigate whether the couplings in
Eq.~\eqref{eq:Leff} can also be constrained at the LHC.
To this end, we propose to use the three-body decays $B_{s,d} \to \mu\mu a$,
where the muon pair originates from an off-shell photon, cf., \FIG~\ref{fig:dia-FF} (left).
With the main goal of measuring the SM decay $B_q \to \mu\mu$, the
ATLAS~\cite{Aaboud:2018mst}, CMS~\cite{Chatrchyan:2013bka} and LHCb~\cite{Aaij:2017vad}
collaborations have collected di-muon events with an invariant mass $q^2$ down to
roughly $(5 \GeV)^2$. As long as no vetos on extra particles in the event are applied,
these datasets can be used to constrain decays with additional particles in the final
state, e.g., the radiative decay $B_q \to \mu \mu \gamma$, as proposed in Ref.~\cite{Dettori:2016zff}. In this paper we focus on the LHCb potential although 
CMS and to some degree ATLAS are capable of a similar study albeit 
less sensitivity as can be inferred from the combined result \cite{CMS:2020rox}.
Whereas Belle II has an exciting physics program it is not competitive in very 
rare decays \cite{Belle-II:2018jsg}.
Here, we point out that the same datasets can be used to constrain the decays
$B_q \to \mu \mu X$, where $X$ is a neutral, beyond-the-SM (BSM) particle with a mass
that is sufficiently small to be kinematically allowed at the tail of $B_q\to\mu\mu$, i.e.,
$m_X \lesssim m_{B_{q}} - 5 \GeV \approx 300 \MeV$.
In this respect, the radiative decay $B_q \to\mu\mu\gamma$ merely constitutes a SM
background, which we take into account in our analysis. 
Note that the axion hypothesis \eqref{eq:Leff} itself has a negligible effect on $B_q\to\mu\mu\gamma$ 
and thus the SM prediction of that decay is considered.
In particular, we suggest that when the measurement of
$B_q\to\mu\mu\gamma$ becomes feasible in the future, it can be directly
interpreted in terms of constraining BSM particles that
replace the final state photon.
A similar strategy can be applied to $s \to d$ transitions, using for example the di-muon
data collected  at LHCb to study $K_S \to \mu \mu$, cf., Ref.~\cite{Junior:2018odx},
and possibly also  to  $c \to u$ transitions, i.e., $D \to \mu \mu$~\cite{Aaij:2013cza}.

In the following we focus on the case of the invisible QCD axion, $a$, but
our analysis can be readily extended to other particles appearing in the final
state, as long as they are not vetoed in the event. In particular
these could be
heavy axions decaying within the detector, i.e.,
axion-like particles (ALPs).
We expect such an analysis to be fully inclusive, that is, independent of the ALP decay mode.
Similarly our proposal can  be extended to constrain light vectors with
flavour-violating couplings, e.g., dark photons or $Z'$s.
We explore these scenarios in 
  Appendix \ref{app:BVllrate}. 
In this article we demonstrate the key elements of the analysis and perform
the first sensitivity studies based on the published dataset of
the LHCb collaboration.
The ATLAS and CMS data can be analysed analogously.

The photon off-shell form factors are  necessary for predicting branching fractions
of $B_q \to \ell\ell X$ where $X$ is any of the above mentioned light BSM particles.
We discuss the complete set of form factors, relevant for the dimension-six effective Hamiltonian,
compute them with QCD sum rules (SRs) and fit them to  a $z$-expansion. In addition
the off-shell basis is shown to be related to the standard
$B \to V = \rho^0,\omega,\phi  \hdots$ basis through a dispersion representation,
which interrelates many properties of these two sets of form factors.

This article is organised as follows: In Section~\ref{sec:BllXrate} we provide the
differential rates for the axionic decay  $B_q\to\mu\mu a$ and the radiative
decay $B_q\to\mu\mu\gamma$.
In Section~\ref{sec:sensLHCb} we provide the tools necessary to perform the
analysis and use available background estimates and data from LHCb's
$B_s\to\mu\mu$ measurement to evaluate the sensitivity to $B_q \to \mu \mu a$
at present and future runs.
We conclude in Section~\ref{sec:conclusions}.
Appendix~\ref{app:FF}  contains the computation of the  $B \to \ga^{(*)}$ form factors
as well as their extension through a dispersion relation  to the region of low lying vector mesons.
In Appendix~\ref{app:BVllrate} we provide the differential rate for $B \to \ga^*(\to \ell^+ \ell^-) \Vc $ 
and apply the proposed analysis.

\section{ Differential Decay Rates\label{sec:BllXrate}}
\begin{figure}[t]
	\centering
	 \includegraphics[]{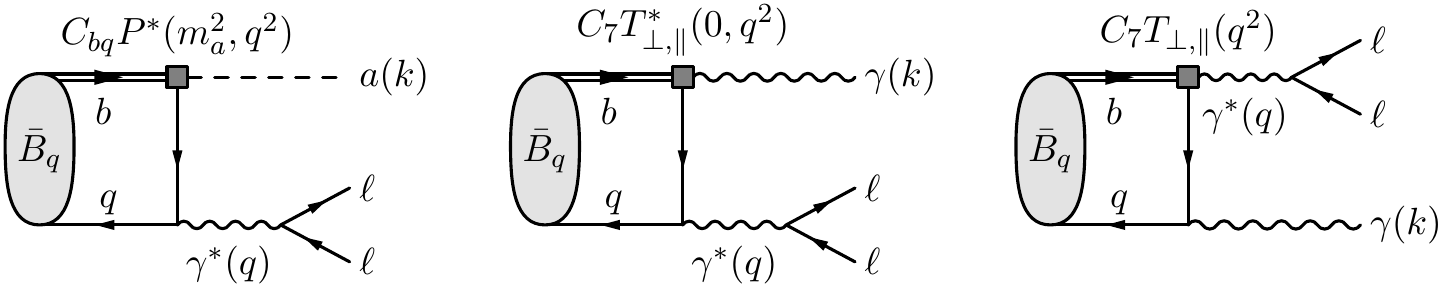}
	\caption{
	The diagram to the left is the main axion process $B_q \to  \ell \ell a$ whereas the
	two diagrams in the centre and the right belong to the $B_q \to  \ell \ell \ga$ background.
	The single and double lines stand for the $q$ and $b$-quark, respectively.
	The left and central diagrams depend on off-shell form factors
	in the sense that the photon that emits the two leptons is off-shell.
	Diagrams in which the photon couples to b-quarks are not shown, but are analogous.
	Also diagrams with $Q_{9,10}$-operator insertions are not shown, and
	resemble the diagram on the right and are proportional to $C_9 \FV$
	and $C_{10} \FA$.
	\label{fig:dia-FF}}
\end{figure}

In this section we calculate the differential rates for
the axionic $B_q\to \ell \ell a$ and radiative $B_q\to \ell \ell \gamma$ decay channels.
In \FIG~\ref{fig:dia-FF}, we show on the left the diagram for the  axionic decay and in the
centre and on the right representative diagrams for the radiative decay.
The rates are differential in the lepton-pair
momentum $q \equiv p_{\ell^+} +  p_{\ell^-}$, and depend on non-perturbative $B_q \to \gamma^{(*)}$
form factors with on- or off-shell photons, which we briefly introduce before presenting
the differential decay rates.
Finally, we evaluate the rates close to the kinematic endpoint $\qsqlow \lesssim q^2 < m_{B_q}^2$,
and compare our prediction for the radiative decay to results in the literature.

\subsection{Summary on $B_q \to \gamma^{*}$ form factors \label{sec:FF}}

We describe $B_q(p_B) \to \gamma^*(k)$ transitions with off-shell photons by a set of form factors
with two arguments $F^*(q^2,k^2) \equiv F^{B \to \ga^*}(q^2,k^2)$.
The first argument (here $q^2$) denotes the momentum transfer at the flavour-violating
vertex while the second argument (here $k^2$) denotes the momentum of the photon.
For on-shell photons, i.e. $k^2 = 0$, these form factors reduce to the well-known on-shell form factors
$F(q^2) \equiv F^*(q^2,0)$ given in Eq.~\eqref{eq:OnS}.
A complete set of form factors is given by\footnote{\label{foot:bar} The scalar form factor, $\matel{ \gamma^*(k,\rho)} {  \bar q  b} {\bar{B}_q (p_B)} $,
vanishes due to parity conservation of QCD.}$^{,}$
\footnote{It is important to keep in mind that these  form factors for the $\bar B$-meson. 
E.g. if one assumes the phase convention  $C\state{\bar{B}} = \state{B}$, where $C$ is the charge transformation, 
then the form factors, 
$(V_{\perp},T_{\perp})^{ B \to \ga^*} = - (V_{\perp},T_{\perp})^{ \bar B \to \ga^*}$, change sign but and all others do not. The same holds true for a CP-transformation 
assuming $CP \state{\bar{B}} = \state{B}$. The adaption of phases under the discrete transformation is straightforward.}
\begin{alignat}{3}
\label{eq:FFsec2}
& M_{5}^\rho (q, k) &\; \equiv\;& \bP \matel{ \gamma^*(k,\rho)} {  \bar q \gamma_5 b} {\bar{B}_q (p_B)}  &\;=\;&  i
\mBq \RR^\rho  \, \PFF^*(q^2,k^2)  \;, \nonumber   \\[0.1cm]
& M^{\mu\rho}_V  (q, k)  &\; \equiv\;& \bV \matel{\gamma^*(k,\rho)} { \bar{q} \gamma^{\mu}   b }{ \bar{B}_q (p_B)}
 &\;=\;&    R_\perp^{\mu\rho}  \, V_\perp ^*(q^2,k^2)  \;, \nonumber   \\[0.1cm]
  & M^{\mu\rho}_A  (q, k) &\; \equiv\;&  \bV   \matel{\gamma^*(k,\rho)} { \bar{q} \ga^{\mu}  \ga_5  b }{ \bar{B}_q (p_B)}
 &\;=\;&      R_\parallel^{\mu\rho}  \, V_\parallel ^*(q^2,k^2)   + \RR^{\mu\rho}_\LL V_\LL^*(q^2,k^2)  +
   \frac{2\RR^{\mu\rho}_P }{\hat{q}^2}   \PFF^*(q^2,k^2) \;, \nonumber  \\[0.1cm]
 & M^{\mu\rho}_T  (q, k) &\; \equiv\;& \bT \matel{\gamma^*(k,\rho)} { \bar{q} iq_\nu \sigma^{\mu \nu}    b }{ \bar{B}_q (p_B)}
 &\;=\;&     R_\perp^{\mu\rho} \, T_\perp ^*(q^2,k^2)  \;,   \nonumber  \\[0.1cm]
  & M^{\mu\rho}_{T_5}  (q, k) &\; \equiv\;& \bT \matel{\gamma^*(k,\rho)} { \bar{q} iq_\nu \sigma^{\mu \nu}  \ga_5   b }{ \bar{B}_q (p_B)}
 &\;=\;&  - (R_\parallel^{\mu\rho}  \, T_\parallel^*(q^2,k^2)  +   \RR^{\mu\rho}_\LL T_\LL^*(q^2,k^2))   \;,
 \end{alignat}
where $\hat{q}^2 \equiv q^2/m_B^2$ throughout,
$q \equiv p_B - k$ denotes the momentum transfer at the flavour-violating vertex, and
we define the off-shell photon state $\langle \ga^*(k,\rho)|$ as in Eq.~\eqref{eq:offshellgamma}.
The coefficients
\begin{align}
\bP & \equiv \frac{m_b + m_q}{s_e e}  \, ,  &
\bV & \equiv  -\frac{\mBq}{s_e e} \, , &
\bT & \equiv  \frac{1}{ s_e e} \, ,
\end{align}
depend on the sign convention, $s_e$, for the covariant derivative $D_\mu = \partial_\mu + s_e i Q_f  e A_\mu$.
The Lorentz tensors $\RR$  are defined by 
 \begin{alignat}{3}
\label{eq:RNEWmain}
& R_\perp^{\mu \rho}  &\;\equiv\;&   \varepsilon^{\mu \rho \be \ga }  q_\be k_\ga
 \; ,& R_\parallel^{\mu \rho }   \equiv\;&  \frac{i}{2}  (1-\hat{q}^2) \,  (\mBq^2  G^{ \mu \rho} - \frac{ (q+ 2k)^{\mu}\RR^{\rho}}{1-\hat{k}^2} ) \; ,
\nonumber \\[0.1cm]
& R_\LL^{\mu\rho} &\;\equiv\;&  \frac{i}{2}  ( q^\mu - \frac{\hat{q}^2(q+ 2k)^{\mu}}{1-\hat{k}^2} ) \, \RR^\rho\; ,  \qquad
&R_P^{\mu\rho} \equiv\;& \frac{i}{2}  q^\mu \RR^\rho   \;, \quad \quad \RR^\rho \equiv  q^\rho - \frac{k\!\cdot\!q}{k^2}k^\rho \;,
\end{alignat}
where $G_{\al\be} \equiv g_{\al \be} - k_\al k_\be/k^2 $ 
 and  (kinematic) hatted quantities are divided by $m_{B_q}^2$, e.g.  $\hat{k}^2 \equiv k^2/m_B^2$.
The matrix element  satisfy  the  QED and the axial Ward identities
\begin{align}
k_\rho M^{\rho}_5 (q, k) = 0 \;, \quad k_\rho M^{\mu\rho}_{\rm V,A, T,T_5} (q, k) & =  0  \, , & q_\mu\, M^{\mu\rho}_{\rm A} (q, k) & = \mBq M_{5}^\rho (q, k) \;.
\end{align}
The latter implies the relation between the pseudoscalar and one of the axial form factors  which
reduces the number of independent form factors down to a total of seven.
At $q^2 =0$ there are two further constraints
\begin{equation}
\label{eq:first}
\PFF^*(0,k^2) = \hat{V}^*_\LL(0,k^2)  \;, \quad
\FTA^*(0,k^2)  = ( 1 -  \hat{k}^2 ) \FTV^*(0,k^2)  \;,
\end{equation}
where $ \hat{V}^*_\LL(q^2,k^2)  \equiv - \hat{q}^2/2  V^*_\LL(q^2,k^2)$
thereby reducing the form factors down to five.
An extensive discussion including dispersion representations in the $q^2$ and $k^2$ variables,
the derivation of Eq.~\eqref{eq:first}, the limit  to photon on-shell  form factors,
and their computation from QCD SRs  are deferred to Appendix~\ref{app:FF}.
As an example let us quote the once-subtracted dispersion representation for the pseudoscalar form factor for
\begin{alignat}{2}
\label{eq:exampleP}
& P^{B_s \to \ga^*}(q^2,k^2) &\;=\;&  P^{B_s \to \ga^*}(q^2,\ksub)  - (k^2-\ksub) \int_{\textrm{cut}}^{\infty}
\frac{du \, \textrm{Im}_u[ P^{B_s \to \ga^*}(q^2,u)   ] } {(u-\ksub)(u-k^2-i0)}   \\[0.1cm]
& &\;\stackrel{\ksub \to 0}{=}\;&
  -k^2 \int_{\textrm{cut}}^{\infty}
 \frac{du \, \textrm{Im}_u[ P^{B_s \to \ga^*}(q^2,u)   ] } {u (u-k^2-i0)}
=  -  k^2  \left( \frac{2}{\mBq}  \frac{  f^{\textrm em}_\phi \, A_0^{B_s \to \phi}(q^2)   }{(m_{\phi}^2 - k^2)}   + \dots  \right)   \;,  \nonumber
\end{alignat}
which simplifies for the subtraction point $\ksub =0$.  Above $f_\phi^{\textrm{em}}$ is the electromagnetic decay constant in the Appendix \ref{app:dispersion} and $A_0$ is the pseudoscalar $B_s \to \phi$ form factors \cite{BSZ15}.
  In the last step the dispersion relation was
evaluated for the lowest term and the dots stand for terms higher in the spectrum
e.g.  $B_s \to \phi'$  and multiple particle configurations.

The off-shell form factors in the limit of small momentum transfer at the flavour-violating vertex,
$T_{\perp,\parallel,\LL}^*(0,q^2)$, $V_{\perp,\parallel,\LL}^*(0,q^2)$ and $P^*(0,q^2)$
are computed in this work for the first time.\footnote{\label{foot:before}
The weak annihilation process, that is $B \to V \ga^*$ four-quark matrix elements where the $B$ valence quarks annihilate,
contain some of these form factors as sub processes.
Weak annihilation has been computed in the SM to leading order (LO) in QCD factorisation \cite{Beneke:2001at}
and including all BSM operators in LCSR  \cite{Lyon:2013gba}.
However, the discussion in our paper is
more complete as even the BSM computation in Ref.~\cite{Lyon:2013gba} does not include all form factors  since the $V$-mesons
do not couple to scalar operators for instance.
Moreover in Ref.~\cite{Kozachuk:2017mdk}, the off-shell form factor $T^*_\perp(0,k^2) = F_{TV}(0,k^2)$
is evaluated using a  vector-meson-dominance approximation.} 
The  $B \to \ga^*(\to \ell^+ \ell^-) \Vc $ differential rate, presented in Appendix~\ref{app:BVllrate}, 
is another example application of these off-shell form factors for searches beyond the SM.
The extension of the off-shell form factors to low $k^2$ into the resonance region, as 
shown in \eqref{eq:exampleP}, requires the matching to the QCD  SR result. The discussion thereof 
is delegated to Appendix~\ref{app:mutlidisp}  and involves, by choice, a multiple subtracted dispersion relation 
where additionally use is made of the on-shell form factors. Plots are shown in \FIG~\ref{fig:FFplots}.
An ancillary Mathematica notebook is added to the arXiv version for reproducing the form factors. 
This is for example relevant for the prediction of $B \to \ga \ell\ell$ as the two
 form factors $T^*_{\perp,\parallel}(0,k^2)$, where $k^2$ is the dilepton mass, enter the description. These two form factors have also been considered in  
\cite{Melikhov:2004mk,Kozachuk:2017mdk,Beneke:2020fot} using 
improved vector meson dominance. 

For the on-shell form factors $B \to \ga$ we use the next-leading-order (NLO) light-cone sum rule (LCSR) computation \cite{Janowski:2021yvz}.
Note that the QCD SR result of the off-shell form factors can be used in the relevant kinematic region
$\qsqlow \lesssim q^2  < \mBq^2$ since thresholds are far away.
The photon on-shell form factors are more challenging in this region because the light-cone expansion
breaks down. They can, however, be extrapolated to this region by using a
$B_q^*$ and $B_{q1}$-pole ansatz, with the residue computed from LCSR \cite{Pullin:2021ebn},
 supplemented with  $z$-expansion corrections to account for further states.

\subsection{The $B_q\to\ell\ell a$ differential rate \label{sec:Bllarate}}

Given the effective Lagrangian in Eq.~\eqref{eq:Leff},
the amplitude for $\bar{B}_q (p_B) \to \ell^+ (p_{\ell^+})~\ell^-(p_{\ell^-})~a(k)$ is\footnote{Notice the interchanged role of $k$ and $q$ with respect to the definition of the form factors in Eq.~\eqref{eq:FFsec2}.}
\begin{align}
{\cal A}_{\mu \mu a} &
= - i \frac{e^2 Q_\ell}{\mBq q^2}~\frac{k_\mu}{F^A_{bq}}~M^{\mu\rho}_A (k, q) ~\bar u^s(p_{\ell^-})\gamma_\rho v^{r}(p_{\ell^+})
= - i \frac{e^2 Q_\ell}{F^A_{bq} q^2}~M^{\rho}_{5}(k, q)~\bar u^s(p_{\ell^-})\gamma_\rho v^{r}(p_{\ell^+}) \, , \end{align}
where ${\cal A}_{\mu \mu a} \equiv \matel{\mu\mu a}{-\Lag_{\textrm{int}} }{\bar{B}_q}$,
$q \equiv p_B - k = p_{\ell^+} +  p_{\ell^-}$ and $Q_\ell = -1$ denotes the lepton charge.
After squaring this amplitude, summing over fermion spins, and integrating over the unobserved axion momentum,
the differential rate in the invariant mass of the final-state leptons, $q^2$, becomes
\begin{equation}
\frac{d \Gamma}{d q^2}(B_q \to  \ell \ell a) =
\frac{\alpha^2 }{48\pi \mBq} \frac{ \la^{1/2}_\ga \, (\la^{(a)}_{B_q} )^{3/2} }{|F_{bq}^A|^2} \frac{2 m_\ell^2 +q^2}{q^8}
|\PFF^*(m_a^2,q^2)|^2\;,
\label{eq:GaBsmumua}
\end{equation}
where
$\la_\ga \equiv \la(q^2,m_\ell^2,m_\ell^2)$, $\la^{(a)}_{B_q} \equiv\la(\mBq^2,  q^2  ,m_a^2)$,
and $\la(x,y,z) \equiv x^2+y^2+z^2 - 2x y -2 x z -2 y z$ is the K\"all\`en function.
For our work it is sufficient to approximate $m_a \to 0$.

\subsection{The $B_q\to\ell\ell\gamma$ differential rate \label{sec:Bllgammarate}}
The relevant part of the effective SM Lagrangian\footnote{By including the factor $s_e$ in the definition of the operators
$Q_7, Q_7^{\prime}$ we ensured that the sign of their Wilson coefficients,
$C_{7}^{\textrm{SM}} < 0$ is independent of the definition of the 
covariant derivative.}
\begin{equation*}
	\Lag_\text{SM} = \frac{4 G_{\rm F}}{\sqrt{2}} \la_t \sum_{i=7,9,10}\left( C_i Q_i + C_i^\prime Q'_i \right)
	+\text{h.c.}\,,
\end{equation*}
 where $\la_t \equiv V_{tb}^*V_{tq} $ and
\begin{align}
	Q_7       	=& \frac{s_e e}{16\pi^2}m_b \bar b \sigma^{\mu\nu} q_L F_{\mu\nu} \, , &
	Q_{9[10]}       	=& \frac{e^2}{16\pi^2} (\bar b \gamma^\mu q_L)(\bar \ell \gamma_\mu [\ga_5] \ell)\, ,
\end{align}
$q_{L,R} \equiv (1\mp \ga_5)/2 q$,  and $Q'_{7,9,10}$ is obtained from $Q_{7,9,10}$ by the replacements $L \to R$ and $m_b \to m_q$.
Upon using  the helicity vector completeness relation \eqref{eq:completeness}, the
$\bar{B}_q (p_B) \to \ell^+ (p_{\ell^+})~\ell^-(p_{\ell^-})~\gamma(k)$ amplitude ($q \equiv p_B -k = p_{\ell^+} +  p_{\ell^-}$, $\la_{B_q} \equiv\la(\mBq^2,  q^2  ,0) $ and $r_ q \equiv  -i s_e e/2 \,  \la^{1/2}_{B_q}$)
\begin{align}
	{\cal A}_{\mu \mu \ga(\la)} & =  \matel{\mu\mu\ga(\la)}{-\Lag_{\textrm{SM}}}{\bar B_q} =
	    \frac{\alpha G_{\textrm{F}} \la_t }{ \sqrt{2} \pi  }  r_q	  \sum_{\la = \perp,\parallel} ( {\cal A}^{V}_\la L^V_\la + {\cal A}^{A}_\la L^A_\la)\;,
\end{align}
can be factored into
a leptonic
 $L^{V,A}_\la = \eps^*_{\nu}(q,\la)  \bar u(p_{\ell^-})  \ga^{\nu} [\ga_5] \bar v(p_{\ell^+})$
and the hadronic helicity amplitude follows from the Hamiltonian with where the leptons are removed
up to a normalisation factor. For the effective axial leptons one finds 
\begin{alignat}{2}
&  r_q {\cal A}^{A}_{\pm} &\;=\;& 
 \matel{ \ga(\pm)   }{\frac{(C_{10} \pm C'_{10})}{2}\bar s \ga^\mu 1(-\ga_5)  b}{\bar B_q} \eps^{\mu}(q,\pm)     
  \nonumber  \\
 &  &\;=\;& \pm \frac{(C_{10} \pm C'_{10})}{2 b_V} M^{V(A)}_{\mu\rho}(q,k) \eps^{*\mu}(k,\pm) \eps^{\rho}(q,\pm)  
  \;,
\end{alignat}
and for the vector ones
\begin{alignat}{2}
& r_q     {\cal A}^{V}_{\pm} &\;=\;& 
r_q   {\cal A}^{A}_{\pm}\Big |_{C_{10} \to C_{9}}  
 - \frac{2 Q_\ell m_b }{q^2}
  \left( \pm \frac{(C_{7} \pm \frac{m_q}{m_b} C'_{7})}{2 b_T} \overline{M}^{T_{(5)}}_{\mu\rho}(q,k) \eps^{*\mu}(k,\pm) \eps^{\rho}(q,\pm) \right)
\end{alignat}
where $ \overline{M}^{T_{(5)}}_{\mu\rho}(q,k)   \equiv  M^{T_{(5)}}_{\mu\rho}(q,k)  + M^{T_{(5)}}_{\mu\rho}(k,q)$ takes into account both types of diagrams in \FIG~\ref{fig:dia-FF}.
 Explicit parameterisations of the polarisation vectors and some more explanations 
can be found in Appendix \ref{app:FFgaon}.
Using the (common) convention $\sqrt{2}{\cal A}_{\perp,\parallel} = {\cal A}_+ \mp {\cal A}_-$ 
 one gets  ($ Q_{\ell} = -1$)
\begin{equation}
\label{eq:ampAV}
\begin{split}
 {\cal A}^V_{{\perp,\parallel}} = &
   \frac{1}{\mBq}  (C_{9} \pm C'_{9})  V_{\perp,\parallel}^*(q^2,0)
+  \frac{2 m_b}{q^2}  (C_7 \pm \frac{m_q}{m_b} C_7') \overline{T}_{\perp,\parallel}(q^2)
\;,\\[0.5em]
 {\cal A}^A_{\perp,\parallel} =&   \frac{1}{\mBq}  (C_{10} \pm C'_{10})  V_{\perp,\parallel}^*(q^2,0) \, .
\end{split}
\end{equation}
 Above
\begin{equation}
\begin{split}
\FTVb(q^2) & =  \FTV^*(q^2,0) +  \FTV^*(0,q^2) \, ,  \\
\FTAb(q^2) & =  \FTA^*(q^2,0) +  \FTA^*(0,q^2)/(1-q^2/\mBq^2) = \FTA^*(q^2,0) +  \FTV^*(0,q^2) \, .
\end{split}
\label{<+label+>}
\end{equation}
The last equality relates $\FTA^*(0,q^2)$ to $\FTV^*(0,q^2)$,
see Appendix~\ref{sec:algebraic} and footnote \ref{foot:JG} just before Eq.~\eqref{eq:NEW}.
Above we omitted the contribution from photons radiated off final-state muons,
because these are obtained from the $B_q \to \mu \mu$ rates using {\small \tt PHOTOS}, cf.,
Ref.~\cite{Bobeth:2013uxa}.
Going slightly lower in $q^2$ would
necessitate the inclusion of broad charmonium resonances \cite{GRZ17,Lyon:2014hpa}.
For an overview of other non form-factor matrix elements see for
instance Refs.~\cite{GRZ17,Kozachuk:2017mdk}.

After integrating over the unobserved photon momentum, the differential rate for the radiative
mode $B_q\to\ell\ell \ga$ reads
\begin{equation}
\label{eq:GaBsmumuga}
\frac{d \Gamma}{d q^2} (B_q \to \ell \ell \ga) =
\frac{\al^3 G_F^2 |\la_t|^2}{ 768 \pi^4}
\frac{ \la^{1/2}_\ga  \la^{3/2}_{B_q} } { \mBq^3 q^2}
\left(
 c_A ( | {\cal A}^A_{\perp}|^2 +  |{\cal A}^A_{\parallel}|^2)
+ c_V  ( | {\cal A}^V_{\perp}|^2 +  |{\cal A}^V_{\parallel}|^2)
\right) \;,
\end{equation}
 and  $c_V \equiv ( q^2+ 2m_\ell^2) $, $c_A \equiv(q^2-4 m_\ell^2)$ are effectively the
 squared leptonic helicity amplitudes.

\subsection{$B_q\to\mu\mu a$ and $B_q\to\mu\mu\gamma$ close to the
kinematic endpoint \label{sec:BmmXratenum}}
\begin{figure}[t]
	\centering
	\includegraphics{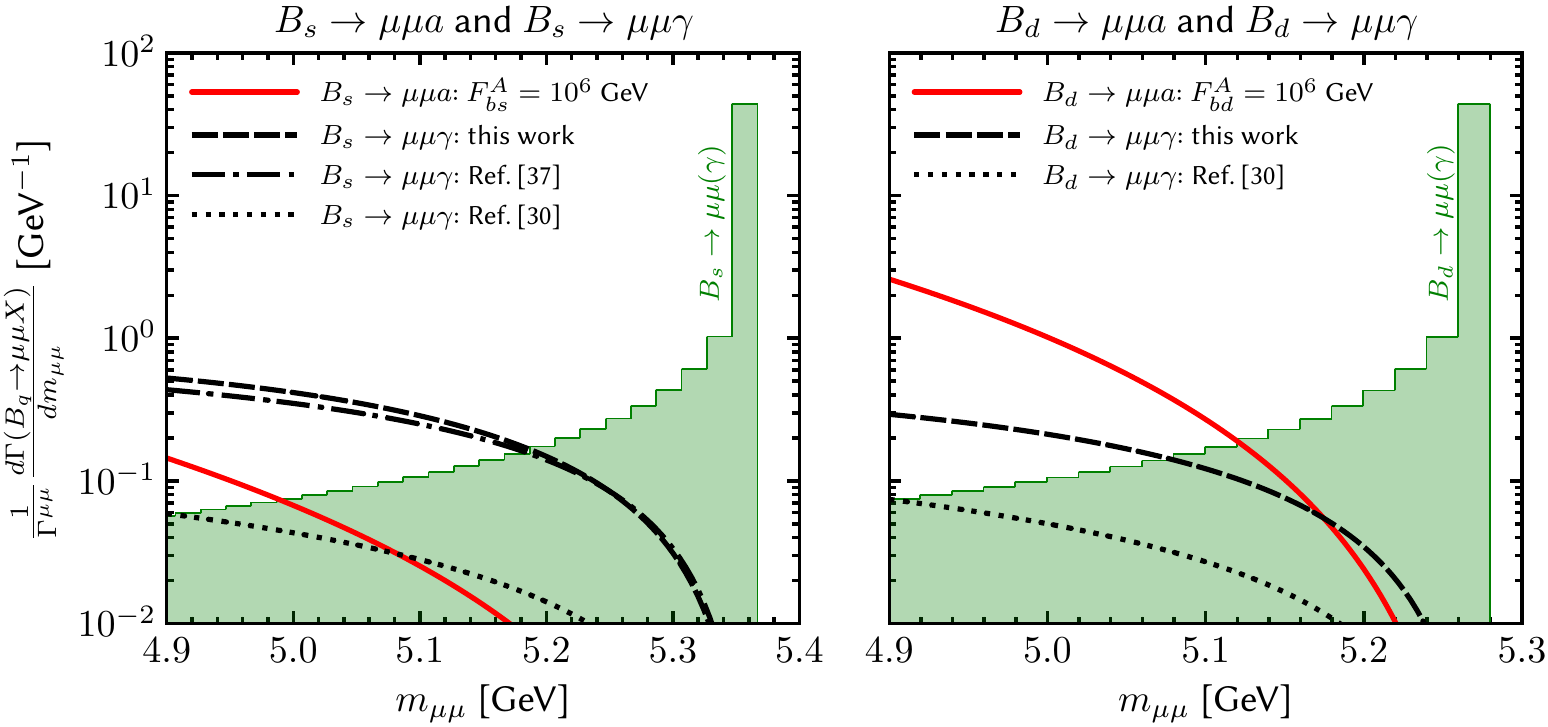}
	\caption{Comparison of the axionic  decay mode $B_q\to \mu\mu a$ (red solid lines)
		and the radiative $B_q\to\mu\mu\gamma$ modes (black lines).
		The left panel shows the $B_s$ case while the right the $B_d$ case.
		For the axion predictions $F_{bq}^A=10^6$~GeV is assumed as a reference value.
		The different black lines are the photon predictions with different form factor
		treatments (see legend and main text).
		In green are bins of the two-body $B_q\to\mu\mu$ rate including
		radiation from final-state muons.
		To better compare the $B_s$ and $B_d$ cases, all rates are normalised to their
		respective two-body decay $B_q\to\mu\mu$, which is why the $B_d\to \mu\mu a$
		line appears enhanced with respect to the $B_s\to \mu\mu a$ one.
		\label{fig:dBRpredictions}}
\end{figure}

To illustrate the relative importance between the SM background $B_q\to\mu\mu\gamma$
and the $B_q\to\mu\mu a$ signal we take as a reference value for the flavour-violating coupling
$F^A_{bq} = 10^6$\,GeV.
In \FIG~\ref{fig:dBRpredictions}, we show the
differential rate normalised with respect to the two-body decay width
\begin{equation*}
	\frac{1}{\Gamma(B_q\to\mu\mu)} \frac{ d \Gamma(B_q\to \mu\mu X)}{d m_{\mu\mu}}\,,
\end{equation*}
where $X=a,\gamma$, $m_{\mu\mu}^2 \equiv q^2$.
In the left panel, we show the predictions
for the $B_s$ decays and in the right the corresponding ones for the $B_d$ case.
The binned (green) predictions are the $B_q\to\mu\mu$ rates including photon radiation
from the final-state muons using {\small \tt PHOTOS} (see Ref.~\cite{Bobeth:2013uxa}).
The red solid lines are the rates from the axion mode for the reference value $F_{bq}^A=10^6$~GeV
(note that the relative enhancement between left and the right panel
is due to the normalization, which carries a different CKM suppression.).
The black lines are the $B_q\to\mu\mu\gamma$ predictions when the photon does not
originate from muon bremsstrahlung.
They depend on the treatment of the non-perturbative input, i.e., the hadronic form factors
introduced in Section~\ref{sec:FF}. In all cases, we use the same perturbative input,
namely the SM Wilson coefficients
$C_7^{\text{eff}}$, $C_9^{\text{eff}}$ and $C_{10}$ evaluated at the hadronic $B_q$ scale.
We obtain $C_{10}$ from Ref.~\cite{Bobeth:2013uxa} and
use {\tt \small flavio} \cite{Straub:2018kue} to evaluate $C_7^{\text{eff}}$ and $C_9^{\text{eff}}$.

We show the results of three different approaches
of estimating the relevant hadronic form factors:
\begin{itemize}
	\item {\bf Dashed line:} the QCD SR form factor computation discussed in Section~\ref{sec:FF} and Appendix~\ref{app:FF},
	\item {\bf Dotted line:} the quark-model approach of Ref.~\cite{Kozachuk:2017mdk},
\item {\bf Dashed-dotted line:} the pole-dominance approach supplemented by experimental
	data and heavy-quark effective theory of Ref.~\cite{Aditya:2012im}.
	It is specific to the $B_s$ case (left panel).
\end{itemize}
The agreement of the predictions is rather crude.
For $q^2 \approx (4.9 \GeV)^2$, our prediction is about a factor of three larger than the quark model \cite{Kozachuk:2017mdk}
and about a factor of two smaller than the pole-dominance approximation \cite{Aditya:2012im}.
The disagreement with the quark model is not surprising
as the method is designed for low $q^2$ and, unlike in our work, no additional input
is employed to constrain the residua of the leading poles near the kinematic endpoint.
The agreement of the form factors themselves at lower $q^2$, which we do not show,
is much better.
The comparison with the pole-dominance approach \cite{Aditya:2012im} has two major components.
The difference in the $B^*_q$-residue and the fact that the effect of $B_{q1}$-resonance is
neglected in  Ref.~\cite{Aditya:2012im} cf.\ Appendix \ref{app:poles}.
While it is important to understand\footnote{
Whereas it will be challenging for lattice QCD to
compute off-shell form factors, the on-shell
ones have gained attention and computations are in progress \cite{Kane:2019jtj,Sachrajda:2019uhh}.}
the origin of the discrepancy in light of a possible measurement of the radiative decay,
the discrepancy does not play a significant role in obtaining a bound on the axion couplings
$F^A_{bq}$, which we derive in the next section.

\section{Sensitivity at LHCb \label{sec:sensLHCb}}
In this section we recast the LHCb analysis of Ref.~\cite{Aaij:2017vad}
to obtain an estimate for the current and future sensitivity of LHCb
to probe the flavour-violating couplings $F^A_{bs}$ and $F^A_{bd}$.
We first discuss, in Section~\ref{sec:rescaling}, how we extract
the backgrounds by rescaling the original LHCb analysis, and derive the
expected number of events in each bin for a given luminosity. We then
describe, in Section~\ref{sec:sensitivity}, our statistical method and provide
the recast of the present data and the sensitivity study
for future runs. Our main results are summarised in Tables~\ref{tab:presentrecastbounds}
and  \ref{tab:futurebounds}.

\subsection{Rescaling the LHCb analysis\label{sec:rescaling}}

The $B_s\to\mu\mu$  analysis of LHCb in Ref.~\cite{Aaij:2017vad} makes
use of datasets collected at different LHC runs, with luminosities
$\overline{{\cal L}}_7 = 1.0$~fb$^{-1}$ from $7$ TeV,
$\overline{{\cal L}}_8 = 2.0$~fb$^{-1}$ from $8$ TeV, and
$\overline{{\cal L}}_{13} = 1.4$~fb$^{-1}$ from $13$ TeV runs.
Under the SM hypothesis, a total number of
$62$ $B_s\to\mu\mu$  events and $6.7$ $B_d\to\mu\mu$ events
are expected in this  analysis
in the full range of  boosted-decision-trees (BDT)
and the signal window ($m_{\mu\mu}\in[5.2, 5.445]$ GeV).
Since the BDT discrimination is flat one expects half of these events to pass the BDT~$>0.5$ selection.
For this BDT selection, LHCb supplies a plot with backgrounds,
which we use to extract their numerical values.
By combining the expected number of $B_q\to\mu\mu$ events in the SM with the SM
branching-fraction predictions, we extract a universal rescaling factor, $r\approx 0.079$, via
\begin{equation}
	\begin{split}
N_{B_d} &= \underbrace{(\epsilon\,2 f_{d})}_{\equiv r} \times ~\overline{\text{BR}}^{B_d\to\mu\mu(n\gamma)}_{[5.2~\text{GeV}-5.445\,\text{GeV}]}\times
\!\!\sum_{i=7,8,13}\sigma_{b,i} \overline{\cal L}_i\,,\\
N_{B_s} &= r \times \frac{f_s}{f_d} \times ~\overline{\text{BR}}^{B_s\to\mu\mu(n\gamma)}_{[5.2~\text{GeV}-5.445\,\text{GeV}]}\times
\!\!\sum_{i=7,8,13}\sigma_{b,i} \overline{\cal L}_i\,.
	\end{split}
	\label{eq:NqLHCb}
\end{equation}
In these equations, $i$ labels the $\sqrt{s}$ run and
$\sigma_i$ is the corresponding $b$-quark production cross section
in the acceptance of LHCb.
The latter has been measured by LHCb for $\sqrt{s}=7,13$ TeV, $\sigma_{b,7}  = 72$~$\mu$b and $\sigma_{b,13}  = 144$~$\mu$b \cite{Aaij:2016avz}.
For $\sigma_{b,8}$ we linearly rescale the $7$\,TeV value ($\sigma_{b,8} = 8/7 \sigma_{b,7}$).
$f_d$ and $f_s$ are the fragmentation ratios of $b$-quarks that are produced at LHCb and fragment
into $B_d$ and $B_s$, respectively.
We absorb $f_d$ in the rescaling factor, $r$, and use the ratio $f_s/f_d$ to obtain
$N_{B_s}$.
This ratio   has been measured by the LHCb collaboration to be $f_s/f_d= 0.259\pm0.015$ \cite{LHCb:2013lka}.
Finally, $\epsilon$ summarises the experimental efficiencies and all other
global rescaling factors, which we absorb into the definition of $r$.

The quantities $\overline{\text{BR}}$'s in Eq.~\eqref{eq:NqLHCb}
are  the respective  branching ratios in the signal window.
This includes the effect of photon radiation from muons \cite{Buras:2012ru,Bobeth:2013uxa},
which LHCb simulates with {\small \tt PHOTOS}.
The overline in the branching-ratio prediction indicates
that the partial width is divided by the width of the heavy mass
eigenstate ($\Gamma^H_{B_s},~\Gamma^H_{B_d}$) to obtain the branching fraction.
In this way the effect of $B_q$-mixing is included \cite{Buras:2013uqa,Bobeth:2013uxa}.
This is  relevant for the $B_s$ system, but much less so for the $B_d$ system.
This is numerically equivalent to LHCb's treatment of the effective lifetime,
cf. Eq.~(1) in Ref.~\cite{Aaij:2017vad}).

LHCb's BDT~$>0.5$ selection covers the $m_{\mu\mu}\in[4.9~\text{GeV}, m_{B_s}]$ region
in bins of $50$~MeV.
We apply the same universal rescaling factor, $r$, to rescale the
predictions of all $B_q\to\mu\mu a$ and $B_q\to \mu\mu\gamma$
branching fractions for all $m_{\mu\mu}$ bins.
This is a good approximation as there are no triggers or similar
thresholds that significantly change the rescaling over this invariant-mass range.
In the next section, we present the sensitivity of this analysis to probe
the flavour-violating $F^A_{bs}$ and $F^A_{bd}$ axion couplings
in future runs of LHCb by rescaling the $13$\,TeV dataset.
We denote the corresponding effective total luminosity by
\begin{equation}
	{\cal L} = \overline{\cal L}_7 +  \overline{\cal L}_8 + {\cal L}_{13}\,.
\end{equation}
At a given total luminosity, ${\cal L}$,
the expected number of events at a given $m_{\mu\mu}$-bin (Bin$_k$) then is
\begin{equation}
	\begin{split}
	N_{k}[F^A_{bs},&F^A_{bd}] = N_{\text{Bin}_k}^{\text{BKG,analysis}} \frac{\text{SL}({\cal L})}{\text{SL}(\overline{\cal L})}\\
	&+ \left(
		  \overline{\text{BR}}_{\text{Bin}_k}[B_d\to\mu\mu(n\gamma)]
		 +\overline{\text{BR}}_{\text{Bin}_k}[B_d\to\mu\mu\gamma]
	         +\overline{\text{BR}}_{\text{Bin}_k}[B_d\to\mu\mu a]\right)\,r\,\text{SL}({\cal L})\\
        &+ \left(
		  \overline{\text{BR}}_{\text{Bin}_k}[B_s\to\mu\mu(n\gamma)]
		 +\overline{\text{BR}}_{\text{Bin}_k}[B_s\to\mu\mu\gamma]
	 +\overline{\text{BR}}_{\text{Bin}_k}[B_s\to\mu\mu a]\right)\,r\,\frac{f_s}{f_d}\text{SL}({\cal L})\,,
 \label{eq:Ntot}
 \end{split}
\end{equation}
with shorthands $\overline{\cal L} \equiv \overline{\cal L}_7 +  \overline{\cal L}_8 + \overline{\cal L}_{13} = 4.4 \, {\rm fb}^{-1}$
and
$\text{SL}({\cal L}) \equiv \sigma_{b,7}   \overline{\cal{L}}_7 +
                      \sigma_{b,8}   \overline{\cal{L}}_8 +
		      \sigma_{b,13} ( {\cal L} - \overline{\cal{L}}_7 -\overline{\cal{L}}_8)$\,.
The quantity $N_{\text{Bin}_i}^{\text{BKG,analysis}}$ is the expected total number of background
events that do not originate from the radiative decay in the given bin.
We obtain $N_{\text{Bin}_i}^{\text{BKG,analysis}}$
by digitising and integrating the plot of LHCb's BDT~$>0.5$ selection.
In Eq.~\eqref{eq:Ntot} we kept separate the rate from photon emission from muons
($B_q\to\mu\mu(n\gamma)$) and the rate from photon emissions from
the initial state ($B_q\to\mu\mu \gamma$).
In principle, the amplitudes interfere but the interference is tiny
close to the $B_q$ threshold and we thus neglect it.

\subsection{Recast and sensitivity analysis\label{sec:sensitivity}}
To compute the sensitivity of the LHCb analysis in probing $F^A_{bs}$ and $F^A_{bd}$,
we must combine the information of all $m_{\mu\mu}$ bins
and include statistical and systematic uncertainties.
We neglect the subdominant experimental systematic uncertainties but will include the
theory uncertainties associated to the form factors entering
the three-body rates.
In what follows we always either turn on $F^A_{bs}$ or $F^A_{bd}$, i.e.,
but will not let them float simultaneously.

Each $m_{\mu\mu}$ bin corresponds to an independent counting experiment
that obeys Poisson statistics.
Exclusion limits on $F^A_{bq}$ are then obtained from a joined Poisson (Log)Likelihood.
For a sufficiently large number of events, Poisson statistics are well described
by Gaussian statistics and
the Poisson (Log)Likelihood is equivalent to a $\chi^2$ function of the NP parameter, i.e., $F^A_{bq}$:
\begin{equation}
	\chi^2(F^A_{bq}) = \sum_{i,j}(N_i - N^{\text{obs}}_i) (V_{\text{cov}}^{-1})_{ij} (N_j-N^{\text{obs}}_j)\,,
\end{equation}
with $i$ numbering the bins and $q=s,\,d$.
$N_i = N_i(F^A_{bq})$ denotes the total number of events (background plus signal)
for the value $F^A_{bq}$ in a given bin,
whereas $N^{\text{obs}}_i$ is the observed number of events.
For the recast we use the actual number of events observed by LHCb, read off from Figure 1 in Ref.~\cite{Aaij:2017vad}.
To project the sensitivity for future LHCb runs we set $N^{\text{obs}}_i$
to the number of events expected in the SM.
The covariance matrix, $V_{\text{cov}}$, incorporates
statistical and systematic uncertainties in a way that we discuss below.
If we neglect systematic uncertainties, this matrix is diagonal and only contains the
squared Poisson variances, $V_{\text{cov}} = V_{\text{stat}}$ with
$(V_{\text{stat}})_{ij} = \delta_{ij} N_i$.
We have explicitly checked, that for the data samples considered here,
the Poisson (Log-)Likelihood is always very well approximated by the $\chi^2$.

To incorporate systematic/theory uncertainties we follow the commonly used approach
of Ref.~\cite{Cousins:1991qz}.
Theory uncertainties are then treated as Gaussian uncertainties smearing
the expectation values of the underlying Poisson probability distribution functions.
We can then obtain the limits on $F^A_{bq}$ by generating Monte-Carlo events
based on the joined Poisson likelihood after smearing the expectation
values by the (correlated) systematic errors.
If the measurement is well-described by Gaussian statistics (as in our case) and
the systematic uncertainties are small with respect to the statistical ones,
this treatment of uncertainties is equivalent to adding
the statistical and systematic errors in quadrature in $V_{\text{cov}}$.

In our case the main systematic uncertainties are due to the
form factors that enter the radiative $B_q\to\mu\mu\gamma$ and
the $B_q\to\mu\mu a$ rate.
Since the uncertainties in the form factors  originate in part from uncertainties
in input parameters like $m_b$ and $\vev{\bar q q } $ that are $q^2$-independent,
the predicted number of events among different bins are correlated.
Therefore, the full covariance matrix for the case in which the axion has a coupling $F^A_{bq}$
is not diagonal and decomposes into
\begin{align}
	V_{\text{cov}} =
	V_{\text{stat}}
	+ V_{\gamma}
	+ \frac{1}{(F^A_{bq})^4} V^q_{a}
	+ \frac{1}{(F^A_{bq})^2} V^q_{a-\gamma}\,.
	\label{<+label+>}
\end{align}
Here, $(V_{\text{stat}})_{ij} = \delta_{ij} N_i$ are the statistical uncertainties,
while the matrices $V_{\gamma}$, $V^q_{a}$, and $V^q_{a-\gamma}$
describe the correlated errors
among the predictions of various rates over the bins.
Aside from trivial functional dependencies on global
rescaling factors, e.g., luminosity, we can
determine them once and for all by generating Monte-Carlo
events in which we vary the parameters on which the form factors depend.
In practice we use the mean values of the $z$-expansion fit (of degree four)
and their covariance matrix (see Appendix~\ref{app:zexpansion})
to determine each piece of $V_{\text{cov}}$.
Using  the  covariance matrices we obtain  the $90\%$ Confidence Level (CL)
exclusion limit on $|F^A_{bq}|$, i.e. $\chi^2(F^A_{bq,90\%}) - \chi^2_{\text{min}} = 1.64$.

\begin{table}[t]
\centering
\begin{tabular}{rrrrr}
	& \multicolumn{2}{l}{$\boldsymbol{B_s\to\mu\mu a}$}
	& \multicolumn{2}{l}{$\boldsymbol{B_d\to\mu\mu a}$}\\
\hline\hline\\[-0.8em]
& sys+stat & stat only & sys+stat & stat only \\
$\chi^2_{\text{min}}$
        & $15.2$ 
        & $15.4$ 
        & $14.8$
        & $15.1$
        \\
$|F^A_{bq,\text{best-fit}}|\times 10^{-5}$ [GeV]
        & $4.3$ 
        & $4.8$
        & $5.0$
		& $5.5$
        \\
$|F^A_{bq,\text{90\%}}|\times 10^{-5}$ [GeV]
		& $>2.4$ 
		& $>2.6$
		& $>3.0$ 
		& $>3.2$ 
		\\\hline
\end{tabular}
\caption{
	The results of recasting LHCb's analysis \cite{Aaij:2017vad}
	to test flavour-violating couplings of the axion to $B_s$ ($F^A_{bs})$ and
	$B_d$ ($F^A_{bd}$). The analysis employs
	a total of $4.4$ fb$^{-1}$ of data from runs at $7$, $8$, and $13$ TeV.
	In the columns labelled ``sys+stat'' we combine statistical and theory uncertainties,
	while in the columns labelled ``stat only'' we neglect the latter.
	We see that presently the bounds are dominated by statistical uncertainties.
	When computing the $\chi^2$ we sum over the ten first bins
	of the analysis, i.e., $m_{\mu^+\mu^-}\in[4.9~\text{GeV}, m_{B_s}]$.
	For every case we list the values of the $\chi^2_\text{min}$
	and the corresponding best-fit
	value for $|F^A_{bq}|$.
	The values of $\chi^2_{\text{min}}$ should be compared with the $\chi^2$ value of the SM,
	$\chi^2_{\text{SM}} = 15.7$.
	The axion best-fit values are thus in roughly $1\sigma$ agreement with the SM.
	$|F^A_{bq,90\%}|$ are the resulting $90\%$ CL exclusion limits.
\label{tab:presentrecastbounds}}
\end{table}

First, we recast the observed data of LHCb's analysis \cite{Aaij:2017vad}
in which ${\cal L} = \overline{\cal L} = 4.4$~fb$^{-1}$.
The measurement is dominated by statistical uncertainties, but for purposes of
illustration we show both the bounds when combining statistical and systematic theory errors
and the bounds when only the statistical uncertainty is included.
In the $\chi^2$ we include the first ten bins of the LHCb analysis.
The observed data are in good agreement with the SM expectation.
Indeed, we find that the $\chi^2$ of the SM
divided by the ten degrees of freedom of the $\chi^2$ (d.o.f.) is
$\chi_{\text{SM}}/{\text{d.o.f.}} = 1.6$.
The best-fit points for the axion lies roughly  $1\sigma$ off the SM.
In \TAB~\ref{tab:presentrecastbounds} we list the best-fit points with their
corresponding $\chi^2_{\text{min}}$, as well as the resulting $90\%$ CL exclusion limits
on $|F^A_{bs}|$ and $|F^A_{bd}|$.

\begin{figure}[t]
	\centering
	\includegraphics{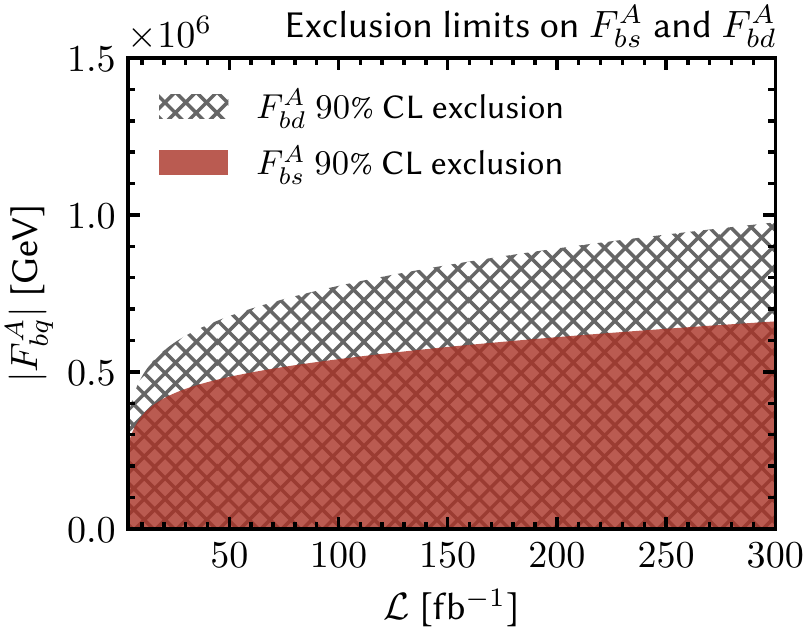}
	\caption{Projected sensitivity of LHCb to probe the flavour-violating axion couplings
		$F^A_{bs}$ (filled red region) and $F^A_{bd}$ (hatched region) as a function
		of the total integrated luminosity.
		Shown are the $90\%$ CL exclusion limits assuming that the observed number of events
		will be the same as predicted in the SM hypothesis.
	 	\label{fig:sensitivity}}
\end{figure}

\begin{table}
\centering
\begin{tabular}{cp{1em}p{5em}p{6em}p{5em}p{5em}}
	&& \multicolumn{2}{l}{$\boldsymbol{B_s\to\mu\mu a}$}
	& \multicolumn{2}{l}{$\boldsymbol{B_d\to\mu\mu a}$}\\
\hline\hline\\[-0.8em]
 	&& \multicolumn{2}{l}{$|F^A_{bs}|\times 10^{-5}$ [GeV]}
& \multicolumn{2}{l}{$|F^A_{bd}|\times 10^{-5}$ [GeV]} \\
${\cal L}$ [fb$^{-1}$] &		& {sys+stat}	& {stat only}	& {sys+stat} 	& {stat only}\\
$10$	&& $3.6$ 	& $3.9$  & $4.9$  & $5.2$  	\\
$30$    && $4.5$	& $5.3$ & $6.2$  & $7.1$  	\\
$50$	&& $4.8$& $6.1$ & $6.8$  & $8.2$ 	\\
$100$	&& $5.4$   & $7.3$  & $7.7$  & $9.8$ 	\\
$300$ 	&& $6.6$& $9.7$& $9.8$& $13$ 	\\\hline
\end{tabular}
\caption{
	Projected $90\%$ CL exclusion limits on the flavour-violating couplings of the
	axion to $B_s$ ($F^A_{bs}$) and $B_d$ ($F^A_{bd}$) as a
	function of the integrated luminosity at LHCb.
	In the columns labelled ``sys+stat'' we combine statistical and theory uncertainties,
	while in the columns labelled ``stat only'' we neglect the latter.
\label{tab:futurebounds}}
\end{table}

Next we make projections for future runs of LHCb.
As discussed in Section~\ref{sec:rescaling}, to this end we rescale
the $13$\,TeV events assuming LHCb will collect a total of  $300$~fb$^{-1}$.
To compute the sensitivity we assume that LHCb will observe
exactly the number of events expected from the SM.
Therefore, the best-fit point always corresponds to observing zero events
from axion decays and $\chi^2_{\text{min}}=0$.
For the projection study we present the results both when only
statistical uncertainties are considered and when they are folded
with the correlated theory uncertainties.
In \FIG~\ref{fig:sensitivity} we show the resulting $90\%$ CL
exclusion limit on $|F^{A}_{bs}|$ (left panel) and $|F^A_{bd}|$ (right panel)
as a function  of the total luminosity.
In addition, the limits for some indicative luminosities are  listed
in \TAB~\ref{tab:futurebounds}.

Note that the limit from the actual recast is weaker than the expected limit
under the background-only hypothesis.
More precisely, if we consider the case
${\cal L} = 4.4$~fb$^{-1}$ and set $N_i^{\text{obs}}= N_{i}^{\text{SM}}$ (as we
do for the projection study) we find for the statistics-only case
$|F_{bs,90\%}^A| < {3.0}\cdot 10^{5}$\,GeV and $|F_{bd,90\%}^A| < {4.0}\cdot 10^{5}$\,GeV.
In comparison, the corresponding exclusion limits of the recast
(table~\ref{tab:futurebounds}) are slightly weaker.
The origin of this difference is mainly an excess of roughly $10$ events
in the first bin of the current LHCb $B_s\to\mu\mu$ analysis, which
can be fitted by the best-fit point of an axion signal.
However, as discussed in the recast the excess is not statistically significant and
the best-fit point of the axion is within $1\sigma$ of the SM.

\section{Summary and Outlook \label{sec:conclusions}}

In this article we have proposed a novel method to probe
flavour-violating couplings of the QCD axion to $b$-quarks
at the LHC, exploiting the di-muon datasets collected for
the $B_q \to \mu \mu$ analyses.
To this end, we have computed the relevant differential decay rates for the
decay of a $B_q$-meson to muons and an axion $B_q \to \mu \mu a$ [Eq.~\eqref{eq:GaBsmumua}]
and the radiative decay $B_q \to \mu \mu \gamma$ [Eq.~\eqref{eq:GaBsmumuga}], which is
a background to the former process.

These rates depend on non-perturbative $B_q \to \gamma^{(*)}$
form factors,  which we have discussed from a general viewpoint, computed with QCD sum rules (at zero flavour-violating
momentum transfer).  
To the best of our knowledge this is the
first discussion of the complete set of form factors, for the
dimension-six effective Hamiltonian ${\cal H}_{\textrm{eff}}^{b \to (d,s)}$, supplemented with an explicit computation of all form factors in Appendix \ref{app:FF}. The  uncertainty of this leading order computation is estimated to be around 
$25\%$ and could benefit from radiative corrections which is, however, an elaborate task. 
We extend these form factors to the low-lying  resonance region of unflavoured vector mesons by using  a multiple subtracted dispersion relation cf. Appendix \ref{app:mutlidisp}. A Mathematica notebook is 
appended to the arXiv version.
 Besides being useful for axion searches these form factors are also the ingredients
for other light BSM particle (e.g. dark photon) searches.
In addition, we have exposed the relation between the introduced basis and the standard $B \to V$ basis
through the dispersion representation in Appendix \ref{app:dispersion},  which interrelates
form-factor properties of the two bases. 
A further application of the off-shell form factor formalism is to consider the 
charged case with $q^2,k^2 > 0$ which is needed to describe 
$B \to \ell^+  \ell^- \ell' \nu$.  We postpone this task to a future study as this involves discriminating charges and dealing with contact terms but 
 note that these decays has recently  
been described by different groups \cite{Danilina:2019dji,Bharucha:2021zay,Beneke:2021rjf,Ivanov:2021jsr} using different approaches.

With these decay rates we performed a recast using available LHCb data and
estimated the sensitivity to $B_q \to \mu \mu a$ at present and future runs, taking into
account the SM background $B_q \to \mu \mu \gamma$.
We find that present data constrain the relevant axion
couplings $F^A_{bd} \, (F^A_{bs}) $ to be larger than $3.0(2.4)  \cdot 10^5 \, {\GeV}$
at 90\% CL [\TAB~\ref{tab:presentrecastbounds}], while the full LHCb dataset will probe  scales
of the order of $10^6 \, {\GeV}$ in both $b \to d$ and $b \to s$ transitions (\TAB~\ref{tab:futurebounds}).

For stable axions, these results should be compared with the ones
derived from $B$-meson decays with missing energy.
In the case of $b\to s$ transitions, the data from the
BaBar collaboration on $B\to K^* \nu \overline{\nu}$  provide
constraints that are roughly two orders of magnitude stronger than the
ones from our LHCb recast of $B_s \to \mu \mu a$, cf. Table~\ref{tab:presentbounds}.
For the case of $b\to d$ transitions, the BaBar constraints are roughly
of the same order than the ones that LHCb can obtain in
upcoming runs.
Nevertheless, the combination with the corresponding ATLAS and CMS
analyses of $B_q \to \mu \mu$ may improve the bounds
significantly. In Appendix \ref{app:BVllrate} we made further use 
of the off-shell form factors to estimate the strength 
of LHCb to  search for  light, Beyond the Standard Model, vectors.

While it is remarkable that the LHC can play a role in constraining
couplings of the QCD axion, the analysis of $B_q \to \mu \mu a$ that
we have presented here can be relevant
for other extensions of the SM with light neutral particles
with flavour-violating couplings.
Since the $B_q \to \mu \mu a$ analysis is inclusive,
it can be extended to search for light BSM particles
even if they decay within the detector.
For example, an ALP that decays promptly to, for instance, photons
may be subject to cuts on additional photons in the analyses of
$B \to K (a \to \gamma \gamma)$ at the $B$-factories
and thus evade detection, while
it would be kept in the $B_q \to \mu \mu (a \to \gamma \gamma)$ samples at
the LHC.
Therefore, the analysis that we have presented here complements axion searches
in rare meson decays with missing energy at B-factories, and can play an
important role in constraining flavour-violating couplings of light particles.

\section*{Acknowledgments}
We are grateful to Tadeusz Janowski for providing the $z$-expansion
fits to the form factors. We  thank Martin Beneke, Alex Khodjamirian, Ben Pullin, Mikolai Misiak, and Uli Nierste for very useful discussions. This research was supported by the Munich Institute for Astro- and Particle Physics (MIAPP) of the DFG Excellence Cluster Origins (www.origins-cluster.de). E.~Stamou and R.~Ziegler thank the Galileo Galilei Institute for Theoretical Physics for the hospitality
and the INFN for partial support during the initial stages of this work. J. Albrecht gratefully acknowledges support of the European Research Council, ERC Starting Grant: PRECISION 714536.
R.\ Zwicky is supported by an STFC Consolidated Grant, ST/P0000630/1.
E.~Stamou was supported by the Fermi Fellowship at the Enrico Fermi Institute and by the U.S.\ Department
of Energy, Office of Science, Office of Theoretical Research in High
Energy Physics under Award No.\ DE-SC0009924 and by the Swiss National Science Foundation under contract 200021--178999.


\appendix

\section{ The  $ B_q \to  \ga^* $ Form Factors  \label{app:FF}}

The standard $B_q \to V$ matrix elements (ME), where $V=   \rho^0, \omega, \phi  \hdots$ is a vector meson, hold some analogy with the   $B_q \to  \ga^* $ ones.
However, the difference is that the analogue of the vector-meson mass is the photon off-shell momentum
which is a variable rather than a constant.
Hence the MEs are functions of two variables and this leads to a more involved analytic structure.
In this paper we restricted ourselves to the kinematic region
$q^2 \in  [ \qsqlow, \mBq^2]$, where the form factors (FFs) can be expected to dominate
over long-distance contributions.

 This appendix is structured as follows. Firstly, we define and state relation and limits of the FFs in Section~\ref{eq:FFdef}, the link with the $B \to V$ basis is discussed in Section~\ref{app:dispersion},
  the QCD SR computation of the off-shell FFs follows
 in Section~\ref{app:FFcomp} and finally we turn to the FF-parametrisation and fits in Section~\ref{app:FFp}.
Note, that sections  \ref{eq:FFdef}, \ref{app:dispersion} and \ref{app:FFp} are independent of the method of computation.

\begin{table}[h]
\centering
\begin{tabular}{l | c | c |  c}
	\multicolumn{1}{c}{Form Factor} &\multicolumn{1}{c}{$\PFF^*(0,k^2)$}   &
	\multicolumn{1}{c}{$\FTA^*(0,k^2) = (1-\hat{k}^2)\FTV^*(0,k^2)$}   &
\multicolumn{1}{c}{$V,T_{\perp,\parallel}(q^2)=V^*,T^*_{\perp,\parallel}(q^2,0)$} 	\\[0.1cm] \hline\hline
Mode        & $B_s \to  \ell \ell a $ & $B_s \to  \ell \ell \ga$ & $B_s \to  \ell \ell \ga $ \\[0.1cm]
 Poles  $q^2,k^2$& $m_\phi^2 , m_\Upsilon^2$& $m_\phi^2 , m_\Upsilon^2$ &  $m_{B^*_q}^2, m_{B_{q1}}^2$  \\[0.1cm]
Defined in Eqs. & (\ref{eq:FFsec2},\;\ref{eq:NEW})  &  (\ref{eq:FFsec2},\;\ref{eq:NEW}) & (\ref{eq:OnS})   \\[0.1cm]
Graph in \FIG~\ref{fig:dia-FF} & (left) & (centre) & (right) \\[0.1cm]
Other notation & $-$ & $  F_{TV}(0,k^2) =
F_{TA}(0,k^2) $\cite{Kruger:2002gf,Kozachuk:2017mdk} & $F_{V,A},\,F_{TV,A}(q^2,0)$\cite{Kruger:2002gf,Kozachuk:2017mdk}  \\\hline
\end{tabular}
\caption{\label{tab:FFoverview}
Overview of FFs referencing definitions, graphs, and analytic structure. The latter defines the region
of validity of the computation. Long-distance contributions are relevant in other kinematic regions
\cite{Kozachuk:2017mdk,GRZ17}.
For $B_d \to \ga^*$,  $m_\phi^2$ is to be replaced by $m_{\rho,\omega}^2$ above.}
\end{table}

\subsection{Definition of $B \to \ga^{(*)}$ form factors\label{eq:FFdef}}

We introduce a complete set of \emph{off-shell} FFs which is related to  the standard $B \to V$ basis
\cite{Wirbel:1985ji,BSZ15}  via
dispersion relations, cf. Section~\ref{app:dispersion}.
On a technical level this appendix  extends previous work  \cite{Kruger:2002gf,Kozachuk:2017mdk},
in that we discuss the full set of seven vector and tensor FFs and not only those needed for the
SM transition.
The complete basis is for example useful for other invisible particle searches such as the  dark photon.
The off-shell FFs are not to be confused with the on-shell FFs  which  have received more attention
in the literature  \cite{Kruger:2002gf,Melikhov:2004mk,Kozachuk:2017mdk,JPZ19}.
An overview of the on- and  off-shell FFs used for this paper  are shown in the diagrams in \FIG~\ref{fig:dia-FF}
and contrasted in \TAB~\ref{tab:FFoverview}.

\subsubsection{The complete basis of seven off-shell form factors $F^*(q^2,k^2)$\label{app:FFoff}}

We introduce the FFs with two momentum squares $q^2$ and $k^2$ collectively as
$F^*(q^2,k^2) \equiv F^{B \to \ga^*}(q^2,k^2)$.
The first argument (here $q^2$) denotes the momentum transfer at the  flavour-violating
vertex while the second argument (here $k^2$)
denotes the momentum of the photon emitted at low energies.

We introduce a new off-shell basis via a dispersion representation based on the standard $B \to V$ basis \cite{Wirbel:1985ji,BSZ15}.  Below we state the basis before
turning to the construction in Section~\ref{app:dispersion}.
The absence of unphysical singularities
in the matrix element enforces relations between FFs which we discuss in some detail.
We will refer to this circumstance as ``regularity" for short.

The complete set of FFs were already introduced in the main text in Eq.~\eqref{eq:FFsec2}
and reproduced here for convenience\footnote{Cf. footnote \ref{foot:bar} in the main text for relevant remarks 
on $B \to \ga$ versus $\bar B \to \ga$ FFs.
The  conventions are $\ga_5 = i \ga^0 \ga^1 \ga^2 \ga^3$, $g = \textrm{diag}(1,-1,-1,-1)$,
$\matel{ 0 }{ \bar{q} \, \gamma^{\mu} \gamma_{5} \, b }{ \bar{B}_q (p_B)}
= i p_B^{\mu} f_{B_q}$, $D_\mu = \partial_\mu + s_e i Q_f  e A_\mu$    and
$\varepsilon_{0123} = 1$.
Together with  $\matel{\ga}{A_\mu}{0} = \epsilon^*_\mu$ this  fixes the phase of the $B_s$- and the $\ga$-state.
 The $B_{q} \to \ga$ FFs  are then positive for $s_e =1$.}$^,$\footnote{\label{foot:JG} Whereas
$M^{\rho\mu}_{T_5}$ in [Eq.~4]
in \cite{GRZ17}, and similarly in \cite{Kozachuk:2017mdk}, is incomplete it remains sufficient within the SM
as there  $\eps^*_\mu(q) M^{\rho\mu}_{T_5}$ and  $q^2 \to 0$ annihilate the $\FTL$-contribution.
However, the correct substitution reads $\FTA(0,q^2)|_{\mbox{\cite{GRZ17} }} \to \FTA(0,q^2)/(1-\hat{q}^2) $ since the normalisation differs slightly.}
 \begin{alignat}{3}
\label{eq:NEW}
& M_{5}^\rho &\; \equiv\;& \bP \matel{ \gamma^*(k,\rho)} {  \bar q \gamma_5 b} {\bar{B}_q (p_B)} &\;=\;& i
\mBq \RR^\rho \, \PFF^*(q^2,k^2)  \nonumber   \\[0.1cm]
& M^{\mu\rho}_V  &\; \equiv\;& \bV \matel{\ga^*(k,\rho)} { \bar{q} \gamma^{\mu}   b }{ \bar{B}_q (p_B)}
 &\;=\;&  +  R_\perp^{\mu\rho} \, V_\perp ^*(q^2,k^2)  \;, \nonumber   \\[0.1cm]
  & M^{\mu\rho}_A  &\; \equiv\;&  \bV   \matel{\ga^*(k,\rho)} { \bar{q} \ga^{\mu}  \ga_5  b }{ \bar{B}_q (p_B)}
 &\;=\;&  +  (  R_\parallel^{\mu\rho} \, V_\parallel ^*(q^2,k^2)   + \RR^{\mu\rho}_\LL V_\LL^*(q^2,k^2)  + \RR^{\mu\rho}_P V^*_P(q^2,k^2) ) \;, \nonumber  \\[0.1cm]
 & M^{\mu\rho}_T  &\; \equiv\;& \bT \matel{\ga^*(k,\rho)} { \bar{q} iq_\nu \sigma^{\mu \nu}    b }{ \bar{B}_q (p_B)}
 &\;=\;&  +   R_\perp^{\mu\rho} \, T_\perp ^*(q^2,k^2)  \;,   \nonumber  \\[0.1cm]
  & M^{\mu\rho}_{T_5}  &\; \equiv\;& \bT \matel{\ga^*(k,\rho)} { \bar{q} iq_\nu \sigma^{\mu \nu}  \ga_5   b }{ \bar{B}_q (p_B)}
 &\;=\;&  - (  R_\parallel^{\mu\rho} \, T_\parallel^*(q^2,k^2)  +    \RR^{\mu\rho}_\LL T_\LL^*(q^2,k^2) )  \;,
 \end{alignat}
  where $\bP \equiv \left( \frac{m_b + m_q}{s_e e}\right)$, $\bV \equiv  \left(-\frac{\mBq}{s_e e} \right)$,  $\bT \equiv \left( \frac{1}{ s_e e}\right)$, the momentum transfer is $q \equiv p_B - k$ and
  the off-shell photon state $\langle \ga^*(k,\rho)|$ is defined through
\begin{equation}
\label{eq:offshellgamma}
\matel{\gamma^*(k, \rho)} { O(0)}{B} \equiv - i e s_e \int d^4 x e^{i k \cdot x } \matel{ 0}{T j^\rho(x) O(0)}{B} \;,
 \end{equation}
 where   $j^\rho = \sum_f Q_f \bar f \ga^\rho f$  is the electromagnetic current.
 \begin{alignat}{3}
\label{eq:RNEW}
& R_\perp^{\mu \rho}  &\;\equiv\;&   \varepsilon^{\mu \rho \be \ga }  q_\be k_\ga
 \; ,& R_\parallel^{\mu \rho }   \equiv\;&  \frac{i}{2}  (1-\hat{q}^2) \,  (\mBq^2  G^{ \mu \rho} - \frac{ (q+ 2k)^{\mu}\RR^{\rho}}{1-\hat{k}^2} ) \; ,
\nonumber \\[0.1cm]
& R_\LL^{\mu\rho} &\;\equiv\;&  \frac{i}{2}  ( q^\mu - \frac{\hat{q}^2(q+ 2k)^{\mu}}{1-\hat{k}^2} ) \, \RR^\rho\; ,  \qquad
&R_P^{\mu\rho} \equiv\;& \frac{i}{2}  q^\mu \RR^\rho   \;, \quad \quad \RR^\rho \equiv  q^\rho - \frac{k\!\cdot\!q}{k^2}k^\rho \;,
\end{alignat}
are Lorentz tensors with convenient properties  (cf. below) and
 hereafter
 \begin{equation}
 \label{eq:nice}
 \hat{k}^2 \equiv \frac{k^2}{ \mBq^2} \;, \quad \hat{q}^2 \equiv \frac{q^2}{ \mBq^2}  \;.
 \end{equation}
The photon transverse tensor, $k^\al G_{\al\be} =0$, is
\begin{equation}
\label{eq:G}
G_{\al \be} \equiv g_{\al \be} - \frac{k_\al k_\be}{k^2} \;,
\end{equation}
and it is noted that $\RR^\rho = q_{\mu} G^{\mu\rho}$ \eqref{eq:RNEW}.

\subsubsection{Constraints for off-shell Form Factors}

The QED Ward identity holds off-shell in the form
 \begin{equation}
 \label{eq:QEDWI}
	 k_\rho M^{\mu\rho}_{\rm V,A, T,T_5} = 0  \;,
 \end{equation}
 without contact term since the weak operator is neutral in the total electric charge.
 Note that Eq.~\eqref{eq:QEDWI} is  automatically satisfied
in our parametrisation since  $k_\rho \RR^{\mu\rho}_{\perp,\parallel,\LL,P} =0$.
The non-singlet axial   Ward identity for $M^{\mu\rho}_{\rm A}$ assumes the form
\begin{equation}
\label{eq:AWI}
q_\mu\, M^{\mu\rho}_{\rm A} = \mBq M_{5}^\rho \;,
	 \end{equation}
	 which in turn holds without contact term  since the electromagnetic current
	 is invariant under non-singlet axial rotations.
 Eq.~\eqref{eq:AWI}, upon using  $q_\mu R^{\mu\rho}_{\perp,\parallel,\LL} = 0$, implies that
 \begin{equation}
 \label{eq:AWIsol}
 \FP^*(q^2,k^2) = \frac{2}{\hat{q}^2} \PFF^*(q^2,k^2) \;.
 \end{equation}

 Regularity enforces constraints on the FFs defined in \eqref{eq:NEW}.\footnote{The two constraints (\ref{eq:A03},\ref{eq:algebraic})
 have well-known analogues
 in $B \to V$ which are stated in Section~\ref{app:dispersion}.
 A similar constraint to \eqref{eq:algebraic} was reported in Ref.~\cite{Kozachuk:2017mdk}
 and we comment in the same section in what way it differs from ours.}
There are two constraints at $q^2=0$ and $k^2 = \mBq^2$ respectively.
The Ward identity  \eqref{eq:AWIsol} enforces
 \begin{equation}
 \label{eq:A03}
 \PFF^*(0,k^2) = \hat{V}^*_\LL(0,k^2)   \;,
  \end{equation}
where $ \hat{V}^*_\LL$ is implicitly defined by
\begin{equation}
\label{eq:Vhat}
V^*_\LL(q^2,k^2) \equiv -  \frac{2}  {\hat{q}^2} \hat{V}^*_\LL(q^2,k^2) \;.
\end{equation}
The second constraint is
 \begin{equation}
 \label{eq:algebraic}
 \FTA^*(0,k^2)  = ( 1 - \hat{k}^2) \FTV^*(0,k^2)  \;.
 \end{equation}
There are two further constraints due to the parametrisation of the form factors
 at $k^2 = \mBq^2$ 
\begin{alignat}{2}
\label{eq:third}
& (1- \hat{q}^2 ) \FA^*(q^2,\mBq^2) + \hat{q}^2 V_\LL^*(q^2,\mBq^2) &\;=\;&  0  \;, \nonumber \\[0.1cm]
& (1- \hat{q}^2 ) \FTA^*(q^2,\mBq^2)+ \hat{q}^2 T_\LL^*(q^2,\mBq^2)  &\;=\;& 0 \;,
\end{alignat}
which are of a similar type as the $A_0^{B \to V}(0)= A_3^{B \to V}(0)$ cf. \eqref{eq:q20BV}.
Whereas the constraints \eqref{eq:A03} and \eqref{eq:third} are  imposed by
the FF-parametrisation (avoiding spurious kinematic singularities), \eqref{eq:algebraic} is of  algebraic origin cf. Section~\ref{sec:algebraic}
for the derivation.

\subsubsection{The four photon on-shell form factors $F(q^2) \equiv F^*(q^2,0)$\label{app:FFgaon}}

We next turn to the case where the low-energy photon is on-shell; $k^2 = 0$.
We introduce the commonly used shorthand
\begin{equation}
F(q^2)  \equiv F^*(q^2,0)   \;, \quad\text{for}\quad F  \in \{P, V_{\perp,\parallel,\LL} , T_{\perp,\parallel,\LL} \}  \;,
\end{equation}
(or   $F^{B \to \ga}(q^2)  \equiv F^{B \to \ga^*}(q^2,0)$).
The basic physics idea is that the absence of the photon's zero helicity component implies the vanishing
the pseudoscalar FF and the zero helicity part of the vector FFs.
We may define the helicity amplitude for $B \to \mu\mu\ga$ by
\begin{equation}
{\cal A}_{\la \la'}^X \propto M_X^{ \mu\rho} \omega^*_\mu(q,\la)  \eps^*_\rho(k,\la')  \;.
\end{equation}
 One then obtains the two $B \to \ga\mu\mu$ helicity amplitudes 
 \begin{alignat}{6}
\label{eq:heli}
& {\cal A}^A_0 &\;=\;&  {\cal A}_{00}^A    \stackrel{k^2 \to 0}{\propto} V_\parallel - V_\LL \;, \quad
& &  {\cal A}^A_t &\;=\;&    {\cal A}_{t0}^A  \stackrel{k^2 \to 0}{\propto} P    \;, \quad
& & {\cal A}^{T_5}_0  &\;=\;&     {\cal A}_{00}^{T_5}    \stackrel{k^2 \to 0}{\propto} T_\parallel - T_\LL  \;,
\end{alignat}
and the $\pm$ direction obey
\begin{alignat}{6}
& {\cal A}^{A}_+ &\;=\;&  {\cal A}_{++}^{A}    \stackrel{k^2 \to 0}{\propto} V_\parallel - V_\perp \;, \quad
& &  {\cal A}^{A}_- &\;=\;&    {\cal A}_{--}^{A}  \stackrel{k^2 \to 0}{\propto} V_\parallel + V_\perp    \;, \quad
& &  & & 
\end{alignat}
and the formulae for the vector $V$ as well as the  tensors $T,T_5$ are similar. The common definition 
\begin{equation}
{\cal A}^A_{\perp,\parallel} \equiv \frac{1}{\sqrt{2}}\left( {\cal A}^A_+ \mp {\cal A}^A_-\right)   \propto V_{\perp,\parallel} \;,
\end{equation}
explains the notation.  
In order to obtain these results it is useful to choose an 
 explicit helicity vector parameterisations. E.g. in the $B_q$-meson restframe
 (e.g. [Eq.15] in \cite{Gratrex:2015hna})
\begin{alignat}{5}
\label{eq:helpol}
&  \eps(k,0)   &\;=\;& (v,0,0 ,\sqrt{k^2 + v^2} )/\sqrt{k^2}  \;, \quad  & &  \eps(k,t) &\;=\;& k/\sqrt{k^2} \;, \quad  
& &
   \!\!\!\!  \!\!\!\! \!\!\!\! k  \;=\;  ( \sqrt{k^2 + v^2},0,0,v)   \;, \nonumber \\[0.1cm]
&  \omega(q,0)  &\;=\;&   (-v,0,0 ,\sqrt{q^2 + v^2} )/\sqrt{q^2} \;, \quad  & &  \omega(q,t) &\;=\;& q/\sqrt{q^2}   \;, \quad & & \!\!\!\!  \!\!\!\! \!\!\!\! q \;=\;  ( \sqrt{q^2 + v^2},0,0,-v) \;,
\nonumber \\[0.1cm]
&  \eps(k,\pm)  &\;=\;&   \omega(q,\mp)  =  (0,\mp 1,i ,0)/\sqrt{2} \;,\quad
 & &     \eps  (\perp,\parallel)  &\;=\;& \frac{ \eps(+)  \mp \eps(-)}{\sqrt{2}} \;, & & 
\end{alignat}
  where  the velocity is given by
  $v \equiv |\vec{k}| = \la^{1/2}(\mBq^2,q^2,k^2)/(2 \mBq) $ and $k \to k^\mu$ is the Lorentz-index convention 
  in the formulae above. This helicity basis is useful since
  \begin{equation}
  \label{eq:completeness}
  g_{\mu\nu} = \sum_{\la = t,0,\pm} \eps_\mu^*(\la) \eps_\nu(\la') G^{\la \la'} 
  = \sum_{\la = t,0,\perp,\parallel} \eps_\mu^*(\la) \eps_\nu(\la') G^{\la \la'}
  \end{equation}
  with $G^{\la\la'} = \textrm{diag}(1,-1,-1,-1)$ and the first entry denotes the $t$-component.
  Second, in the limit $k^2 \to 0$,
  $\eps^*(k,0) \propto k $, and this enforces,
  \begin{equation}
  \label{eq:lim}
   \lim_{k^2 \to 0}\eps^*_\rho(k,0)M^{\mu\rho}_{A,T_5} =0 \;,
   \end{equation}
     since it is
equivalent to the QED Ward identity \eqref{eq:QEDWI}.
Eqs.~(\ref{eq:heli},\;\ref{eq:lim})  lead to the  constraints
\begin{equation}
\label{eq:FFzero}
 \FA (q^2) =    \FL(q^2)  \;, \quad  \FTA(q^2) =   \FTL(q^2) \;,  \quad  \PFF(q^2)  = 0 \;,
\end{equation}
and reduces the seven FFs of Eq.~\eqref{eq:NEW} to four.
Alternatively one can infer the constraints \eqref{eq:FFzero} from the regularity of
the matrix elements as $k^2 \to 0$. The regularity condition and the helicity arguments
 are clearly related.

For completeness we give the explicit $k^2 \to 0$ basis \cite{GRZ17,JPZ19}\footnote{
 The charged FF $B_u \to \ga^{(*)}$ is similar but comes with a non gauge invariant contact term
 for the axial vector structure.  This contact term
 is canceled by the photon
emission of the lepton \cite{Janowski:2021yvz}.}
\begin{alignat}{3}
\label{eq:OnS}
& M^{\mu}_V  &\; \equiv\;& \bV \matel{\gamma(\eps(k))} { \bar{q} \gamma^{\mu}   b }{ \bar{B}_q (p_B)}
 &\;=\;&  +  P_\perp^{\mu} \, V_\perp (q^2)  \;, \nonumber   \\[0.1cm]
  & M^{\mu}_A  &\; \equiv\;&  \bV   \matel{\gamma(\eps(k))} { \bar{q} \ga^{\mu}  \ga_5  b }{ \bar{B}_q (p_B)}
 &\;=\;&  +   P_\parallel^{\mu} \, V_\parallel (q^2)    \;, \nonumber  \\[0.1cm]
 & M^{\mu}_T  &\; \equiv\;& \bT \matel{\gamma(\eps(k))} { \bar{q} iq_\nu \sigma^{\mu \nu}    b }{ \bar{B}_q (p_B)}
 &\;=\;&  +   P_\perp^{\mu} \, T_\perp (q^2)  \;,   \nonumber  \\[0.1cm]
  & M^{\mu}_{T_5}  &\; \equiv\;& \bT \matel{\gamma(\eps(k))} { \bar{q} iq_\nu \sigma^{\mu \nu}  \ga_5   b }{ \bar{B}_q (p_B)}
 &\;=\;&  -   P_\parallel^{\mu} \, T_\parallel(q^2)  \;,
 \end{alignat}
where
\begin{alignat}{3}
\label{eq:PNEW}
 P_\perp^{\mu }  &\;\equiv\;&    \eps^*_\rho \RR_\perp^{\mu \rho} =  \varepsilon^{\mu \rho \be \ga }   \eps^*_\rho  q_\be k_\ga
 \;  \quad  P_\parallel^{\mu }   \equiv\;& \eps^*_\rho ( \RR_\parallel^{\mu\rho}   + \RR_\LL^{\mu\rho} ) =  i  \,  (   q \! \cdot \! k \,  \eps^{*\,\mu} -   q\! \cdot \!\eps^* \, k^{\mu}  )  \; .
\end{alignat}

\paragraph{Relation to the standard $B \to V$ basis}

We consider it worthwhile to comment on some aspects in the standard basis
of $B \to V$ FFs e.g. \cite{BSZ15}. The $k^2 \to 0$ limit is then akin  to $m_V \to 0$.
The relations  $\FA (q^2)-   \FL(q^2)  = \FTA(q^2) -  \FTL(q^2)= 0$  implies
\begin{alignat}{2}
\label{eq:inter}
&  \V^{B \to V}_2(q^2) = (1  -  \hat{q}^2) \V^{B \to V}_3(q^2) + {\cal O}(m_V) \;, \nonumber \\[0.1cm]
&  T^{B \to V}_2(q^2) = (1  -  \hat{q}^2)T^{B \to V}_3(q^2) + {\cal O}(m_V) \; .
\end{alignat}
Such relations were noted previously.
Firstly, in the $B \to V$ context in  Ref.~\cite{Dimou:2012un} in Appendix A
and around Eq.~[5] in Ref.~\cite{Lyon:2013gba}, where it is argued that the relation
has to hold in order to cancel a kinematic $1/m_V$-factor.
Second for $B \to \ga$ ($m_V =0$) they were previously reported in Ref.~\cite{Kruger:2002gf}
as a consequence of regularity.

\subsubsection{The five form factors $F^*(0,k^2)$  at zero flavour-violating momentum transfer
\label{sec:q20}}

In the process $B_q\to\ell\ell X$, with $X$ a light BSM particle,
the limit in which the flavour-violating  momentum transfer goes to zero, i.e., $q^2 = 0$,
corresponds to the case of zero or small mass of $X$. In this limit the two constraints in Eqs.~\eqref{eq:A03} and \eqref{eq:algebraic} reduce
the number of independent FFs from seven to five.

The matrix elements, at $q^2 =0$,the become
\begin{alignat}{3}
\label{eq:q20}
& M^{\mu\rho}_V  &\; \equiv\;& \bV \matel{\gamma^*(k,\rho)} { \bar{q} \gamma^{\mu}   b }{ \bar{B}_q (p_B)}
 &\;\to\;&  +  \RR_\perp^{\mu\rho} \, \FV ^*(0,k^2)  \;, \nonumber   \\[0.1cm]
  & M^{\mu\rho}_A  &\; \equiv\;&  \bV   \matel{\gamma^*(k,\rho)} { \bar{q} \ga^{\mu}  \ga_5  b }{ \bar{B}_q (p_B)}
 &\;\to\;&   +( \RR_\parallel^{\mu\rho} \, \FA ^*(0,k^2)   + i   \frac{(2k+q)^\mu}{1 - \hat{k}^2 }  \RR^\rho  \,\PFF^*(0,k^2))  \;, \nonumber  \\[0.1cm]
 & M^{\mu\rho}_T  &\; \equiv\;& \bT \matel{\gamma^*(k,\rho)} { \bar{q} iq_\nu \sigma^{\mu \nu}    b }{ \bar{B}_q (p_B)}
 &\;\to\;&  +   \RR_\perp^{\mu\rho} \, \FTV ^*(0,k^2)
 \;,\nonumber\\[0.1cm]
   & M^{\mu\rho}_{T_5}  &\; \equiv\;& \bT \matel{\gamma^*(k,\rho)} { \bar{q} iq_\nu \sigma^{\mu \nu}  \ga_5   b }{ \bar{B}_q (p_B)}
  &\;\to\;&  - (   \RR_\parallel^{\mu\rho} \, (1-\hat{k}^2)\FTV^*(0,k^2)  + \frac{i}{2} q^\mu \RR^\rho \FTL^*(0,k^2) ) \;,
  \end{alignat}
where Eqs.~ (\ref{eq:A03},\ref{eq:algebraic})
 and $2 k \!\cdot \!q|_{ q^2 =0} = \mBq^2 - k^2$  have been used.
 At $q^2=0$ the constraints \eqref{eq:third} imply
 \begin{equation}
 \label{eq:third2}
  \FA^*(0,\mBq^2)  =  2\PFF^*(0,\mBq^2)  \;, \quad \FTA^*(0,\mBq^2) = 0 \;.
 \end{equation}
With $T_\LL^*(0,\mBq^2)$ finite, the  last constraint is obeyed trivially by \eqref{eq:algebraic}.

\subsubsection{Counting form factors \label{sec:counting}}

\begin{table}[h]
 \renewcommand*{\arraystretch}{1.3}
\centering
    \begin{tabular}{| l | l   | l l l l  | }
     \rowcolor{gray!7}
        \hline
 type$\backslash$  $ J^P$ &  $\#$  & $1^-$  & $1^+$ & $1^+$ & $0^-$       \\ \hline
\DOS  $F^*(q^2,k^2) $ &  \DOS     $\!\!7\!$      & \DOS $\FV^*(q^2,k^2)$ & \DOS $\FA^*(q^2,k^2)$ & \DOS $\hat{V}_\LL^*(q^2,k^2)$ & \DOS$ \PFF^*(q^2,k^2)$ \\
\DOS       &  \DOS  & \DOS $\FTV^*(q^2,k^2)$ & \DOS $\FTA^*(q^2,k^2)$ & \DOS $\FTL^*(q^2,k^2)$ & \BACK  \\ \hline
 \OS  $F(q^2) \equiv $            &   \OS  $\!\!4\!$  & \OS $\FV(q^2)     $
 & \OS $\FA(q^2)   $
 & \BACK $ V_\LL(q^2) = V_\parallel(q^2) $
 &\BACK  $ \PFF(q^2)=0$ \\
  \OS $F^*(q^2,0) $    & \OS              & \OS $\FTV(q^2)    $ & \OS $\FTA(q^2)   $  & \BACK $\FTL(q^2)=\FTA(q^2) $  & \BACK  \\
                   \hline
 \SOS  $F^*(0,k^2)$  & \SOS $\!\!5\!$  & \SOS $\FV^*(0,k^2)$ & \SOS $\FA^*(0,k^2)$   &    \BACK
 $\hat{V}_\LL^*(0,k^2) = \PFF^*(0,k^2)$   & \SOS $\PFF^*(0,k^2)$ \\
   \SOS                              & \SOS                                                               &\SOS $\FTV^*(0,k^2)$
& \BACK $ \FTA^*(0,k^2)  = (1\!-\!\hat{k}^2)\FTV^*(0,k^2)  $   & \SOS $\FTL^*(0,k^2)$    & \BACK        \\
   \hline
     \end{tabular}
\caption{ The $J^P = 0^+$ FF vanishes
by parity conservation of QCD.
Generally, there are seven independent $F^*(q^2,k^2)$ FFs (light-blue) with two constraints
$ \hat{V}_\LL^*(0,k^2)= \PFF^*(0,k^2) $ \eqref{eq:A03} and
$T^*_\parallel(0,k^2) = (1- \hat{k}^2)T^*_\perp(0,k^2) $ \eqref{eq:algebraic}.
 For the photon on-shell case,
 $F(q^2) \equiv F^*(q^2,0)$,
there are four independent FFs (light-red) and the reduction is due to the absence of the photon $0$-helicity component.
At zero flavour-violating momentum there are five independent FFs (light-green),
 due to the two constraints mentioned above. For the computation of the $B \to \ell\ell\ga$ SM rate, the following five  FFs are sufficient
$\{ V_{\perp,\parallel}(q^2), T_{\perp,\parallel}(q^2) ,\FTV^*(0,k^2)\}$. }
\label{tab:FFsummary}
\end{table}

Since the last few sections were a bit involved with many steps we summarise
the classification in
\TAB~\ref{tab:FFsummary}.
In general there are seven FFs for the $B \to 1^-$ transition.
In the photon on-shell case this reduces to four because the photon comes with
two polarisations only.
In the case of zero flavour-violating momentum transfer the two general constraints
(\ref{eq:A03},\;\ref{eq:algebraic}) reduce this number from seven to five.

\subsubsection{Derivation of $\FTA^{B \to \ga^*}(0,k^2)  = ( 1 - \hat{k}^2) \FTV^{B \to \ga^*}(0,k^2)$}
\label{sec:algebraic}

At last we turn to the derivation of the relation \eqref{eq:algebraic}. 
We may choose to proceed by uncontracting the $B \to V$ matrix element, 
first in   $q^\nu$  \eqref{eq:NEW}  \begin{equation}
\label{eq:BVpara}
\bT \matel{V(\eta(k))} { \bar{q}  \sigma^{\mu \nu} \ga_5  b }{ \bar{B}_q (p)}  =
 x_0 \eta^* \!\cdot\! p \frac{ k^{[\mu} \pB^{\nu]}}{q \cdot k} +    x_1 \eta^{*\,[\mu} k^{\nu]} + x_2 \eta^{*\,[\mu} p^{\nu]} = 
  \eta^{*\,\al} x_\al^{\mu \nu} \;,
\end{equation}
with shorthands $x_i  = x_i(q^2,k^2)$, $p = p_B$, $\eta$ is the polarisation vector of a massive vector boson
and square brackets
denote antisymmetrisation in the respective indices.
 The corresponding uncontracted $B \to \ga^*$ matrix element then reads
\begin{equation}
\label{eq:MT5three}
M^{\mu\nu\rho }_{T_5} \equiv \bT  \matel{\ga^*(k,\rho)} { \bar{q}  \sigma^{\mu \nu}  \ga_5   b }{ \bar{B}_q (\pB)} =
X_0   \RR^\rho  \frac{ k^{[\mu} \pB^{\nu]}}{q \cdot k}  + X_1 g^{\rho[\mu} k^{\nu]} + X_2  (
g^{\rho[\mu} \pB^{\nu]}- \frac{k^\rho}{k^2} k^{[\mu} \pB^{\nu]}) \;,
\end{equation}
where 
\begin{equation}
 M^{\mu\nu\rho }_{T_5} =  c \,G^{\rho \al} x_\al^{\mu \nu} \;,
\end{equation}
with $c$ some $i$-independent kinematic function ($X_i =  c x_i $) which is irrelevant for our purposes.
The appearance of the tensor $G^{\rho \al}$ can be understood from the viewpoint of a dispersion relation
cf. Section~\ref{app:dispersion} or and footnote \ref{foot:LSZ}.
Regularity  enforces at $k \cdot q \propto 1 -\hat{k}^2 - \hat{q}^2 \to 0$\;,
\begin{equation}
\label{eq:X0}
X_0(q^2,\mBq^2 - q^2) = 0 \;,
\end{equation}
and at $k^2 \to 0$ we have
\begin{equation}
\label{eq:X02}
X_0(q^2,0) + X_2(q^2,0) = 0 \;.
\end{equation}
These two constraints are generally valid.

We may make the connection with our basis by identifying
\begin{equation}
 M^{\mu\rho}_{T_5} = i q_\nu M^{\mu\nu\rho }_{T_5} \;, \quad
 M^{\mu\rho}_{T} = - \frac{i}{2} (iq_\nu ) \eps^{\mu\nu}_{\phantom{\mu\nu}\mu'\nu'}  M^{\mu'\nu'\rho }_{T_5}    \;,
 \end{equation}
  to
obtain
\begin{alignat}{2}
& \FTV^* (q^2,k^2)&\;=\;& - (X_1(q^2,k^2)+X_2(q^2,k^2)) \;,  \nonumber  \\[0.1cm]
&  \FTA^*(q^2,k^2) &\;=\;& -  \frac{1- \hat{k}^2}{1- \hat{q}^2} \,(X_1(q^2,k^2)+X_2(q^2,k^2)) + \frac{\hat{q}^2}{1- \hat{q}^2} \, (X_1(q^2,k^2)-X_2(q^2,k^2)) \;,  \nonumber  \\[0.1cm]
& \FTL(q^2,k^2) &\;=\;& (X_2(q^2,k^2) - X_1(q^2,k^2)) + 2 \frac{1- \hat{k}^2}{1- \hat{k}^2 + \hat{q}^2} \, X_0(q^2,k^2)\;.
\end{alignat}
There are two consequences of this equation.
 Since $X_1$ and $X_2$ are free from poles at  $q^2 =0$ one gets \eqref{eq:algebraic},
\begin{equation}
\label{eq:algebraic2}
\FTA^*(0,k^2)  = ( 1 - \hat{k}^2) \FTV^*(0,k^2) \;,
\end{equation}
and by inserting \eqref{eq:X02} into $\FTL^*$ one deduces that $\FTL(q^2) = \FTA(q^2)$, which we derived earlier cf.~\eqref{eq:FFzero}. This confirms the earlier observation that the regularity conditions in $k^2 \to 0$ are equivalent
to the previously mentioned helicity argument.
The derivation of relations \eqref{eq:algebraic2}
achieves the purpose of this section. The relation  \eqref{eq:algebraic2} 
appears for any set of FF and reads $T_1^{B \to V}(0) =T_2^{B \to V}(0)  $ in the notation of \cite{Wirbel:1985ji,BSZ15}  and has been derived in the off-shell FF context in reference \cite{Kruger:2002gf}.

\subsection{Relation of the $B \to \ga^*$- and $B \to V$-basis through the dispersion relation}
\label{app:dispersion}

In this appendix we make the link  between the $B \to V$- and the $B \to \ga^*$-FFs
through the dispersion relations. This is an instructive exercise and we will be able to recover properties
of the $B \to \ga^*$ FFs from the $B \to V$-ones. Our argumentation remains true if
one considers any  intermediate state (e.g.   two pseudoscalar particles in a $P$-wave)  as long as its quantum number, $J^{PC} = 1^{--}$, is equal to the one of the photon.
This is the case since the properties follows from the general decomposition and the fact that any such state
can be interpolated by the electromagnetic current in the LSZ formalism.
In addition the dispersion representation may be useful for improving the fit ansatz of these FFs.

For our purposes it is convenient to first write the $B \to V$ FFs  
\cite{Wirbel:1985ji,BSZ15}
in the following form\footnote{Below $\Vp = (-\mBq) \V_i$ absorbs the factor on the left-hand side into the definition. This renders the $\Vp$ FFs  dimensionless.}
 \begin{alignat}{2}
 &    c^{(q)}_V \matel{V(k,\eta)}{\bar q \gamma^\mu(1 \mp \gamma_5) b}{\bar B(p_B)}  (-\mBq)
 &\;=\;&   \;\;   P_1^\mu \, \Vp_1^{B \to V}(q^2) \pm   \sum_{i=2,3,P}  P_i^\mu \Vp_i^{B \to V}(q^2)
   \; ,\nonumber  \\[0.1cm]
  & c^{(q)}_V \matel{V(k,\eta)}{\bar q iq_\nu \sigma^{\mu\nu} (1 \pm \gamma_5) b}{\bar B(p_B)}
  &\;=\;& \;\;  P_1^\mu  T_1^{B \to V}(q^2)   \pm  \sum_{i = 2,3} P_i^\mu  T_i^{B \to V}(q^2)
   \; ,
   \label{eq:ffbasis}
\end{alignat}
where $\eta$ is the vector-meson  polarisation,  $P_i^\mu$ are
Lorentz vectors\footnote{\label{foot:LSZ} Formally one should write $\eta^\rho(k) \to G_V^{\rho \mu} \eta_\mu(k)$ where
$G_V^{ \rho  \mu} = (g_{\rho \mu} - k_{\rho}k_{\mu}/m_V^2)$ such that the matrix elements
are invariants under $\eta \to \eta + k$ for on-shell $k^2 = m_V^2$. This follows from the LSZ formalism.}
\begin{alignat}{2}
\label{eq:Vprojectors}
& P_P^\mu = i (\eta^* \cdot q) q^\mu \; ,&P_1^\mu  =&  2 \epsilon^{\mu}_{\phantom{x} \alpha \beta \gamma} \eta^{*\alpha} k^{\beta}q^\gamma  \; , \\
& P_2^\mu = i (1 \mi \hat{k}^2) \{ \mBq^2 \eta^{*\mu} \mi
\frac{(\eta^*\!\cdot\! q)}{1- \hat{k}^2}(k+p_B)^\mu\} \; , \qquad
&P_3^\mu =&  i(\eta^*\!\cdot\! q)\{q^\mu \mi  \frac{\hat{q}^2 }{1\mi \hat{k}^2} (k+p_B)^\mu \}   \;,
\nonumber
\end{alignat}
where in \eqref{eq:ffbasis} but not \eqref{eq:Vprojectors} $k^2  = m_V^2$ is assumed, and
\begin{equation}
\label{eq:cV}
 c^{(d)}_{\omega} =      -c^{(d)}_{\rho^0}  =  \sqrt{2}  \;, \quad  c^{(s)}_\phi=1   \;,
 \end{equation}
 ($ c^{(u)}_{\omega}   = c^{(u)}_{\rho_0}  = \sqrt{2}$ are not used)
  take into account the composition
 of the vector mesons' wave-functions, $| \rho_0[\omega]  \rangle \approx ( | \bar u u \rangle \mp |\bar d d\rangle)/\sqrt{2}$
 and $|\phi \rangle \approx | \bar ss\rangle $.
The correspondence of    $\Vp^{B \to V}_{1,2,3,P}$  with the more traditional FFs $A^{B \to V}_{0,1,2,3}$ 
(e.g. \cite{BSZ15}) and $V^{B \to V}$
is as follows
\begin{alignat}{1}
& \Vp_P^{B \to V}(q^2) =  \frac{ 2 \hat{m}_{V}}{\hat{q}^2} A_0^{B \to V}(q^2) \;,  \quad \Vp_1^{B \to V}(q^2) =  \frac{ V^{B \to V}(q^2)}{1+\hat{m}_{V}} \;, \quad     \Vp_2^{B \to V}(q^2) =    \frac{A_1^{B \to V}(q^2)}{1-\hat{m}_{V}}
\;, \nonumber  \\[0.1cm]
 &  \Vp_3^{B \to V}(q^2) =    \frac{1-\hat{m}_{V}}{\hat{q}^2}    A_2^{B \to V}(q^2) -  \frac{1+\hat{m}_{V}}{\hat{q}^2}    A_1^{B \to V}(q^2) \equiv \frac{- 2 \hat{m}_{V}}{\hat{q}^2} A_3^{B \to V}(q^2) \; ,
 \label{eq:VAs}
\end{alignat}
where $\hat{m}_{V} \equiv m_V /\mBq$.
The analogue of the two ($q^2\!=\!0$) constraints (\ref{eq:A03},\;\ref{eq:algebraic})  are
 \begin{equation}
 \label{eq:q20BV}
 A^{B \to V}_3(0) = A^{B \to V}_0(0)  \;, \quad T^{B \to V}_1(0) = T^{B \to V}_2(0) \;,
\end{equation}
respectively. The constraint \eqref{eq:third} does not apply since $m_V^2 \neq \mBq^2$.

As stated above the relation between the FFs becomes apparent in the dispersion representation (cf.
the textbook \cite{Weinberg:1995mt} or the recent review \cite{Zwicky:2016lka}).
A specific example is chosen for illustration,\footnote{In order to distinguish the various dispersion representations
throughout this paper, we
use the variables $(s,t,u)$  for the momenta $(p_B^2,q^2,k^2)$.}
\begin{alignat}{2}
\label{eq:example}
& M_{T_5}^{\mu \rho} &\;=\;& - i \bT  \int d^4 x e^{i k \cdot x } \matel{ 0}{T j^\rho(x) \bar{q} iq_\nu \sigma^{\mu \nu}
\ga_5   b(0)}{\bar{B}_q}   \nonumber \\[0.1cm]
& &\;=\;&  \sum_{i=2,3} \, \int^{\infty}_{u_{\textrm{low}}}   \frac{\rho_{T_i}(q^2,u)\, du}{u- k^2-i0}
  (- G^{\rho}_{\phantom{\rho}\al} ) P_i^{\mu \al}  + \textrm{subtractions}\;,
\end{alignat}
where $u_{\textrm{low}} = (m_{P_1} +m_{P_2})^2$, and
$V \to P_1 + P_2$ is the lowest decay channel (e.g. $\rho^0 \to \pi^+ + \pi^-$ for $B_q = B_d$).
Note, that the appearance of the tensor $G^{\rho}_{\phantom{\rho}\al}$ \eqref{eq:G}
 goes hand in hand with the QED Ward identity constraint.

In order to further illustrate we resort to the narrow width approximation (NWA) which
can be improved by introducing a finite decay width or better multiparticle states of stable particles
(cf. remark at beginning this section).
With the NWA
\begin{equation}
\label{eq:why}
\sum_{\la = -1,0,1}  \eta_\rho^*(k,\la)\eta_\al(k,\la)  = \left( - g_{\rho\al} + \frac{k_\rho k_\al}{m_V^2} \right)
 \big|_{m_V^2 = k^2}  =  (-G_{\rho\al}) \;,
\end{equation}
and the  spectral or discontinuity function
 $\rho_{T_i}(q^2,u)$ assumes the simple form
\begin{equation}
\rho_{T_i}(q^2,u) = \de(u-m_\rho^2) r^{\rho}_{T_i}(q^2) +  \de(u-m_\omega^2) r^{\omega}_{T_i}(q^2)  + \dots \;,
\end{equation}
where the dots stand for higher states in the spectrum.
The residua  $r^{V}_{T_i}$  are then given by
\begin{equation}
\label{eq:residua}
r^{V}_{T_i} = - m_V  f^{\textrm{em}}_V/|c^{(d,s)}_V|^2 \,  T^{B \to V}_i(q^2)  \;,
\end{equation}
where $f^{\textrm{em}}_V$ is a conveniently normalised  matrix element
\begin{equation}
 (c^{(d,s)}_V)^* \matel{0}{j_\mu}{V(k,\eta)} =   m_V f^{\textrm{em}}_V \eta_\mu \;,
\end{equation}
of the electromagnetic current and the vector meson.
In particular
\begin{equation}
\label{eq:fem}
f^{\textrm{em}}_{\rho_0} =  (Q_d -Q_u)  f_{\rho_0} = -  f_{\rho_0} \;, \quad
 f^{\textrm{em}}_{\omega} = (Q_d+Q_u) f_\omega =  \frac{1}{3}   f_{\omega}  \;, \quad
 f^{\textrm{em}}_{\phi} =  Q_s f_\phi = -  \frac{1}{3}  f_{\phi}  \;.
 \end{equation}
Rewriting our parametrisation \eqref{eq:NEW}, in compact form,
\begin{equation}
M_{T_5}^{\mu \rho} = \sum_{J = \parallel ,\LL} R_J^{\mu\rho} T^{B\to \ga^*}_J(q^2,k^2)  \;,
\end{equation}
and equating with \eqref{eq:example} we are able to identify the two bases
\begin{equation}
  R_J^{\mu \rho} \kmat_{Ji}(q^2,k^2) =  P_i^{\mu \al} (-  G^{\rho}_{\phantom{\rho}\al}) \;,
\end{equation}
where  $\kmat_{Ji}$ is, by construction, a matrix with diagonal entries 
\begin{equation}
\label{eq:kmatrix}
\kmat_{\perp 1}(q^2,k^2)  =  2 \;, \quad \kmat_{\parallel 2}(q^2,k^2)  = 2\frac{1 - \hat{k}^2}{1- \hat{q}^2} \;, \quad
\kmat_{\LL 3}(q^2,k^2)  = 2\;,
\quad \kmat_{PP}(q^2,k^2)  = 2 \;,
\end{equation}
 only. Of course we could have chosen any other basis at the cost of having
a non-diagonal $\kmat$-matrix but we feel that this is an economic way.

Let us make the dispersion representation more concrete by clarifying  what the subtraction terms mean in \eqref{eq:example}.
Whether or not the $B \to \ga^*$ FFs does  require a subtraction is practically not important
since it is advantageous to do so anyway as the on-shell FF, which provides the subtraction information, is
well-known and also physical.\footnote{\label{foot:asymptotic}
The asymptotic behaviour of the triangle function at LO, cf. \FIG\;\ref{fig:dia-Cor} (left, center),
is  ${\cal O}(\ln k^2)$ as can be inferred from the explicit expressions in \eqref{eq:dens}.
This continues to hold  when  resumming the leading-log expressions,
cf. Ref.~\cite{Prochazka:2016ati}, for the correlation functions in question (with $N_c =3$).
The dispersion relation in $p_B^2$  and its restriction to the finite interval $[m_b^2,s_0]$
leads to a $F_{\textrm{LOPT}} \propto 1/k^2$ asymptotic behaviour. It is possible that this changes at NLO.
However, there are condensate terms of order $F_{\textrm{LO}\vev{\bar qq} } \propto {\cal O}(1)$ as can be inferred from \eqref{eq:dens} originating from
 diagrams where the photon is emitted from the $b$-quark.
It might be  that at NLO this turns out to be ${\cal O}(\ln k^2)$.
Note that, reassuringly,  in the limit $q^2,k^2 \to - \infty$ all condensate terms vanish as one would expect 
since  perturbation theory dominates in this regime.  
At least when investigating the mixed quark-gluon condensate term one
can see that it is suppressed relative to the condensate terms suggesting that the OPE itself converges.
In passing let us note that the difference to   $B \to V$ or $B \to \ga$ FFs is the off-shellness of the photon
which does not require a double cut for the FF and can lead to a more divergent expression.}
One may write
\begin{alignat}{2}
\label{eq:FFk2}
& T_J^{B \to \ga^*}(q^2,k^2)  &\;=\;&
T_J^{B \to \ga^*}(q^2,\ksub) +
\kmat_{Ji} \, (k^2-\ksub)  \int^{\infty}_{u_{\textrm{low}}}  \frac{ \, \rho_{T_i}(q^2,u)\, du}{(u - \ksub -i0)(u- k^2-i0)}  \;,
\end{alignat}
where $J = \perp,\LL,P$ and $i = 1,3,P$ with $\kmat_{Ji}$ defined above and
the same formula applies for $T^*_J \to V^*_J$ and $\rho_{T_i} \to \rho_{\Vp_i}$.
The $T,V_\parallel$-FFs are a bit more involved since the
matching  factor $\kmat_{\parallel 2} \propto (1-\hat{k^2}) $  contain a non-trivial $k$ dependence.
In order to avoid the artificial pole we may define an expression
\begin{equation}
\Delta T^{B \to \ga^*}_{\parallel}(q^2,k^2) =  \frac{T^{B \to \ga^*}_{\parallel}(q^2,k^2) - T^{B \to \ga^*}_{\parallel}(q^2,\mBq^2)}{\kmat_{\parallel 2 }(q^2,k^2) } \;,
\end{equation}
which establishes regularity at $k^2 = m_{B_q}^2$ and the corresponding subtracted dispersion relation reads
\begin{alignat}{2}
\label{eq:FFk2p}
& \Delta T_\parallel^{B \to \ga^*}(q^2,k^2)  &\;=\;&
\Delta T_\parallel^{B \to \ga^*}(q^2,\ksub) +
\, (k^2-\ksub)  \int^{\infty}_{u_{\textrm{low}}}  \frac{ \, \rho_{T_2}(q^2,u)\, du}{(u - \ksub -i0)(u- k^2-i0)}  \;.
\end{alignat}
 The same applies again for $V_\parallel^*$ with the substitutions  $T^*_\parallel \to V_\parallel^*$ and $\rho_{T_2} \to \rho_{\Vp_2}$.
The analogy with \eqref{eq:FFk2} is restored if one divides the latter equation by $\kmat_{Ji}$.


For the sake of clarity, we give a few examples of FFs in the  $k^2$-dispersion representation \eqref{eq:FFk2}
which illustrates some of its properties:
\begin{alignat}{2}
\label{eq:examples}
 & V_\perp^{B_d \to \ga^*}(q^2,k^2) &\;=\;& V_\perp^{B_d \to \ga^*}(q^2,\ksub) -
\frac{\kmat_{\perp 1}}{2} (k^2 - \ksub)  \left(
\frac{ m_{\rho} f^{\textrm em}_\rho \, \Vp_1^{B_d \to \rho}(q^2)   }{(m_{\rho}^2 - k^2)(m_\rho^2-\ksub)}  +
\frac{ m_{\omega}  f^{\textrm em}_\omega  \, \Vp_1^{  B_d \to \omega}(q^2) }{(m_{\omega}^2 - k^2)(m_\omega^2-\ksub)}  + \dots  \right) \;,
\nonumber \\[0.1cm]
 & \hat{V}_\LL^{B_s \to \ga^*}(q^2,k^2) &\;=\;&  \hat{V}_\LL^{B_s \to \ga^*}(q^2,\ksub) -
 \kmat_{\LL 3} (k^2-\ksub)  \left(\frac{m_\phi}{\mBq}  \frac{ m_\phi f^{\textrm em}_\phi  \,
 A_3^{B_s \to \phi}(q^2)   }{(m_{\phi}^2 - k^2)(m_\phi^2-\ksub)}   + \dots  \right) \;, \nonumber \\[0.1cm]
 & P^{B_s \to \ga^*}(q^2,k^2) &\;=\;&  P^{B_s \to \ga^*}(q^2,\ksub)  -
 \kmat_{PP} (k^2-\ksub)  \left( \frac{m_\phi}{\mBq}  \frac{ m_\phi f^{\textrm em}_\phi \, A_0^{B_s \to \phi}(q^2)   }{(m_{\phi}^2 - k^2)(m_\phi^2-\ksub)}   + \dots  \right) \;,\nonumber \\[0.1cm]
 & T_\perp^{B_s \to \ga^*}(q^2,k^2) &\;=\;&  T_\perp^{B_s \to \ga^*}(q^2,\ksub)-
\kmat_{\perp 1} (k^2-\ksub)   \left( \frac{ m_\phi f^{\textrm em}_\phi \, T_1^{B_s \to \phi}(q^2)   }{(m_{\phi}^2 - k^2)(m_\phi^2-\ksub)}   + \dots  \right) \;,
\end{alignat}
where the minus sign is due to the minus sign in \eqref{eq:residua} which in turn comes
from the electromagnetic interaction term.
Above the $k^2 +i0$ prescription has been dropped for brevity and
$|c_{\rho^0}|^2 = |c_{\omega}|^2 = 2$ has been used.

These formulae show that properties of the $B \to \ga^*$- and $B \to V$-FFs imply each other.
For example, the
$B \to V$ constraint $A^{B \to V}_0(0) = A^{B \to V}_3(0)$ \eqref{eq:q20BV} implies
the constraint $ \PFF^*(0,k^2) = \hat{V}^*_\LL(0,k^2)$ (\ref{eq:A03}).
The algebraic relation \eqref{eq:algebraic} follows from  \eqref{eq:FFk2}, if  $T_\parallel(0,\mBq^2) =0$ holds
which in turn follows from \eqref{eq:third}.
In the $SU(3)_F$ limit $m_u  = m_d = m_s$, $m_V$ and $f_V$ are degenerate and
\eqref{eq:fem}, $f^{\textrm em}_\rho/|c_\rho|^2 +  f^{\textrm em}_\omega/|c_\omega|^2  = f^{\textrm em}_\phi/|c_\phi|^2 $ and
$F^{B_d \to \rho} = F^{B_d \to \omega} = F^{B_s \to \phi}$ which finally  implies $F^{B_d \to \ga^*} = F^{B_s \to \ga^*} $ as expected. These relations can be turned around since they hold for any $k^2$, they necessarily hold
at each point of the spectrum and thus for the $B \to \ga^*$ properties imply the $B \to V$ FF properties.

Moreover, the examples  reveal that the slope of the FF are positive which is the choice by convention.
This is the case since
$r_\phi > 0$ and $r_\rho > |r_\omega| > 0$.
At last let us note that a particularly convenient form for $P^*$ can be obtained
\begin{equation}
\label{eq:Pnice}
P^{B_s \to \ga^*}(q^2,k^2) \;=\;
- 2 k^2 \left( \frac{1}{\mBq}  \frac{ f^{\textrm em}_\phi \, A_0^{B_s \to \phi}(q^2)   }{(m_{\phi}^2 - k^2)}   + \dots  \right) \;,
\end{equation}
if ones chooses the subtraction point
$\ksub =0$ where the pseudoscalar FF vanishes.
This corresponds to \eqref{eq:exampleP} in the main text given as an illustration.

\subsection{Explicit results of the off-shell form factors \label{app:FFcomp}}

\subsubsection{QCD sum rule for the off-shell form factors $\PFF^*(0,k^2)$,  $T^*_{\perp,\LL}(0,k^2)$
and $\FVA^*(0,k^2)$}

The FFs are computed using QCD SRs \cite{Shifman:1978bx}. The starting point is the correlation function
of the form
 \begin{eqnarray}
\label{eq:corrPT}
\Corr_{\mu \rho} ^{V-A}(p_B,q)
&\equiv &\!\! -  i^2 (b_V s_e e  )
 \int_{x,y} e^{-ip_B \cdot x} e^{i k \cdot y} \matel{0}{ T j_\rho(y)  J_{B_q}(x)
 \bar{q} \gamma^{\mu}(1- \ga_5)   b
 (0)}{0}  \nonumber  \\[0.1cm]
 &=& \!\!
  \RR^\perp_{\mu \rho} \, \Corr^V_\perp
- (\RR^\parallel_{\mu \rho} \, \Corr^A_\parallel + \RR^\parallel_{\mu \rho} \, \Corr^A_\parallel
+\RR^P_{\mu \rho} \, \Corr^A_P)  + C_{\mu \rho }(q^2) \;,
\end{eqnarray}
where $(b_V s_e e  ) = - \mBq$, $\Corr^{V,A} = \Corr^{V,A}(q^2,p_B^2,k^2)$ are  analytic functions in three  variables
and
the Lorentz structures $\RR_{\mu \rho}$ are defined in \eqref{eq:RNEW}.
Gauge invariance, again, holds in the simplest form $k^\rho \Corr_{\mu \rho} ^V(p_B,q)=0$ since
we work  with electrically neutral states. The term $C_{\mu \rho }(q^2)$ is a contact term but of no relevance
for our purposes since they are $p_B^2$-independent. It is the correction to the naive non-singlet axial
Ward identity \eqref{eq:AWI}. The operator $J_{B_q} \equiv (m_b + m_q) \bar{b} i \ga_5 q $ is the interpolating operator for the
$B_q$-meson with matrix element $ \matel{\bar{B}_q}{J_{B_q}}{0} = m_{B_q}^2 f_{B_q}$.

\begin{figure}[t]
	\centering
	\includegraphics[scale=0.56,trim={0 0 0 0},clip]{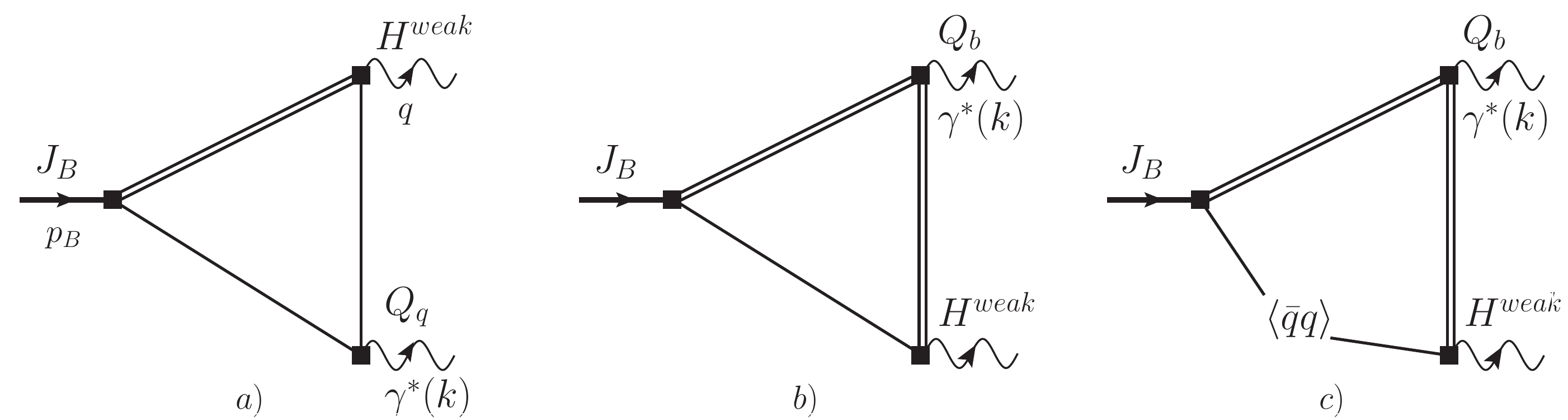}
	\caption{
	Figures for off-shell form factors $V,T_{\perp,\parallel,\LL}^*$ and $\PFF^*$. The single/double
         lines denote the $q/b$-quark respectively.
	The diagrams on the left and right are the perturbative- and the one
	on the left is the quark condensate-type. The  quark condensate diagram corresponding
	to fig a) is not shown. It is proportional to $\vev{\bar q q}/k^2$ and implicitly assumes
	$k^2 \neq 0$. In the $k^2 \to 0$ limit these diagrams are replaced by the photon
	distribution amplitude e.g. \cite{Janowski:2021yvz}. In the SR method the $B$-meson
	is projected out via a dispersion relation in the variable $p_B^2$ giving access  to the
	matrix element of the off-shell form factor $F^*(q^2,k^2)$. The momentum assignment corresponds to
	the convention in the Appendix which differs from the phenomenological discussion in the main text.
	\label{fig:dia-Cor}
	}
\end{figure}

The QCD SR is then obtained by evaluating \eqref{eq:corrPT} in the  operator product expansion (OPE) (cf. \FIG~\ref{fig:dia-Cor})
and equating it to the dispersion representation.  The OPE consists of a perturbative part
and a condensate part for which we include only the quark condensate.
The OPE is convergent, in a pragmatic sense, for momenta $p_B^2, q^2 < {\cal O}(m_b \Lambda)$
and $k^2 <  - \Lambda^2$ with $\Lambda \approx 500\MeV $ a typical hadronic scale.
The perturbative part is evaluated with the help of FeynCalc \cite{FeynCalc1,FeynCalc2}.
We neglect light-quark masses i.e. $ m_d = m_s = 0$.

The dispersion representation of $\Corr^V_\perp$  reads
\begin{equation}
\Corr^V_\perp(p_B^2,q^2,k^2) = \frac{1}{\pi} \int_{0}^\infty  \frac{\textrm{Im}[\Corr^V_\perp(s,q^2,k^2)] \, ds}{s-p_B^2- i0} =
\frac{  \mBq^2 \fBq  \FV^{B \to \ga^*}(q^2,k^2) }{m_{B_q}^2 - p_B^2 - i0}  + \dots  \;,
\end{equation}
where the dots stand for higher  resonances and multiparticle states. Moreover
 the NWA for the $B$-meson has been assumed. The FFs are then extracted via
 the standard procedures of Borel transformation and by approximating the ``higher states" contribution
 by the perturbative integral \cite{Shifman:1978bx}.
  The latter is exponentially suppressed
 \begin{equation}
 \label{eq:FFSR}
 \FV^{B \to \ga^*}(q^2,k^2)  = \frac{1}{ m_{B_q}^2 f_{B_q }}
  \int_{m_b^2}^{s_0}    e^{(m_{B_q}^2-s)/M^2} \rho_{V^*_\perp}(s,q^2,k^2) \, ds  \;,
 \end{equation}
 due to the  Borel transform in $p_B^2$. Note, that the contact term $C_{\mu \rho }(q^2)$, which can appear
 as a subtraction constant in the dispersion relation, vanishes  under the Borel transform.
 Above $ \pi \rho_\perp^V(s,q^2,k^2) = \textrm{Im}[\Corr^V_\perp(s,q^2,k^2)]$ and $M^2$ is the Borel mass.
 If we were able to compute $\rho_\perp^V$
 exactly then $\FV(q^2)$, obtained from \eqref{eq:FFSR}, would be independent of the Borel mass and it therefore serves as a quality measure
 of the SR. Other FFs are obtained in exact analogy with the exception of $F_{\parallel}$ $(F=V,T)$
 
 Let us turn to a technical point. Namely on how to avoid spurious kinematic singularities.
 The constraints \eqref{eq:third} avoid these constraints and in the computation 
 they are satisfied provided that  $s=m_{B_q}^2$. However,  since that is only satisfied within ${\cal O}(1\%)$ in a SR some care needs to be taken. 
This procedure is equivalent to considering the FF combination in \eqref{eq:FFsec2} 
proportional to $1/(1-\hat{k}^2)$ as a single FFs and thus avoids this spurious pole which 
is the correct treatment. More concretely, let us introduce  
 $F_\textrm{aux} = \frac{1}{1-\hat{k}^2} (F_\parallel + \frac{\hat{q}^2}{1-\hat{q}^2} F_L)$ and then define 
 $F_\parallel = (1- \hat{k}^2) F_{\textrm{aux}} -    \frac{\hat{q}^2}{1-\hat{q}^2} F_L$.  
 This procedure is equivalent to defining the improved density as follows:
 \begin{equation}
 \label{eq:improvement}
\tilde{ \rho}_{F_\parallel} =    \left[ \rho_{F_\parallel}
 + \frac{\hat{q}^2}{1- \hat{q}^2}  \rho_{F_{\mathbb{L}}}  \right] U(s,k^2)  - \frac{\hat{q}^2}{1- \hat{q}^2}  \rho_{F_{\mathbb{L}}}  \; , \quad  U(s,k^2) =
  \frac{1-k^2/{m_{B_q}^2} }{1-k^2/s} \;.
 \end{equation}
Crucially, the extra $1/(1-k^2/s)$ does not render the density singular at that point 
which is  equivalent to the statement that there are  no $1/(1-k^2/m_B^2)$ poles 
in the matrix elements. Let us emphasise that if the results carry statistical errors, such as in lattice Monte Carlo simulations then it might be advantageous to directly fit $F_{\textrm{aux}}$ and $F_\LL$ and then regain 
$F_\parallel$ from the formula above. This ensures cancellation of the pole in that case.

Before stating the results of the computations let us turn  to the issue of analytic continuation.
We would like to employ our FFs in the Minkowski region  $k^2 > 0$, whereas the OPE is convergent
for $k^2 <  - \Lambda^2$. The convergence is broken by thresholds at $k^2 = 4 m_q^2$ which
signal long-distance effects corresponding
to $\rho/\omega$ ($\phi$)-like resonances  cf. \TAB~\ref{tab:FFoverview}.
The standard procedure is to analytically continue into the Minkowski region and use the
FF for say $k^2 > 4 \GeV^{\,2}$ which is far enough from the lowest lying narrow resonances.
For  $k^2 > 4 \GeV^{\,2}$ the resonances are broad and disappear into the continuum.
Under such circumstances  local quark-hadron duality is usually assumed to be a reasonable approximation.
In our region of use  $k^2 \in [\qsqlow, m_{B_{d,s}}^2]$ there are no narrow resonances in the
$k^2$-channel.\footnote{In fact one should be able to use these computations up
to $k^2 \approx 4 m_b^2 - {\cal O}(10 \GeV^{\,2})$ at least.}
 On a pragmatic level it is best to implement the $ \FV^{B \to \ga^*}(q^2,k^2+i0)$-prescription in the process
 of analytic continuation by deforming the path in \eqref{eq:FFSR} from $ \int_{m_b^2}^{s_0} ds \to  \int_\ga ds $, where
 $\ga$ is a path in the lower-half plane starting at $m_b^2$ and ending at $s_0$. One may for instance choose
 a semi-circle in the lower half-plane. This prescription leads to numerical stability.
 Clearly our computation remains valid and useful for $D^0 \to \ga^*$ FFs with replacements
 $B_q \to D^0$ and $m_b \to m_c$.

\subsubsection{Explicit results for the off-shell form factors from QCD sum rules}
\label{app:densities}

The explicit FFs are found to be
\begin{alignat}{2}
\label{eq:dens}
& \PFF^*(0,k^2) &\;=\;& \frac{e^{\mBq^2/M^2}}{\fBq \mBq^2}
Q_b \left(\, \int_{m_b^2}^{s_0} e^{-s/M^2}   \rho_{\PFF^*}(s,0,k^2) ds -   \frac{2}{\bar{m}_{B_q}} \vev{ \bar q q} e^{-m_b^2/M^2} \right)  + {\cal O}(\alpha_s,m_q) \;,  \\
& T_{\perp}^*(0,k^2) &\;=\;& \frac{e^{\mBq^2/M^2}}{\fBq \mBq^2} Q_b \left(\,  \int_{m_b^2}^{s_0} e^{-s/M^2} \rho_{\FTV^*}(s,0,k^2) ds  + \left( 1 - \frac{1}{\bar{k}^2} \right)     \vev{ \bar q q} e^{-m_b^2/M^2} \right)  + {\cal O}(\alpha_s,m_q)  \;, \nonumber \\
& T_{\LL}^*(0,k^2) &\;=\;& \frac{e^{\mBq^2/M^2}}{\fBq \mBq^2} Q_b \left(\,  \int_{m_b^2}^{s_0} e^{-s/M^2} \rho_{\FTL^*}(s,0,k^2) ds +   \left( 1 - \frac{1}{\bar{k}^2} \right)   \vev{ \bar q q} e^{-m_b^2/M^2} \right)  + {\cal O}(\alpha_s,m_q)  \;, \nonumber \\
& \FV^*(0,k^2) &\;=\;& \frac{e^{\mBq^2/M^2}}{\fBq \mBq^2} Q_b \left(\,  \int_{m_b^2}^{s_0} e^{-s/M^2} \rho_{\FV^*}(s,0,k^2) ds +  \bar{m}_{B_q} \left(  1 - \frac{1}{\bar{k}^2} \right)   \vev{ \bar q q} e^{-m_b^2/M^2} \right)  + {\cal O}(\alpha_s,m_q)  \;, \nonumber \\
& \FA^*(0,k^2) &\;=\;& \frac{e^{\mBq^2/M^2}}{\fBq \mBq^2} Q_b \left(\,  \int_{m_b^2}^{s_0} e^{-s/M^2} \tilde{\rho}_{\FA^*}(s,0,k^2) ds +   \bar{m}_{B_q}   \left(  1 + \frac{1}{\bar{k}^2} \right)  \vev{ \bar q q} e^{-m_b^2/M^2} \right)  + {\cal O}(\alpha_s,m_q)  \;, \nonumber
\end{alignat}
where $\bar{m}_{B_q} \equiv \mBq/m_b$,  $\bar{s} \equiv s/ m_b^2$, $\bar{k}^2 \equiv k^2/m_b^2$ and
the perturbative densities
\begin{equation}
\rho_i   \equiv  \frac{N_c m_b}{ 4 \pi^2 (\bar{s} \!- \!\bar{k}^2)^3}  \hat{\rho}_i\;,
\end{equation}
are given by
\begin{alignat}{2}
& \hat{\rho}_{\PFF^*}(s,0,k^2)  &\;=\;&    2 \bar{k}^2  \bar{m}_{B_q} /s  \Big(  \big( (\bar{s} - \bar{k}^2)-1\big)\, L_q +L_b \Big)  \;,  \nonumber \\[0.2cm]
&  \hat{\rho}_{\FTV^*}(s,0,k^2) &\;=\;&  \big( \bar{k}^4 (1-\bar{s})+ 2\bar{k}^2  (\bar{s}-1)-(\bar{s}-1) \bar{s}^2 \big)  +
   \bar{k}^2 \, L_q  - \bar{k}^2\, L_b    \;, \nonumber \\[0.2cm]
&  \hat{\rho}_{\FTL^*}(s,0,k^2) &\;=\;&
(\bar{s}-1) (5 \bar{k}^2-\bar{s}) (\bar{s} - \bar{k}^2)  +
       \big(  \bar{k}^2/ (\bar{s} - \bar{k}^2)  (4 \bar{k}^4+7 \
\bar{k}^2+(5-4 \bar{s}) \bar{s}) \big)\, L_q   + \nonumber \\[0.1cm]
& &\;\phantom{+}\; &    \big(\bar{k}^2/  (\bar{s} - \bar{k}^2)  ( \bar{k}^2 (8 \
\bar{s}-7)-  \bar{s} (8 \bar{s}+5))  \big) \, L_b \;, \nonumber \\[0.2cm]
&  \hat{\rho}_{\FV^*}(s,0,k^2) &\;=\;&
\bar{m}_{B_q} \{  \big( \bar{k}^4 - \bar{k}^2 (\bar{s}-2) \big)\, L_q   +  \big( \bar{k}^2 (\bar{s}-2) -\
\bar{s}^2 \big)\,L_b \} \;, \nonumber \\[0.2cm]
&  \hat{\rho}_{\FA^*}(s,0,k^2) &\;=\;&  \bar{m}_{B_q}
\{ 2 (\bar{k}^2/ \bar{s}^2) (\bar{s}-1) (\bar{s} - \bar{k}^2)^2   +   \big( (\bar{k}^2/ \bar{s})\, (
\bar{k}^4-2 \bar{k}^2 (\bar{s}-1)+\bar{s}^2-2 \bar{s}+2 ) \big)  \, L_q + \nonumber \\[0.1cm]
& &\;\phantom{+}\;&   \big(  ( \bar{k}^4 \bar{s}-2 \bar{k}^2
(\bar{s}^2-\bar{s}+1)+(\bar{s}-2) \bar{s}^2)/s \big) \, L_b  \}    \;,
\end{alignat}
and the
improvement discussed around \eqref{eq:improvement} reads in the case at hand
\begin{equation}
\tilde{\rho}_{V^*_{\parallel}} =   \left[ {\rho}_{V^*_{\parallel}} - 2  \rho_{P^*} \right] U(s,k^2)
+  2  \rho_{P^*} \;.
\end{equation}
Further note that the $1/k^2$ factor is of no concern since $P^*(0,k^2) \propto k^2$.
The logarithms $L_q$ and $L_b$
\begin{equation}
\label{eq:logs}
 L_q \equiv  \ln
   \left(\frac{ \bar{k}^2}{\bar{s}
   \left(1+\bar{k}^2-\bar{s}\right)}\right) \;, \quad      L_b \equiv  \ln
   \left(\bar{s}-\bar{k}^2+\frac{\bar{k}^2}{\bar{s}} \right)  \;,
\end{equation}
  lead to imaginary parts in the FFs for
$k^2 > 4 m_q^2 = 0$ and $k^2 > 4 m_b^2$ respectively. From the viewpoint of the original function these
singularities are anomalous thresholds which consist of putting all the propagators on-shell.
These expression are consistent  with
the $B \to V\ell\ell$ weak annihilation computation detailed in appendix of Ref.~\cite{Lyon:2013gba}
cf. footnote \ref{foot:before} for further remarks.
Note that, there is no singularity at $k^2 = s$ when expanded properly.
The condensate contributions could be written in terms of the densities $\rho_i$ as well.
The backward substitution   $e^{-m_b^2/M^2} \vev{\bar q q} \to \vev{\bar q q} \de( s- m_b^2)$
achieves this task.

A few comments on interpreting the results.
The $k^2 \to 0$ limit is not well-defined for the condensates.
In that limit the condensates originating from quark lines attached to the photon are replaced
by a photon distribution amplitude  which makes the FFs computation more involved.
However, the perturbative part remains well-defined in that limit.
Hence the latter must contribute positively to the $\{T_{\perp,\parallel}(0,0), \,V_{\perp,\parallel}(0,0)\}$
by convention which can be verified indeed by using $Q_b = -1/3$ and sending $\hat{k}^2 \to 0$.
The $q^2$ constraints (\ref{eq:A03},\ref{eq:algebraic}) are obeyed exactly by the SRs and are assumed
as we do not show $\hat{V}_\LL(0,k^2)$ and $T_\parallel(0,k^2)$;  they are simply redundant.
The constraints at $k^2 = \mBq^2$ \eqref{eq:third2} are obeyed for the correlation functions with
$k^2 = p_B^2$. However they do not hold exactly for the FFs as $p_B^2 \approx \mBq^2$ within the approximation
of the Borel procedure. We have checked that these relations hold to within $2\%$ where for the last one
we compare to a value of the FF at $q^2 = 10\GeV^{\,2}$. In the fits we have implemented these constraints
as they are important to cancels the poles present in the Lorentz structures $\RR_\parallel^{\mu\rho}$
and $\RR_\LL^{\mu\rho}$.

The expressions could be improved by $m_q \neq 0$, adding the gluon
condensate and radiative corrections. The first two are expected to be rather small effects
since $1 \gg m_q/\Lambda_{\textrm{QCD}}$ and  $m_b \vev{\bar qq} \gg \vev{G^2}$. 
On the other hand, radiative corrections could be sizeable and would  reduce the scale uncertainty considerably.  This leads us naturally to the sum rule input and error discussion. 

\subsubsection{Numerical input to sum rules and rough uncertainty estimates}

For a LO computation of the sum rule type discussed here the most important uncertainty will 
be due to the $b$ quark mass and its associated scheme.  Since the sum rule is effectively a ratio of two sum rules,
\begin{equation}
F(0,k^2) = \frac{ [f_B  F(0,k^2)]_{\textrm{SR}}}{[f_B]_{\textrm{SR}}} \;,  \quad F = P,T_{\perp,L} V_{\perp,\parallel} \;,
\end{equation}
 various uncertainties cancel. Some guidance can be taken 
from the NLO computation of the $B$-meson decay constant.  For the latter it is well-known that 
NLO corrections are moderate in the MS-bar mass-scheme \cite{Jamin:2001fw,PZ19} it is therefore advisable 
to use that scheme  anticipating the smallest scale uncertainty.  
The LO expression are $f_{B_q},f_{B_s} \approx ( 222, 244 ) \MeV$ with 
the MS-bar mass $m_b = 4.18(4)\textrm{GeV}$ \cite{PDG18}.
 In fact the NLO corrections slightly reduce these values, unlike in other schemes, by less than $10\%$. Hence one can anticipate an  uncertainty of that order.  The other input are the condensates for which 
 $\vev{\bar q q}_{\mu = 2 \textrm{GeV}} =  -(0.269(2) \textrm{GeV})^3$ (e.g. \cite{PDG18}) is well-known
 from the Gell-Mann Oakes Renner relation and the strange quark condensate is taken from a lattice computation $\vev{\bar ss}_{\mu = 1 \textrm{GeV}} = 1.08(16) \vev{\bar qq}_{\mu = 1 \textrm{GeV}} $
\cite{McNeile:2012xh}.
The SR specific parameters, the Borel mass and the continuum threshold, are determined by a series 
of constraints.  First and foremost we impose these constraints at Euclidean values $k^2  < 0$ 
as for positive $k^2$ the logarithms can lead to large cancellations and invalidate the prcedure. We then proceed by  
analytic continuation $k^2 \to - k^2$
 to obtain the values for $k^2 >0 $ as found in the results of this paper. 
This is generally believed to be a reasonable approximation as long as $k^2$ is not 
in a resonance region per se (smooth QCD curves).
An example where this works well is $e^+ e^- \to \textrm{hadrons}$ in  the region above the broad regions
as can be inferred from the $R \propto \sigma (e^+ e^- \to \textrm{hadrons})/\sigma(e^+ e^- \to \mu^+\mu^-)$-plots in  \cite{PDG18}.

The  first constraint comes from the formally exact relation, 
\begin{equation}
\label{eq:daughter}
	\mBq^2 =  -  e^{\mBq^2/M^2}  \frac{d}{d(1/M^2)} e^{-\mBq^2/M^2}  \ln \PFF^*(k^2,M^2,s_0)\,,
\end{equation}
often referred to as the daughter sum rule, which we require to be  satisfied within ${\cal O}(2\%)$. 
Next $s_0$ ought to be in the window $(m_B + 2 m_\pi)^2$ and $(m_B + m_\rho)^2$ which is 
compatible with \eqref{eq:daughter}. The Borel parameter is further constrained by requiring two 
standard  criteria i) the condensate not to exceed  $10\%$  assuring convergence of the OPE 
ii)  the continuum contribution  not exceed $30\%$ which assures that the $B$-meson is projected out rather than an entire set of states.  
Note that the later two conditions are exclusive in that the former requires a large and the latter a smaller 
Borel parameter. The values adapted for the thresholds are $(s_0^{B_q}, s_0^{B_s})  = (35(2),35.5(2.0))\GeV^2$.  The $B_d$ Borel parameter for $P,T_{\perp,L}, V_\perp$ is taken to be 
$M^2 = 9(2)\GeV^2$ for $|k^2|  <  10\GeV^2$ and 
$M^2 = 7(2)\GeV^2$ for $|k^2|  >  15\GeV^2$  with smooth interpolation  and finally $M^2_{V_\parallel} = 6(2) \GeV^2$. A global shift is applied for the  $B_s$ Borel parameters 
 $M^2_{B_s} = M^2_{B_q}+ 0.5\GeV^2$, which respect the mass ratios of the meson roughly.
 
Let us turn to the uncertainty  analysis.
For the main section we use a $z$-expansion fit with similar uncertainty analysis as in
 Ref.~\cite{BSZ15} with some more detail delegated to the fit-section. This analysis and fit is 
 restricted to  high $q^2$. In addition we append a Mathematica notebook to the arXiv version 
 for reproducing the plots. The brief comments on uncertainties apply to this version.  
 The main sources of uncertainty are the Borel parameter and the scale uncertainty. 
 The threshold uncertainty is negligible and the quark condensate uncertainty is only relevant for 
 the $B_s$-mode where its impact is still moderate in almost all cases and regions. 
 The uncertainty due to NLO corrections is estimated to be $15\%$ on grounds of the known 
 radiative corrections to the $B$-meson decay constant as previously discussed in this section. 
The uncertainty of the Borel parameter is roughy $20\%$, which is on the conservative side,  
and when added in quadrature this  amounts to a $\approx 25\%$ uncertainty.  An NLO 
computation would  presumably not only reduce the scale but also the Borel uncertainty. 
We refrain from a more elaborate error analysis altogether including the extension in the resonance region.
This is in principle not difficult to achieve in that one can at first just vary the input into the dispersion relation. 
This procedure ought to give a reasonable error estimate.

\subsection{Extending the off-shell form factors below  the ${\cal O}(\kOPE \GeV^{\,2})$-region}
\label{app:mutlidisp}

\begin{figure}[t]
	\includegraphics[scale=0.85]{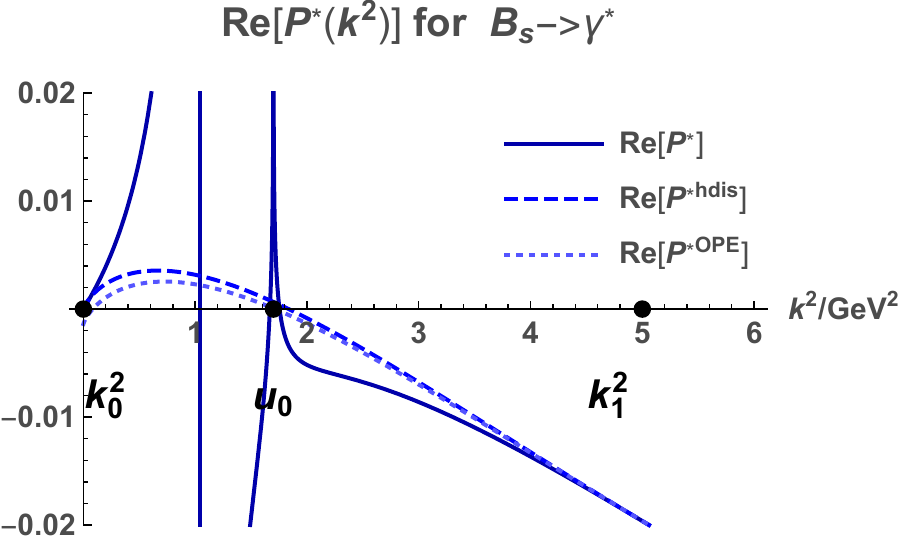}
	\includegraphics[scale=0.85]{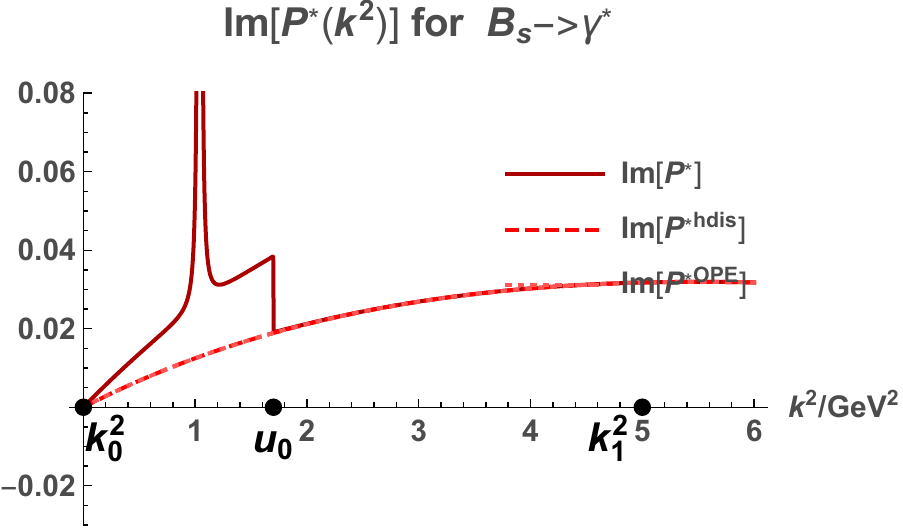} \\
	\includegraphics[scale=0.85]{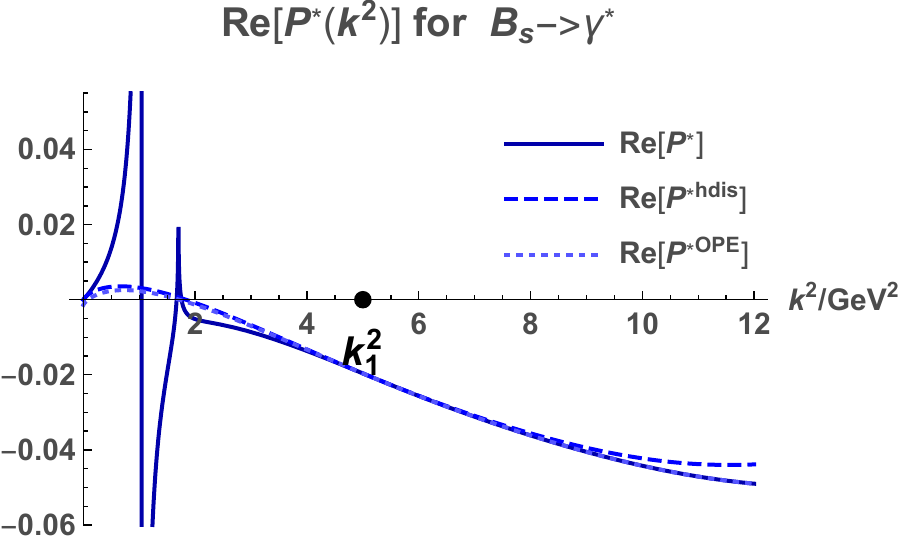}
	 \includegraphics[scale=0.55]{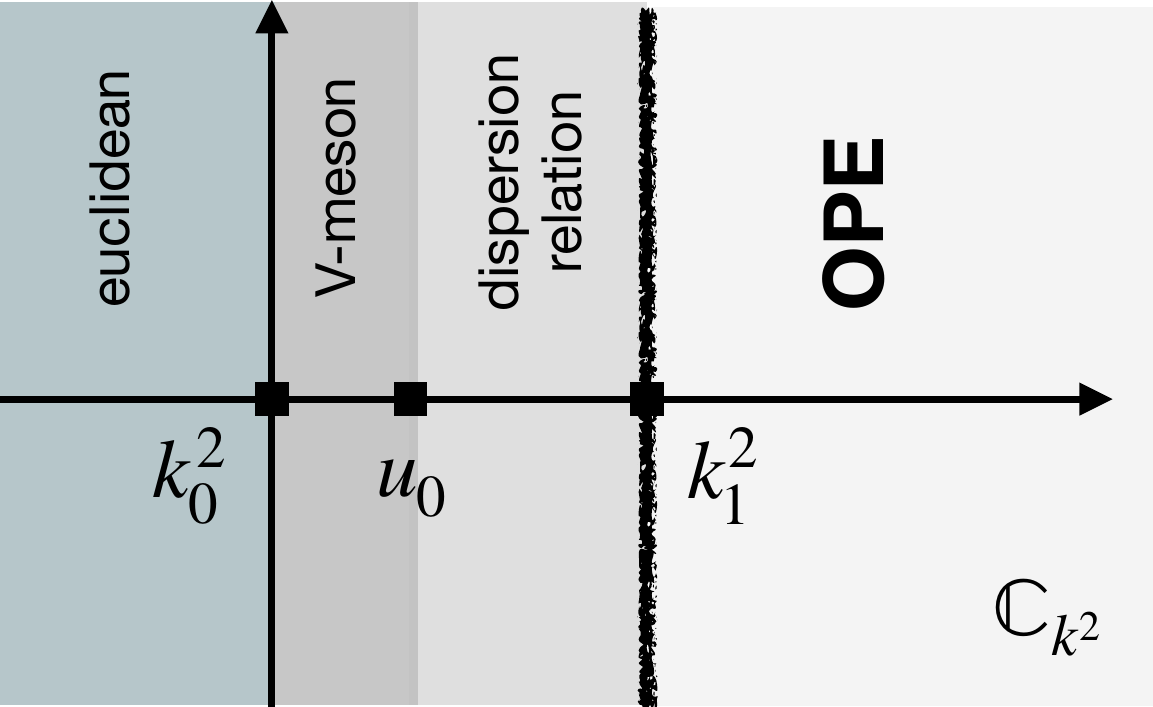}
	\caption{At the top left and right we plot the $P^*(k^2)$ FF as an example in the region
	between the first and the second subtraction point $\ksub=0 \GeV^{\,2}$ and $\ksubtwo=5\GeV^{\,2}$
	for the improved FF ($P^*$), the subtracted dispersion relation ($P_{12}^{*\textrm{dis}}$)
	and the QCD-SR version ($P^{*\textrm{OPE}}$).  As expected 
	the QCD-SR version, which is based on the
	asymptotically valid OPE, does not satisfy the constraint $P^*(0)=0$.  Other than at this point
	the QCD-SR and the dispersion relation version are rather similar where the improved version is very different of course because of the inclusion of $\phi$-meson. 
	 On the bottom part we show the matching region on the left and  a plot showing how the dispersion relation differs from the QCD-SR version for $k^2 \gg 5\GeV^{\,2}$ which
is again expected since the polynomial nature of the subtraction points will invalidate its use as well as the on-set of the condensate terms. On the right we show the imaginary part where by construction 
$\textrm{Im} (  P^{*\textrm{OPE}})  =  \textrm{Im} (P_{12}^{*\textrm{dis}})$  and 
the improved version is identical above the second matching point.
	\label{fig:regions}}
\end{figure}

The QCD SR results \eqref{eq:dens} of the last section are valid for $k^2 > {\cal O}(\kOPE \GeV^{\,2})$.
They are not valid below, since the low-lying vector meson resonances such as the $\phi$-meson,
in the $B_s$-case, distort the amplitude considerably. Conversely the $1/k^2$-factors of the condensates,
which mimic these effects in the region of validity become singular in this limit.
On the other hand a fair amount is known on the FFs in that region, as  previously discussed,
cf.  appendix \ref{app:dispersion} and \eqref{eq:examples} in particular, including  the on-shell $B\to \ga$
FFs and the $B_s \to \phi$ FFs.
We advocate that optimal use of this knowledge can be made by using a multiple subtracted dispersion
relation  with input of this hadronic data and the use of the OPE from the QCD SR for sufficiently
large $k^2$. This is  illustrated in \FIG\;\ref{fig:regions}.

We first begin by discussing the dispersion relation.
We write the generic FF, denoted by $F$,  as the sum of subtraction terms and the dispersion integral
\begin{eqnarray}
F(k^2) =  F_{12}^{\textrm{sub}}(k^2)+ F^{\textrm{dis}}_{12}(k^2)
\end{eqnarray}
where the subscript $``12"$ refers to a single subtraction point at $\ksub$ and a double subtraction
point at $\ksubtwo$. The double subtraction ensures that even the derivatives is continous at the matching point. The subtraction term is given by
\begin{equation}
\label{eq:F12sub}
 F_{12}^{\textrm{sub}}(k^2)  = U(k^2,\ksubtwo,\ksub)^2 F(\ksub) +U(k^2,\ksub,\ksubtwo)\big( \frac{\ksub - 2 \ksubtwo+k^2}{\ksub-\ksubtwo} F(\ksubtwo) + (k^2-\ksubtwo) F'(\ksubtwo) \big)  \;,
\end{equation}
where the prime denotes the derivative, $U(a,b,c) \equiv \frac{a-b}{c-b}$, and the dispersion integral reads
\begin{equation}
\label{eq:F12dis}
  F^{\textrm{dis}}_{12}(k^2) =  \frac{1}{\pi} \int^\infty_{\textrm{cut}} \frac{du \,\textrm{Im}[ F(u)]  }{u-k^2-i0}
  w_{12}( k^2,u)\;, \quad  w_{12}( k^2,u) = U(k^2,\ksub,u)  U(k^2,\ksubtwo,u)^2 \;,
\end{equation}
where it was used  that $F$ is  real somewhere on the real axis so that Schwartz's reflection principle applies.
We emphasise that these  expressions
are exact as it is in essence based on partial fraction decomposition.

Let us  turn to the specific treatment  proposed.
The first step is to write, in close analogy with \eqref{eq:examples},\footnote{This time we write the finite
width as we are interested in numerics and not only the formal relations between FFs. This presentation
could be further improved by writing the finite $\phi$-meson as resulting from a dispersion integral, the effects are though negligible as the width is rather narrow.}
\begin{equation}
\label{eq:dis-QHD}
 F^{\textrm{dis}}_{12}(k^2) \to F^{\textrm{dis-QHD}}_{12}(k^2)  = \frac{ r^\phi_F}{ m_\phi^2-  k^2  - i m_\phi \Gamma_\phi}  w_{12}( k^2,m_\phi^2) +
\frac{1}{\pi}  \int^\infty_{u_0} \frac{du \, \textrm{Im}  F^{\textrm{PT}}(u) }{u-k^2-i0}  w_{12}( k^2,u) \;,
\end{equation}
where the the acronym QHD stands for quark-hadron duality in the sense
of semi-global quark-hadron duality, used in QCD SR,
\begin{equation}
  \int^\infty_{\textrm{cut}} \frac{du \, \textrm{Im}  F(u)}{u-k^2-i0}  w_{12}( k^2,u)  \approx
    \int^\infty_{u_0} \frac{du \, \textrm{Im}  F^{\textrm{PT}}(u)}{u-k^2-i0}  w_{12}( k^2,u) \;,
\end{equation}
with the difference that there is no Borel transformation but subtraction terms instead.
The superscript PT stands for perturbation theory whereby we mean the SR computation
where the condensates are dropped.
The quantity $u_0$  is an effective threshold marking the onset of multiple particle states and
the  $\phi(1680)$-meson (and the $\rho'/\omega'$ etc in the $B_d$-case). The residue $r^\phi_{F}$ is specific to the FF and can be inferred from
the equations in appendix \ref{app:dispersion}.  Finally we can make the outline sketched at the beginning of the section  concrete\footnote{If one wanted to continue the representation  into the euclidean region
then one could do a further matching at say $k^2 = -3\GeV^{\,2}$ and use the OPE SR result for lower values.
For the  extension above $4 m_b^2$ one would need to introduce the $\Upsilon$-resonances.}
\begin{equation}
\label{eq:improvedFF}
\renewcommand{\arraystretch}{1.5}
F(k^2) =  \left\{  \begin{array}{ll}  F^{\textrm{hdis}}(k^2)
    &  0 < k^2 < \ksubtwo  \\[0.5cm]
F^{\textrm{OPE}}(k^2)   &  \ksubtwo  <    k^2 \ll 4 m_b^2      \end{array}   \right. \;,
\end{equation}
where $F^{\textrm{hdis}}(k^2)  \equiv F_{12}^{\textrm{sub}}(k^2) + F_{12}^{\textrm{dis-QHD}}(k^2)$,
and the acronym ``hdis" stands for hadronic dispersion relation (which is  the correct approach when used in the region of discontinuities), and the OPE superscript  corresponds to the QCD-SR expressions found in \eqref{eq:dens} which is based
on the perturbative and the condensate contributions.
In the following we refer to \eqref{eq:improvedFF} as the improved FF.


\subsubsection{Some detail on the double dispersion relation including the numerical input}

So far we have been somewhat formal on how to obtain  the concrete dispersion relations
within our specific computation. We first give the recipe for the explicit integrals before stating
the sources of numerical input.

We may parameterise the densities in \eqref{eq:dens} as follows
$\rho_F = - q_F L_q  - b_F L_b + \textrm{rest} $, and then  its discontinuity $\rho_D$, which is
formally a double discontinuity, is defined and given by\footnote{As previously stated the condensates are not to be included. They enter in the asymptotic formula in the OPE-region and for the matching at $\ksubtwo$ of course.}
\begin{eqnarray}
\label{eq:rhoD}
\rho^D_F(s,u) &\;=\;&  \frac{1}{2\pi i}  \big( \rho(s,0,u+i0) -\rho(s,0,u-i0 ) \big)   \nonumber \\[0.1cm]
 &\;=\;& q_F \big( \Theta(u) - \Theta[u - (s-m_b^2)] +
b_F \Theta[u-s^2/(s-m_b^2) \big) \;,
\end{eqnarray}
where $q^2=0$ is not shown explicitly for brevity. The subscript $q$ and $b$ indicate whether
the photon is emitted from a $q=d,s$ or  a $b$ quark. The purely dispersive $F^{\textrm{dis}}_{12}$  \eqref{eq:F12dis}
reads
\begin{alignat}{2}
& F^{\textrm{dis}}_{12}(k^2) &\;=\;&   p \int_0^{\infty} \frac{du}{u - k^2-i0} \int_{u+m_b^2}^{s_0}ds   e^{-\frac{s}{M^2}}  \rho^D_F(s,u) w_{12}( k^2+i0,u)  \\[0.1cm]
& &\;=\;& p \int_{m_b^2}^{s_0}ds  e^{-\frac{s}{M^2}} \left(
 \int_0^{s-m_b^2 }  \!\!\!  \frac{du\, w_{12}( k^2+i0,u)}{u - k^2-i0}  q_F(s,u)  +
  \int_{\frac{s^2}{s-m_b^2} }^{\infty}  \!\!\!  \frac{du\,  w_{12}( k^2+i0,u)}{u - k^2-i0}  b_F(s,u)  \right) \;,
  \nonumber
  \end{alignat}
where  $p \equiv \frac{e^{\mBq^2/M^2}}{\fBq \mBq^2} Q_b$ is a common prefactor. We have verified numerically that these relations work for all five off-shell FFs.
The improved version in the resonance region is then given by
\begin{equation}
\label{eq:hdisB}
 F^{\textrm{hdis}}(k^2) = F_{12}^{\textrm{sub}}(k^2) + F^{\textrm{dis}}_{12}(k^2)  +   [\de F_{12}(k^2) ]_{u_0}
\end{equation}
with the  improvement term
\begin{equation}
\label{eq:deu0}
 [\de F_{12}(k^2) ]_{u_0} =   \frac{ r^\phi_F}{ m_\phi^2-  k^2  - i m_\phi \Gamma_\phi}  w_{12}( k^2,m_\phi^2)  - [F^{\textrm{dis}}_{12}]_{u_0} (k^2)  \;,
\end{equation}
where
\begin{alignat}{2}
& [F^{\textrm{dis}}_{12}]_{u_0} (k^2)  &\;=\;&   p \int_0^{u_0} \frac{du}{u - k^2} \int_{u+m_b^2}^{s_0}ds   e^{-\frac{s}{M^2}}  \rho^D_F(s,u) =  p \int_{m_b^2}^{s_0}ds  e^{-\frac{s}{M^2}}  \!\!\!  \!\!\! \!\!\!  \!\!\!
 \int_0^{\textrm{min}(u_0,s-m_b^2) }  \!\!\!  \!\!\! \!\!\!  \frac{du\,w_{12}( k^2,u) }{u - k^2}  q_F(s,u)  \;,
  \end{alignat}
  where the emission from the $b$-quark has not support in the $[0,u_0]$ interval since the cut only starts
  at $4m_b^2 \gg u_0$.
This completes the formal recipe. However, for the numerical implementation, replacing \eqref{eq:hdisB} by
 the following expression
\begin{equation}
\label{eq:better}
 F^{\textrm{hdis}}(k^2)  \;=\;
 \de F_{12}^{\textrm{sub}}(k^2) + F^{\textrm{PT}}(k^2) +  [\de F_{12}(k^2) ]_{u_0}   \;,
\end{equation}
is even better
suited as it avoids an integral to infinity.
Above $F^{\textrm{PT}}$ is the SR expression without condensates, $[\de F_{12}]_{u_0}$ as in
\eqref{eq:deu0} and $\de F_{12}^{\textrm{sub}}$ is given by
\begin{equation}
\de F_{12}^{\textrm{sub}}(k^2) =
F_{12}^{\textrm{sub}}(k^2)\big|^{F(\ksubtwo)\to F^{\bar qq}(\ksubtwo) }_{F(\ksub) \to F^{B \to \ga}(\ksub)  - F_{\textrm{PT}}(\ksub)} \;,
\end{equation}
with $F_{12}^{\textrm{sub}}$ given in \eqref{eq:F12sub}.  The expression \eqref{eq:better},
which is formally equivalent to the former one, make the continuity of the matching condition
up to the first derivative manifest,
\begin{alignat}{2}
&  F^{\textrm{hdis}}(\ksubtwo)    &\;=\;&
F^{\textrm{PT}}(\ksubtwo)  + F^{\bar qq} (\ksubtwo)  = F^{\textrm{OPE}}(\ksubtwo)  \;,  \nonumber \\[0.1cm]
& F^{'\textrm{hdis}}(\ksubtwo)  &\;=\;&
 F^{'\textrm{PT}}(\ksubtwo)  + F^{'\bar qq} (\ksubtwo)  = F^{'\textrm{OPE}}(\ksubtwo)  \;.
\end{alignat}
Above we used $[\de F_{12}(\ksubtwo)]_{u_0} = 0$ and the prime denotes again the derivative.

The improved version is absolutely necessary 
when considering a decay rate as then the resonance, cf. \FIG\,\ref{fig:regions},
is effectively squared and one will
otherwise not get a correct expression. There is no quark-hadron duality at the level of exclusive decay rates!
However,  quark-hadron duality applies at the level of amplitudes since they obey dispersion relations. 
One can therefore evaluate them away from the resonance region. Within the resonance region, a
reasonable approximation is obtained when averaged over a sufficiently large interval that does not end 
in a resonance region. 
Thus averaging  the FF  over   $[0,\ksubtwo]$ in $d k^2$ should provide reasonable agreement 
between the QCD-SR and improved FF.  Indeed we find that  in all cases this average well within a factor of two. This  is remarkable when one inspects the shapes and considers the uncertainties of all the numerical input per se. We stress that it is the improved FF that is expected to give the reliable average and 
the difference to the QCD-SR result is another reason for introducing the improved FF.

We now turn to the required numerical input.
For $F_{12}^{\textrm{sub}}(k^2)$ we just need the matching points.
For the first subtraction point we choose $\ksub = 0$ since the FFs are known at this point from $B \to \ga$ at zero momentum
transfer in the weak matrix element. For the five FFs at hand they are given by
\begin{alignat}{1}
& P^*(0,0) = 0 \;, \quad T^*_\perp(0,0) = T_{\mathbb{L}}(0,0) =  T_\perp^{B \to \ga}(0)    \;, \quad
 V_{\perp,\parallel}^*(0,0) = V^{B \to \ga}_{\perp,\parallel}(0)  \;,
\end{alignat}
where  concretely  $T_\perp^{B_{(d,s)}  \to \ga}(0) = (0.130(13),0.160(15))$,
$V^{B_{(d,s)} \to \ga}_{\perp}(0)   = (0.079(9),0.105(8))$ and
$V^{B_{(d,s)} \to \ga}_{\parallel}(0)   = (0.129(13),0.153(14))$  \cite{Janowski:2021yvz} are used.
The second matching point is taken at $\ksubtwo = 5 \GeV^{\,2}$ which is far enough above $u_0 = 1.7(2) \GeV^{\,2}$ for the $B_s$-case  ($u_0 = 1.5(2)\GeV^{\,2}$ for the $B_d$-case) as well as high enough to trust the OPE even when analytically continued to Minkowski space.

The term  $F^{\textrm{dis-QHD}}_{12}(k^2)$ in \eqref{eq:dis-QHD} consists of the pole
term and the dispersion integral. The data entering the pole part is, of course, the same as in
\eqref{eq:examples} and consists of the pole data: masses and decay widths as well
as the residue which is a product of the decay constant and the $B \to \phi [\rho,\omega]$ FFs at zero momentum transfer. The former are determined in large from experiment and taken from the analysis
in appendix C in \cite{BSZ15}. The FFs are taken from the same reference which consists of a NLO
LCSR analysis. 

\subsubsection{Plots of the off-shell form factors}

\begin{figure}
	 \includegraphics[scale=0.8]{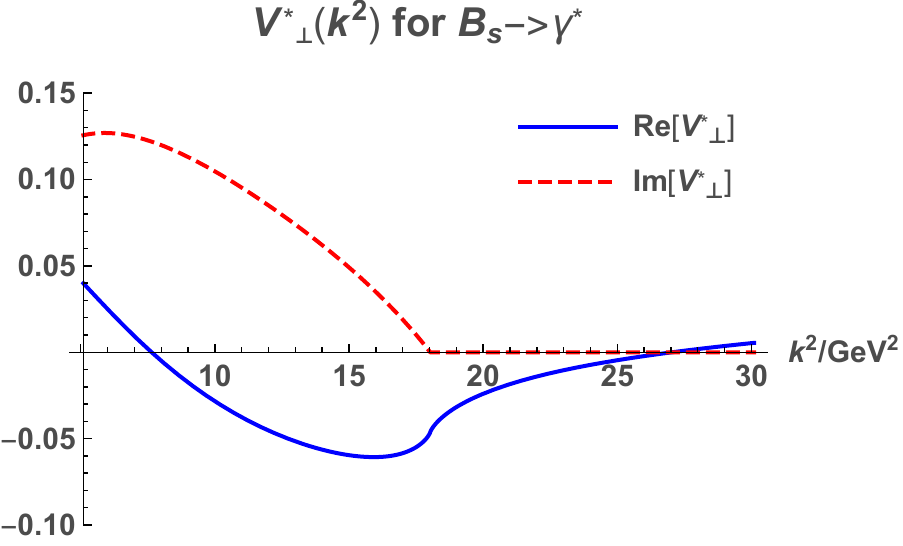}
	\includegraphics[scale=0.8]{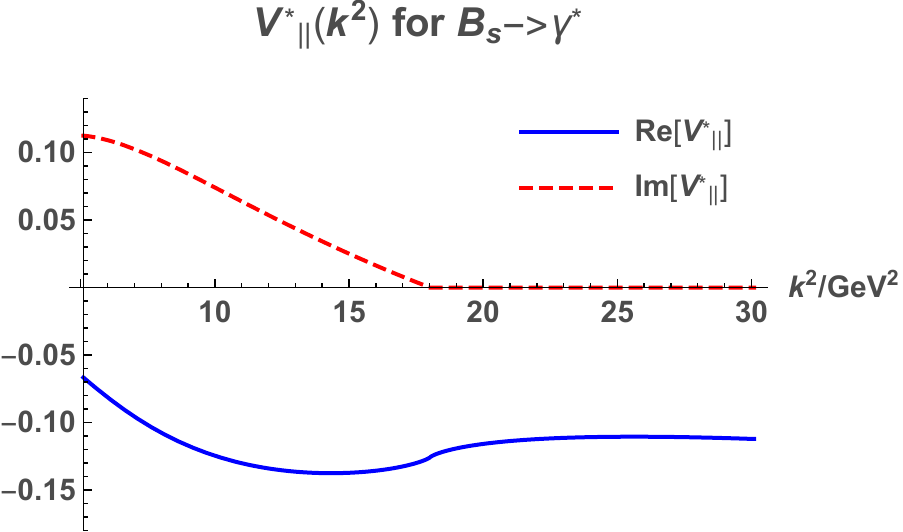}  \\
		\includegraphics[scale=0.8]{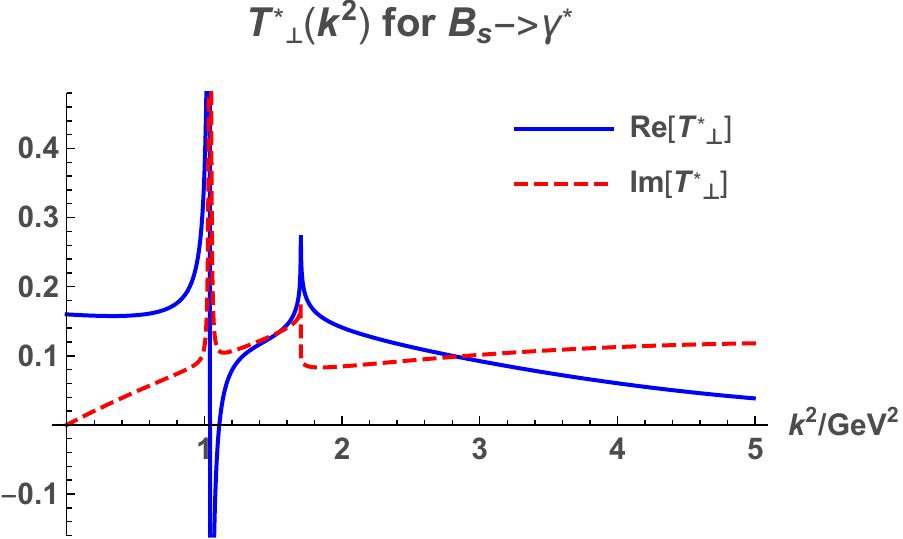}
	\includegraphics[scale=0.8]{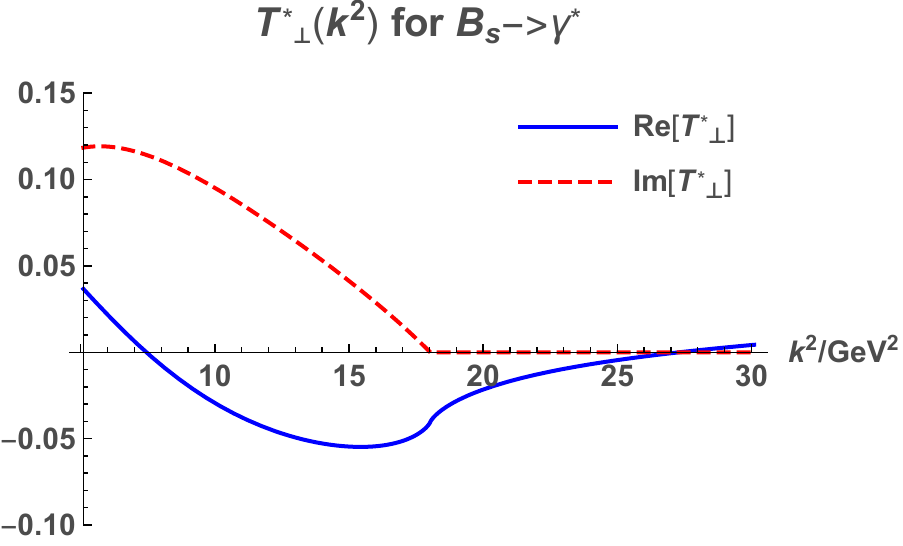}  \\
	\includegraphics[scale=0.8]{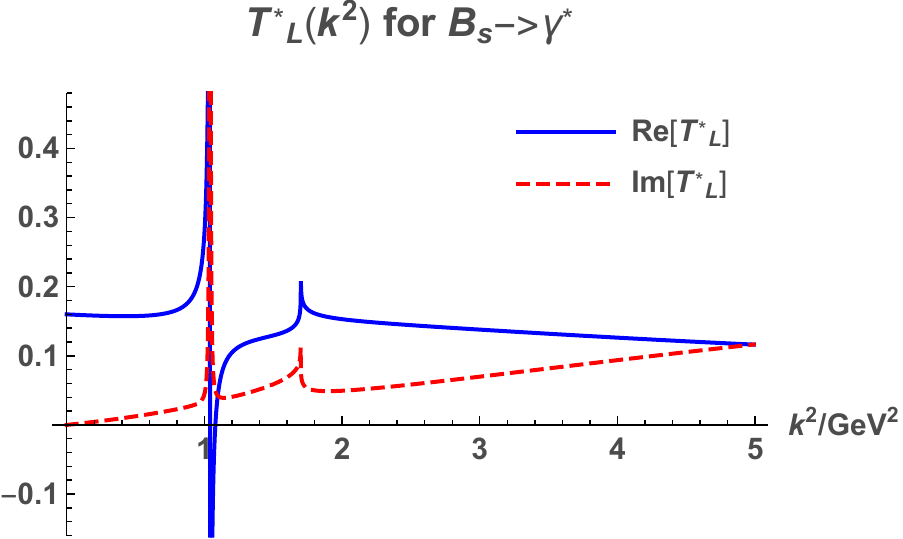}
	\includegraphics[scale=0.8]{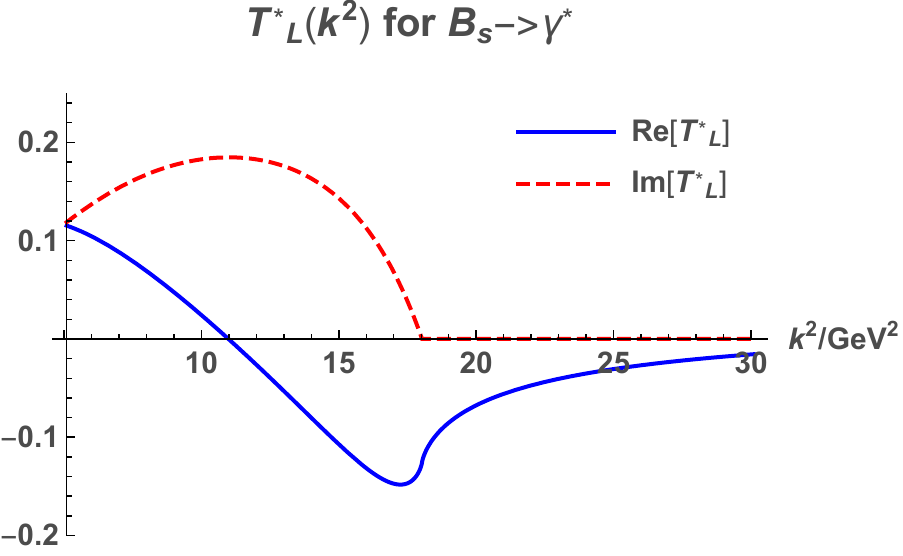}   \\
	\includegraphics[scale=0.8]{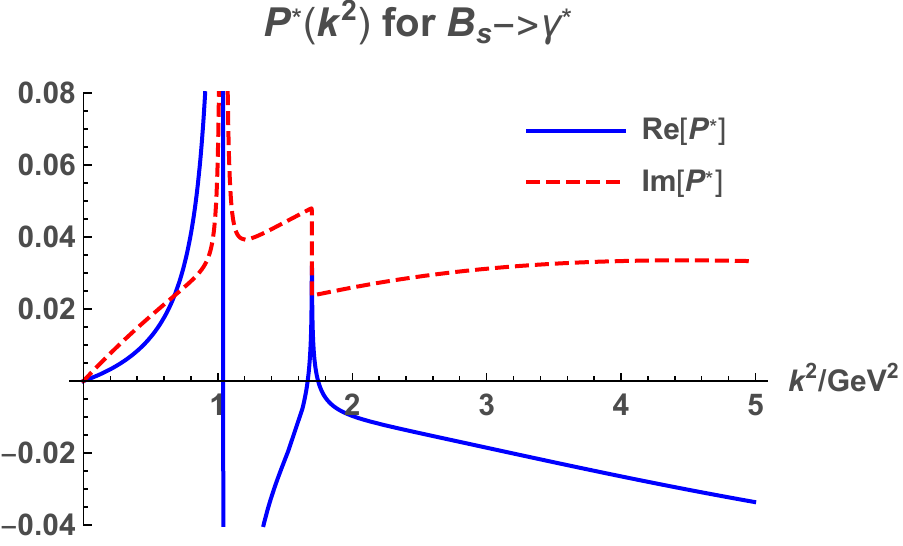}
	\includegraphics[scale=0.8]{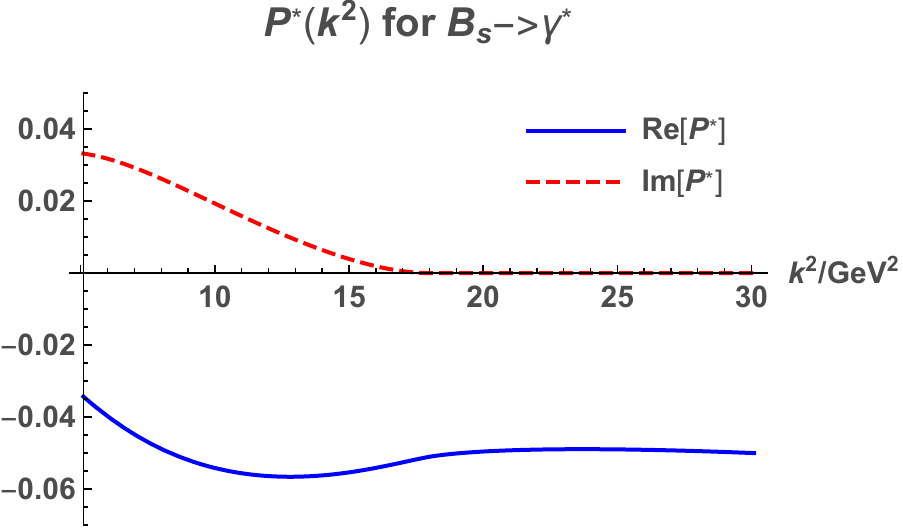}
	\caption{Plots of the five FFs in the OPE region $[5,30]\GeV^{\,2}$ with
	formulae in \eqref{eq:dens} and denoted by $F^{\textrm{OPE}}$.
	In three cases (the $V_{\perp,\parallel}$ case are similar to
	$T_{\perp,L}$ in that region)  we plot the $[0,5]\GeV^{\,2}$-region   where the vector meson
	resonance is visible.
		For $P^*$ the reader can find more plots in \FIG\;\ref{fig:regions}.
	\label{fig:FFplots}}
\end{figure}

Plots of the five FFs are shown in \FIG\;\ref{fig:FFplots} in the non-resonant region 
and three examples in the resonant region.  
Whereas all of these FFs asymptotically tend to a constant due to the condensate term,
as discussed in footnote \ref{foot:asymptotic}, $T^*_{\perp,L}$ and $V_{\perp}$ tend towards
zero since there is a cancellation for $k^2 \approx m_b^2$. The imaginary
part drops to zero at around $k^2 = s_0 - m_b^2$ as can  be inferred from \eqref{eq:rhoD}.
Both effects will presumably be lifted at NLO. The partonic computation acquires an imaginary part 
at $k^2 > 4 m_b^2$ which is however far to the left of the plot-window and thus not visible.  
In the hadronic region the single resonance is used together with a continuum threshold. 
The later results in a $\ln (u_0-k^2)$ and is representative of two particle thresholds
and effects from higher resonances and the $\phi'$. Obviously in the vicinity of $u_0$
the result is to be used in the sense of bins only.

\subsection{Dispersion relation and fit ansatz for form factors \label{app:FFp}}

\subsubsection{Extending the $B \to \ga$  on-shell   form factors into the
$q^2 \approx m_B^2$-region\label{app:poles}}

The  $B \to \ga$   on-shell FFs $F(q^2) \equiv F^*(q^2,0)$, cf. \eqref{eq:OnS}, taken from the NLO
LCSR analysis \cite{Janowski:2021yvz}. The region of validity of the computation is the previously mentioned
 $q^2 < m_b \Lambda$ which is just outside our region of interest $q^2 \in [\qsqlow, m_{B_{d,s}}^2]$.
Progress can be made with the help of the generally valid dispersion representation
in the flavour-violating momentum transfer $q^2$
\begin{alignat}{2}
\label{eq:disp}
& \FV(q^2)  &\;=\;&  \frac{1}{\pi} \int_{\textrm{cut}}^\infty \frac{  \textrm{Im} [ \FV(t) ]\, dt}{t-q^2 -i0}  =
\frac{r_{\FV}}{1- q^2/m_{B^*_q}^2} + \dots  \;, \nonumber \\[0.1cm]
& \FA(q^2)  &\;=\;&  \frac{1}{\pi} \int_{\textrm{cut}}^\infty \frac{   \textrm{Im} [ \FA(t) ]\, dt}{t-q^2 -i0}  =
\frac{r_{\FA}}{1- q^2/m_{B_{q1}}^2} + \dots  \;,
\end{alignat}
where the dots stand for higher resonances and multiparticle states. The values
and quantum numbers of the resonances are collected in \TAB~\ref{tab:Res}.
The dispersion relations
of the other FFs are analogous.  The residua are related to the $B_q^* \to B_q \ga$  and
$B_{q1} \to B_q \ga$
on-shell matrix elements respectively. Unfortunately they are not known from
experiment.\footnote{The width of the $B_{d,s}^*$-mesons are unknown and the $B_{(d,s)1}$ mesons
are dominated by the strong decays to $B_{(d,s)} \pi$.}
They  can be extracted from the same SR as the FFs themselves by
applying a double dispersion relation
to interpolate for the $B_q^*$- and $B_{q1}$-meson respectively. We take the NLO result of this residue from
\cite{Pullin:2021ebn}, collected  in \TAB~\ref{tab:residue}.
\begin{table}
\addtolength{\arraycolsep}{3pt}
\renewcommand{\arraystretch}{1.4}
$$
\begin{array}{ l   r r   r r  }
                     & r_{V_\perp}  \propto g_{B_q^* B_q \ga}  & r_{T_\perp} \propto g_{B_q^* B_q \ga}&   r_{V_\parallel} \propto g_{B_{q1} B_q \ga}&   r_{T_\parallel} \propto g_{B_{q1} B_q \ga}    \\ \hline \hline
B_d \to \ga     &  \Verr{\phantom{-}0.179}{0.019}{0.019} & \Verr{0.171}{0.018}{0.020}  & \Verr{0.076}{0.015}{0.016}  & \Verr{0.104}{0.015}{0.015}    \\[0.1cm]
B_s   \to \ga   &    \Verr{\phantom{-}0.235}{0.024}{0.025} & \Verr{0.224}{0.023}{0.024}  & \Verr{0.114}{0.016}{0.018}  & \Verr{0.146}{0.017}{0.017}  \\
\hline
\end{array}
$$
\caption{The residua of the poles at $k^2 = m_{B_q^*}^2 , m_{B_{q1}}^2$ proportional to the
on-shell matrix  of an NLO computation \cite{Pullin:2021ebn}.
}
\label{tab:residue}
\end{table}

Here, we make the link to the predictions of Ref.~\cite{Aditya:2012im},
for which  a single $m_{B_s^*}$-pole approximation was employed to
estimate the FFs for the radiative decay.
The single-pole approximation is expected to give a reasonable
approximation around the pole provided the residue is known sufficiently well.
By identifying the defining matrix elements of the residue (cf. Eq.(6) in Ref.~\cite{Aditya:2012im})
we find the relation
\begin{equation}
\label{eq:poleId}
   |r_{\FV^{B_s \to \ga}}|      =   |\mu| f_{B_s}|_{\mbox{\cite{Aditya:2012im}}}  \approx 0.265 \;,
\end{equation}
with $f_{B_s} = 227 \textrm{MeV}$ \cite{PDG18} the standard decay constant and $|\mu|$ defines the strength
of the on-shell matrix element in Ref.~\cite{Aditya:2012im}.
The authors of Ref.~\cite{Aditya:2012im}  determine $|\mu| = 1.13\, \GeV^{\,-1}$
in an effective-theory approach valid at leading order in $1/m_{b,c}$ using experimental
data from $D^{*+} \to D^+ \gamma$ and $D^{*0} \to D^+ \pi^-$.
They neglect the pole of the $B_{s1}$ meson (cf. \TAB~\ref{tab:Res}) and thus we
cannot compare the $|r_{\FA}|$ residua to theirs.
Given the methods employed on both sides the agreement  of $0.235(23)$ and
$0.265$ is presumably somewhat accidental. Whereas the former is LO in the coupling with preliminary error analysis,
the latter is subject to $1/m_c$ corrections which might well be sizeable.

At last let us mention that we performed a non-trivial test of the identification in Eq.~\eqref{eq:poleId}.
Approximating our FF-expression to the pole part, inserting it into our rate in
Eq.~\eqref{eq:GaBsmumuga}, and then comparing to the rate in
Ref.~\cite{Aditya:2012im} (cf. their Eq.~(25)),
we can confirm that Eq.~\eqref{eq:poleId} is consistent with both rates.
This is a strong hint of the correctness of the treatment in our work and theirs.

\subsubsection{The dispersion representation of the $B \to \ga^*$  off-shell   form factors \label{app:poles2}}

The assumed $q^2 =0$ is well below the various  $m_B^2$-type poles and does not affect the computation.
However,  in the variable $k^2$ there are  the previously mentioned
$\rho/\omega$ ($\phi$)- and $\Upsilon$-resonances (cf. \TAB~\ref{tab:Res}) which are far away from our region of interest $k^2 \approx m_B^2$ and therefore have little impact. If one wanted to fit the FFs at  lower $k^2$ then
a dispersion ansatz, e.g. \eqref{eq:examples}, could be combined with the $z$-expansion.

\begin{table}
\addtolength{\arraycolsep}{3pt}
\renewcommand{\arraystretch}{1.4}
$$
\begin{array}{ l  rrr  }
 &  \mBq = m_{0^-}&       m_{B^*} = m_{1^-}  = m_\perp &   m_{B_{q1}} = m_{1^+} = m_\parallel \\ \hline\hline
{b \to s}       &  5.367\, \textrm{GeV}   &   5.415\, \textrm{GeV}& 5.829\, \textrm{GeV} \\
{b \to d,u}      &   5.280 \, \textrm{GeV}  & 5.325\, \textrm{GeV} & 5.726\, \textrm{GeV} \\
\hline
\end{array}
$$
\caption{Lowest resonance masses for FFs   \cite{PDG18}. The mass $m_{0^-}$
is the $B$-meson mass in  $B_q \to \ga$.
The Form Factors $V,\FTV$ and $V,T_{\parallel}$ are associated with the resonances $J^P = 1^-$
and $1^+$ respectively.
}
\label{tab:Res}
\end{table}

\subsubsection{Fit ansatz and $z$-expansion\label{app:zexpansion}}

The procedure to fit the FFs and how to include the correlation of uncertainties
largely follows Ref.~\cite{BSZ15}. Based on the previous part of this section let us
first motivate the fit-ansatz before  summarising the essence of the $z$-expansion.
There are four on-shell FFs and at $q^2 =0$ there are five off-shell FFs,
\begin{alignat}{2}
& \textrm{on-shell: }  \quad &  &\{V^{B \to \ga} _{\perp,\parallel}(q^2),T^{B \to \ga} _{\perp,\parallel}(q^2)\} \;, \nonumber
\\[0.1cm]
& \textrm{off-shell: } \quad   &  &\{P^{B \to \ga^*}(0,k^2)\;, V^{B \to \ga^*} _{\perp,\parallel}(0,k^2),\;
T^{B \to \ga^*} _{\perp,\LL}(0,k^2)\}  \;.
\end{alignat}

\begin{itemize}
\item The on-shell FFs are parameterised
\begin{equation}
\label{eq:Fon}
    F_n^{B \to \ga}(q^2) = \frac{1}{1 -q^2/m_R^2}\left( \alpha_{n0}  + \sum_{k=1}^N \alpha_{nk}(z(q^2)-z(0))^k \right) \;,
\end{equation}
using the knowledge of the presence of the first  pole $m_R$ \eqref{eq:disp}, cf., \TAB~\ref{tab:Res}.
The remaining part in brackets are supposed to take into account higher states in the spectrum.
Specifically the $\al_{nk}$-coefficients are to be determined from a fit and
$z(q^2)$ is defined further below.
The  constraint of  the residue, cf.  \eqref{eq:disp} and \TAB~\ref{tab:residue}, is implemented by
\begin{equation}
r_{V_\perp} =   \al_{V_\perp 0}  + \sum_{k=1}^N \al_{V_\perp k}(z(m_{B^*_q}^2)-z(0))^k \;,
\end{equation}
and similarly for other FFs. Further to that the  constraint \eqref{eq:algebraic} is imposed by
\begin{equation}
T^{B \to \ga}_{\perp}(0) = T^{B \to \ga}_{\parallel}(0) \;\; \Leftrightarrow  \;\; \al_{T_{\perp}0} = \al_{T_{\parallel}0} \;.
\end{equation}

\item The off-shell FFs are simply parameterised by
\begin{equation}
\label{eq:Foff}
    F_n^{B \to \ga^*}(0,k^2) =  \alpha_{n0}  + \sum_{k=1}^N \alpha_{nk}(z(k^2)-z(0))^k  \;,
\end{equation}
The  constraint $ \FA^*(0,\mBq^2)  =  2\PFF^*(0,\mBq^2)$ \eqref{eq:third2} is imposed
\begin{equation}
\al_{V_\parallel^* 0}  + \sum_{k=1}^N \al_{V_\parallel^* k}(z(m_{B^*_q}^2)-z(0))^k =
2\big( \al_{P^* 0}  + \sum_{k=1}^N \al_{p^* k}(z(m_{B^*_q}^2)-z(0))^k\big) \;.
\end{equation}
The fit-ansatz \eqref{eq:Foff} could easily be improved including the information on
the $\rho/\omega$ ($\phi$)-like resonances from the dispersion representation \eqref{eq:examples}.\footnote{The extension to fit
the two-variable FF
$F_n^{B \to \ga^*}(q^2,k^2)$ is not straightforward but one would best proceed by building
an ansatz from a double dispersion relation in $q^2$ and $k^2$ and in addition force
the constraints (\ref{eq:A03},\ref{eq:algebraic},\ref{eq:third}).}
\end{itemize}

Let us now describe the $z$-expansion in order to remain self-consistent.
The function $z(t)$ is defined by
\begin{equation}
    z(t) = \frac{\sqrt{t_+ - t} - \sqrt{t_+ - t_0}}{\sqrt{t_+ - t} + \sqrt{t_+ - t_0}} \;,
\end{equation}
where  $t_0\equiv t_+(1-\sqrt{1-t_-/t_+})$ and $t_\pm \equiv (\mBq \pm m_\rho)^2$.
The $\rho$-mass, $m_\rho = 770\MeV$, is just an arbitrary reference scale  and the values of $\mBq$ are given in \TAB~\ref{tab:Res}.

 The coefficients $\al_{nk}$  are determined by fitting
$N=200$ random points at each integer value of $q_i^2$ (in $\textrm{GeV}^2$-units)
in a specific interval.
Uncertainties in input parameters, $p\pm \delta p$,  as for example $m_b$, are accounted for
by sampling them with a normal distribution $N(p, \delta p)$, which  accounts for the same
correlations as in Ref.~\cite{BSZ15}.
The $N = 200$ random samples of $F_I=F_i(q_j^2)$, where $I=(i,j)$ denotes the
collective index for the FF-type and the momentum,
determine the $(ij) \times (ij)$ covariance matrix
\begin{equation}
	C_{IJ} = \langle F_I F_J \rangle - \langle F_I \rangle \langle F_J \rangle\,.
\end{equation}
Angle brackets denote the average over random samples.
The coefficients $\alpha_{nk}$ are then found by minimising the function
\begin{equation}
	\chi^2(\{\alpha\})=\sum_{IJ}(F_I^{\text{sample}} - F_I^{\text{fit}}(\{\alpha\})) C_{IJ}^{-1} (F_J^{\text{sample}} - F_J^{\text{fit}}(\{\alpha\})) \;,
\end{equation}
for each random sample, where the correlation matrix remains constant for all samples.
The fitted values of $\al$ are then averaged over all the samples and errors are calculated
from the standard deviation, which is justified because each of the samples
are statistically independent.

\begin{itemize}
\item The computation of the  four  on-shell FFs \cite{Janowski:2021yvz} are
limited to roughly $q^2 < 14 \GeV^{\,2} $.
The $200$ sample points  are generated for each integer interval in
$q^2 \in [- 5 , 14]\GeV^{\,2}$ to which the $\al_n$'s are then fitted to the ansatz \eqref{eq:Foff}.
\item Since we only need the off-shell FFs in the region $k^2 \in [\qsqlow , m_{B_{d,s}}^2]\GeV^{\,2}$
we restrict our fitting procedure to this region.
\end{itemize}

\section{The $ B_q  \to  \ell  \ell \Vc$  for searches beyond the Standard Model}
\label{app:BVllrate}

\begin{figure}[t]
	\centering
    \includegraphics[]{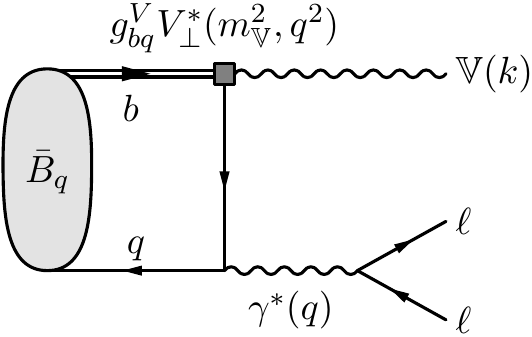}
		\caption{
	Illustration of the three-body decay of a $B_q$ to a pair of leptons and
    a new, light, beyond-the-standard-model vector $\Vc$
    induced by a flavour-violating vector coupling ($\bar b\slashed{\Vc} q$, 
    see Lagrangian in Eq.~\eqref{eq:LeffBVll}).
	\label{fig:BVll}}
\end{figure}

 The main focus of this work has been to propose a search for the QCD axion via
 the three-body decay $B_q\to \ell\ell a$ and the computation of the 
 FFs contributing to this beyond-the-SM decay channel as well as to the radiative 
 $B_q\to \ell\ell\gamma$ decay.
 However, the computation of the off-shell FFs allows us to extend 
 this analysis to searches for light, beyond-the-SM vector bosons ($\Vc$) 
 with flavour violating couplings.
 In this appendix, we present the $B_q  \to  \ell  \ell \Vc$ rate, which
 is relevant for a search via $B_q\to\mu\mu$, and briefly discuss 
 the search's sensitivity.
 
 We parametrise the relevant flavour violating couplings via the 
 effective Lagrangian

\begin{align}
\label{eq:LeffBVll}
\Lag_\Vc & = \left(   \Vc _\mu 
\bar q
\ga^\mu \left( g^V_{b q} \frac{m_\Vc^2}{\Lambda^2} + g^A_{b q} \frac{m_\Vc^2}{\Lambda^2}  \ga_5  \right) b +  
\partial_\mu \Vc _\nu \bar q
			\sigma^{\mu\nu} 
          \left( \frac{g_{bq}^T}{\Lambda}  +  \frac{g_{bq}^{T_5}}{\Lambda}\gamma_5\right) b
 +\text{h.c.} \right)
- \frac{1}{2}m_\Vc^2 \Vc^\mu \Vc_\mu \;,
\end{align}
where $\Lambda$ is the heavy NP scale.
Above, we opted to factorise out the term $m_\Vc^2 / \Lambda^2$ in the dimension-four vector- and axial-vector couplings
to demonstrate the ``restoration'' of gauge invariance in the limit $m_\Vc\to 0$.
We further introduce the quantities 
\begin{align}
  &\hat{g}_{bq}^V     \equiv \frac{m_\Vc^2}{\Lambda^2} g_{bq}^V\,,&
  &\hat{g}_{bq}^A     \equiv \frac{m_\Vc^2}{\Lambda^2} g_{bq}^A\,,&
  &\hat{g}_{bq}^T     \equiv \frac{m_{B_q}}{\Lambda}   g_{bq}^T\,,&
  &\hat{g}_{bq}^{T_5} \equiv \frac{m_{B_q}}{\Lambda}   g_{bq}^{T_5}\,.&
  \label{<+label+>}
\end{align}
which simplify some of the intermediate formulae.

Based on Eq.~\eqref{eq:LeffBVll} we compute the $B_q  \to  \ell  \ell \Vc$ rate
using the helicity formalism.
 The $B_q  \to  \ell  \ell \Vc$ decay differs from the semileptonic and flavour-changing-neutral-current 
 decays $B \to \rho \ell \nu$ and $B \to K^* \ell \ell$ in that the vector meson is emitted 
 from the weak vertex and the lepton pair originates from an off-shell photon 
 cf. \FIG\;\ref{fig:BVll}.
The helicity amplitude for the decay reads
\begin{equation}
{\cal A}_{\la}  \propto
	 \omega^*_{\mu}(\Vc, k ,\la)  \eps_{\rho}^*(\ga^*, q, \la)
	~
	\left(
			\hat{g}^V_{bq} M^{\rho\mu}_{V}
		       +\hat{g}^A_{bq} M^{\rho\mu}_{A}
		     +  \hat{g}^{T}_{bq}M^{\rho\mu}_{T}
		       + \hat{g}^{T_5}_{bq} M^{\rho\mu}_{T_5}
	\right)
\end{equation}
where $g^{T}_{bq} \equiv  m_{B_q}/\Lambda^{T}_{bq}$ and analogous for $T_5$
as this leads to transparent formulae. 
Explicit  polarisation vectors are given in Eq.~\eqref{eq:helpol}.
The helicity amplitudes are then given by
\begin{equation}
  \begin{split}
 {\cal A}_\perp     &\;=\;   (\la^{(\Vc)}_{B_q})^{1/2} F_\perp  \;,  \\[0.1cm]
 {\cal A}_\parallel &\;=\;  (   m_{\Vc}^2-m_{B_q}^2)  F_\parallel  \;, \\[0.1cm]
 {\cal A}_0         &\;=\;  \frac{m_{\Vc} } {2\sqrt{2 q^2}  ( m_{B_q}^2 - q^2)  }  \left( \la^{(\Vc)}_{B_q}  F_\LL
- ( m_{B_q}^2 - m_{\Vc}^2)( m_{B_q}^2 - m_{\Vc}^2+ 3 q^2)  F_\parallel \right)    \;,
  \end{split}
\end{equation}
with $\la^{(\Vc)}_{B_q} = \la(m_B^2,q^2,m_\Vc^2)$ and the FFs
\begin{equation}
F_\perp = \hat{g}^V_{bq} V^*_\perp(m_\Vc^2,q^2) + \hat{g}^T_{bq} T^*_\perp(m_\Vc^2,q^2) \;, \quad
F_{\parallel,\mathbb{L}} = \hat{g}^A_{bq} V^*_{\parallel,\mathbb{L}}(m_\Vc^2,q^2) - \hat{g}^{T_5}_{bq} T^*_{\parallel,\mathbb{L}}(m_\Vc^2,q^2) \;.
\end{equation}
The minus sign in the last expression originates from the minus sign in Eq.~\eqref{eq:FFsec2}.

Our FFs provide a good description in the case where $m_\Vc$ is much smaller than 
all other scales since we have evaluated them for $m_\Vc=0$. 
Three observations on the helicity amplitudes are in order.
Firstly, in the $m_\Vc \to 0$ limit the pseudoscalar part in the longitudinal component survives 
since $V_\mathbb{L}^*(m_\Vc^2,q^2)  = - 2m_{B_q}^2/m_\Vc^2 \hat{V}^*(m_\Vc^2,q^2)$ in general 
and $\hat{V}^*(0,q^2)= P^*(0,q^2)$ in particular.
Secondly, in the $m_\Vc \to 0$ limit the part of the amplitude
proportional to $\hat{g}_{bq}^T$ happens to be identical to the part proportional
to $\hat{g}_{bq}^{T_5}$, i.e., both are proportional to the same FF, 
namely $T^*_{\perp}(0,q^2)$ (see Eq.~\eqref{eq:algebraic2}).
Finally, it is observed that at the 
kinematic endpoint $q^2 \to (m_B-m_\Vc)^2$ one has 
${\cal A}_0 = {\cal A}_+ = {\cal A}_-$ 
(where $\sqrt{2} {\cal A}_{\perp(\parallel)} = {\cal A}_+ \mp {\cal A}_-$)
as a result of the restoration spherical symmetry \cite{Hiller:2013cza}.

For the  differential rate  we find
 \begin{equation}
\frac{d \Gamma}{d q^2}(\bar B_q  \to  \ell  \ell \Vc) = \frac{\al^2 Q_\ell^2 (\la^{(\Vc)}_{B_q})^{1/2} \la_\ga^{1/2}}{24 \pi m_{B}^5 q^6     } (q^2+ 2 m_{\ell}^2)  \left( |{\cal A}_\perp|^2 +  |{\cal A}_\parallel|^2
+ |{\cal A}_0|^2 \right) \;,
 \end{equation}
 with $\la^{(\Vc)}_{B_q}$ defined above and $\la_\ga \equiv \la(q^2,m_\ell^2,m_\ell^2) = q^2(q^2-4 m_\ell^2)$.

 The sensitivity study for searching for such light, flavour-violating vectors at the tail
 of $B_q\to\mu\mu$ is analogous to the axion study presented in Section \ref{sec:sensLHCb}.
 As an illustration we show here the expected $90\%$ CL exclusion limits 
 with the full LHCb data set of $300$\,fb$^{-1}$ for the different cases
 \begin{align*}
 \frac{m_\Vc^2}{\Lambda^2}       {g}^V_{bs}  &< 9\times 10^{-7}\,,&                      \frac{m_\Vc^2}{\Lambda^2}       {g}^V_{bd} &< 4\times 10^{-6}\,,&\\[0.5em]
 \frac{m_\Vc\,m_{B_s}}{\Lambda^2} {g}^A_{bs} &< 2\times10^{-7}\,,&                       \frac{m_\Vc\,m_{B_d}}{\Lambda^2} {g}^A_{bd}&< 6\times 10^{-7}\,,&\\
 \frac{g^{T/T_5}_{bs}}{\Lambda}              &< \frac{10^{-6}}{\text{GeV}}\,,&           \frac{g^{T/T_5}_{bd}}{\Lambda}             &< \frac{3\times 10^{-6}}{\text{GeV}}\,.& 
 \end{align*}
 The bounds are computed for the case of $m_{\Vc}\to 0$ so contain only the leading in $m_{\Vc}/m_{B_q}$ term.

\addcontentsline{toc}{section}{References}
\bibliographystyle{JHEP}
\bibliography{references}

\end{document}